\documentclass{LMCS}

\def\dOi{10(2:5)2014}
\lmcsheading%
{\dOi}
{1--49}
{}
{}
{Mar.~13, 2013}
{May\phantom.~29, 2014}
{}

\ACMCCS{[{\bf Theory of computation}]: Logic}

\usepackage{hyperref}
\usepackage[latin1]{inputenc}
\usepackage{wasysym} % for \leadsto
\usepackage[all,dvips]{xy}

\usepackage{latexsym}  %\usepackage{amsmath} creates conflict
\usepackage{amssymb}
\usepackage{graphicx} % for inporting eps files
\usepackage{graphics}
\usepackage[dvips]{color}
\usepackage{psfrag}

\def\sysS {{\script S}}%
\def\merge {\mathbin{\bigdiamond}}%

\let\colder=<

\def\grammareq {\mathrel{\raise.4pt\hbox{::}{=}}}%

\def\permover{\mathchoice
             {\displaystyle
              \mathrel{\lower.493\fontdimen5\textfont3
                       \hbox{$\lhook$}%
                       \mkern-3mu{\rightarrow}}}%
             {\textstyle
              \mathrel{\lower.493\fontdimen5\textfont3
                       \hbox{$\lhook$}%
                       \mkern-3mu{\rightarrow}}}%
             {\scriptstyle
              \mathrel{\lower.54\fontdimen5\scriptfont3
                       \hbox{$\scriptstyle\lhook$}%
                       \mkern-3.2mu{\rightarrow}}}%
             {\scriptscriptstyle
              \mathrel{\lower.55\fontdimen5\scriptscriptfont3
                       \hbox{$\scriptscriptstyle\lhook$}%
                       \mkern-3.2mu{\rightarrow}}}%
             }%
\let\neg=\bar

\def\nsy #1#2#3#4{\nsysilent{#1}{#2}{#3}{#4}%
   \proofreadingcolor{$\csname #1\endcsname$}}%
\def\nsysilent #1#2#3#4{%
   \expandafter\let\csname #1\endcsname=#2%
   \sidx{#3@${#2}$, #4}}%

\def\boxtherule #1#2#3#4{\hbox{%
      \vtop{\kern0pt\hbox to0pt{\hss\structcolor{#1}\strut\enspace}}%
      \vtop{\kern0pt\structcolor{\boxit{\hbox to#2{\hfil\vbox to#3{%
                  \vfil\hbox to0pt{\hss\Black{$#4$}\hss}\vfil}\hfil}}}}}}%
\def\lowerdertobase #1{\lower
                       \ifsmallprint
                          6.89164pt
                       \else\ifverysmallprint
                          6.14815pt
                       \else
                          7.63518pt
                       \fi\fi\hbox{$#1$}}%
\def\InvisibleMark {\White{\vbox to0pt{\vss
   \hbox to0pt{\hss\vrule height1sp depth0pt width1sp}}}}%
\def\InvisibleMarkDown  #1{\kern-.#1pc\vbox to0pt{\kern.#1pc\InvisibleMark\vss}}%
\def\InvisibleMarkDDown #1{\kern-#1pc\vbox  to0pt{\kern#1pc\InvisibleMark \vss}}%
\def\InvisibleMarkUp    #1{\vbox to0pt{\vss\InvisibleMark\kern.#1pc}\kern-.#1pc}%
\def\InvisibleMarkUUp   #1{\vbox to0pt{\vss\InvisibleMark\kern#1pc}\kern-#1pc}%
\def\lv #1{\underline{#1\phantom{\hbox to0pt{,\hss}}}{}\mkern-1mu\lower1ex\hbox
                                                   {$\scriptscriptstyle\Vsss$}}%
\def\vl #1{\underline{#1\phantom{\hbox to0pt{,\hss}}}{}\lower1ex\hbox
                                                   {$\scriptscriptstyle\Lsss$}}%
\def\ls #1{\underline{#1\phantom{\hbox to0pt{,\hss}}}{}\lower1ex\hbox
                                                   {$\scriptscriptstyle\Ssss$}}%
\def\lg #1{\underline{#1\phantom{\hbox to0pt{,\hss}}}{}\lower1ex\hbox
                                                   {$\scriptscriptstyle\Gsss$}}%
\let\sl=\vl

\def\conldel {\{}%
\def\conrdel {\}}%
\def\lrgldel {\mathchoice{(}{(}{\langle}{\langle}}%
\def\lrgrdel {\mathchoice{)}{)}{\rangle}{\rangle}}%
\def\aprldel {\mathchoice
   {\mathopen {\setbox0=\hbox{$\displaystyle     \lrgldel$}\hbox to\wd0
                        {\hfil$\displaystyle     (       $\hfil}}}%
   {\mathopen {\setbox0=\hbox{$\textstyle        \lrgldel$}\hbox to\wd0
                        {\hfil$\textstyle        (        $\hfil}}}%
   {\mathopen {\setbox0=\hbox{$\scriptstyle      \lrgldel$}\hbox to\wd0
                        {\hfil$\scriptstyle      (        $\hfil}}}%
   {\mathopen {\setbox0=\hbox{$\scriptscriptstyle\lrgldel$}\hbox to\wd0
                        {\hfil$\scriptscriptstyle(        $\hfil}}}}%
\def\aprrdel {\mathchoice
   {\mathclose{\setbox0=\hbox{$\displaystyle     \lrgrdel$}\hbox to\wd0
                        {\hfil$\displaystyle     )       $\hfil}}}%
   {\mathclose{\setbox0=\hbox{$\textstyle        \lrgrdel$}\hbox to\wd0
                        {\hfil$\textstyle        )        $\hfil}}}%
   {\mathclose{\setbox0=\hbox{$\scriptstyle      \lrgrdel$}\hbox to\wd0
                        {\hfil$\scriptstyle      )        $\hfil}}}%
   {\mathclose{\setbox0=\hbox{$\scriptscriptstyle\lrgrdel$}\hbox to\wd0
                        {\hfil$\scriptscriptstyle)        $\hfil}}}}%
\def\seqldel {\mathchoice
   {\mathopen {\setbox0=\hbox{$\displaystyle     \lrgldel$}\hbox to\wd0
                        {\hfil$\displaystyle     \langle  $\hfil}}}%
   {\mathopen {\setbox0=\hbox{$\textstyle        \lrgldel$}\hbox to\wd0
                        {\hfil$\textstyle        \langle  $\hfil}}}%
   {\mathopen {\setbox0=\hbox{$\scriptstyle      \lrgldel$}\hbox to\wd0
                        {\hfil$\scriptstyle      \langle  $\hfil}}}%
   {\mathopen {\setbox0=\hbox{$\scriptscriptstyle\lrgldel$}\hbox to\wd0
                        {\hfil$\scriptscriptstyle\langle  $\hfil}}}}%
\def\seqrdel {\mathchoice
   {\mathclose{\setbox0=\hbox{$\displaystyle     \lrgrdel$}\hbox to\wd0
                        {\hfil$\displaystyle     \rangle  $\hfil}}}%
   {\mathclose{\setbox0=\hbox{$\textstyle        \lrgrdel$}\hbox to\wd0
                        {\hfil$\textstyle        \rangle  $\hfil}}}%
   {\mathclose{\setbox0=\hbox{$\scriptstyle      \lrgrdel$}\hbox to\wd0
                        {\hfil$\scriptstyle      \rangle  $\hfil}}}%
   {\mathclose{\setbox0=\hbox{$\scriptscriptstyle\lrgrdel$}\hbox to\wd0
                        {\hfil$\scriptscriptstyle\rangle  $\hfil}}}}%
\def\parldel {\mathchoice
   {\mathopen {\setbox0=\hbox{$\displaystyle     \lrgldel$}\hbox to\wd0
                        {\hfil$\displaystyle     [       $\hfil}}}%
   {\mathopen {\setbox0=\hbox{$\textstyle        \lrgldel$}\hbox to\wd0
                        {\hfil$\textstyle        [        $\hfil}}}%
   {\mathopen {\setbox0=\hbox{$\scriptstyle      \lrgldel$}\hbox to\wd0
                        {\hfil$\scriptstyle      [        $\hfil}}}%
   {\mathopen {\setbox0=\hbox{$\scriptscriptstyle\lrgldel$}\hbox to\wd0
                        {\hfil$\scriptscriptstyle[        $\hfil}}}}%
\def\parrdel {\mathchoice
   {\mathclose{\setbox0=\hbox{$\displaystyle     \lrgrdel$}\hbox to\wd0
                        {\hfil$\displaystyle     ]       $\hfil}}}%
   {\mathclose{\setbox0=\hbox{$\textstyle        \lrgrdel$}\hbox to\wd0
                        {\hfil$\textstyle        ]        $\hfil}}}%
   {\mathclose{\setbox0=\hbox{$\scriptstyle      \lrgrdel$}\hbox to\wd0
                        {\hfil$\scriptstyle      ]        $\hfil}}}%
   {\mathclose{\setbox0=\hbox{$\scriptscriptstyle\lrgrdel$}\hbox to\wd0
                        {\hfil$\scriptscriptstyle]        $\hfil}}}}%
\def\aprs #1{\aprldel #1\aprrdel}%
\def\cons #1{\conldel #1\conrdel}%
\def\pars #1{\parldel #1\parrdel}%
\def\sqn  #1{{\turnstile #1}}%

\def\quadcm {\rlap{\quad,}}%
\catcode`@=11
\def\upsmash{\relax % \relax, in case this comes first in \halign
  \ifmmode\def\next{\mathpalette\mathupsm@sh}\else\let\next\makeupsm@sh
  \fi\next}
\def\makeupsm@sh#1{\setbox\z@\hbox{#1}\finupsm@sh}
\def\mathupsm@sh#1#2{\setbox\z@\hbox{$\m@th#1{#2}$}\finupsm@sh}
\def\finupsm@sh{\ht\z@\z@ \box\z@}
\def\downsmash{\relax % \relax, in case this comes first in \halign
  \ifmmode\def\next{\mathpalette\mathdownsm@sh}\else\let\next\makedownsm@sh
  \fi\next}
\def\makedownsm@sh#1{\setbox\z@\hbox{#1}\findownsm@sh}
\def\mathdownsm@sh#1#2{\setbox\z@\hbox{$\m@th#1{#2}$}\findownsm@sh}
\def\findownsm@sh{\dp\z@\z@ \box\z@}
\catcode`@=12

\def\rdx #1{\ColorRGB{0.0 0.4 0.0}{#1}}%
\input rotate.tex

\def\hexnumber #1{\ifcase #10\or 1\or 2\or 3\or 4\or 5\or 6\or 7\or 8\or
   9\or A\or B\or C\or D\or E\or F\fi}%

\newfam\eufmfam
   \font\teneufm=eufm10
   \font\nineeufm=eufm10 at 9pt
   \font\eighteufm=eufm10 at 8pt
   \font\seveneufm=eufm7
   \font\sixeufm=eufm7 at 6pt
   \font\fiveeufm=eufm5
   \textfont\eufmfam=\teneufm \scriptfont\eufmfam=\seveneufm
      \scriptscriptfont\eufmfam=\fiveeufm

\newfam\eurmfam
   \font\teneurm=eurm10
   \font\nineeurm=eurm10 at 9pt
   \font\eighteurm=eurm10 at 8pt
   \font\seveneurm=eurm7
   \font\sixeurm=eurm7 at 6pt
   \font\fiveeurm=eurm5
   \textfont\eurmfam=\teneurm \scriptfont\eurmfam=\seveneurm
      \scriptscriptfont\eurmfam=\fiveeurm

\newfam\eusmfam
   \font\teneusm=eusm10
   \font\nineeusm=eusm10 at 9pt
   \font\eighteusm=eusm10 at 8pt
   \font\seveneusm=eusm7
   \font\sixeusm=eusm7 at 6pt
   \font\fiveeusm=eusm5
   \textfont\eusmfam=\teneusm \scriptfont\eusmfam=\seveneusm
      \scriptscriptfont\eusmfam=\fiveeusm

\newfam\msamfam
   \font\tenmsam=msam10
   \font\ninemsam=msam10 at 9pt
   \font\eightmsam=msam10 at 8pt
   \font\sevenmsam=msam7
   \font\sixmsam=msam7 at 6pt
   \font\fivemsam=msam5
   \textfont\msamfam=\tenmsam \scriptfont\msamfam=\sevenmsam
      \scriptscriptfont\msamfam=\fivemsam

\newfam\msbmfam
   \font\tenmsbm=msbm10
   \font\ninemsbm=msbm10 at 9pt
   \font\eightmsbm=msbm10 at 8pt
   \font\sevenmsbm=msbm7
   \font\sixmsbm=msbm7 at 6pt
   \font\fivemsbm=msbm5
   \textfont\msbmfam=\tenmsbm \scriptfont\msbmfam=\sevenmsbm
      \scriptscriptfont\msbmfam=\fivemsbm

\newfam\scriptfam
      \font\tenfs=rsfs10 \skewchar\tenfs='177
      \font\ninefs=rsfs10 at 9pt \skewchar\ninefs='177
      \font\eightfs=rsfs10 at 8pt \skewchar\eightfs='177
      \font\sevenfs=rsfs7 \skewchar\sevenfs='177
      \font\sixfs=rsfs7 at 6pt \skewchar\sixfs='177
      \font\fivefs=rsfs5 \skewchar\fivefs='177
      \textfont\scriptfam=\tenfs \scriptfont\scriptfam=\sevenfs
         \scriptscriptfont\scriptfam=\fivefs
\newfam\stmaryrdfam
   \font\tenstmaryrd=stmary10
   \font\ninestmaryrd=stmary10 at 9pt
   \font\eightstmaryrd=stmary10 at 8pt
   \font\sevenstmaryrd=stmary7
   \font\sixstmaryrd=stmary7 at 6pt
   \font\fivestmaryrd=stmary5
   \textfont\stmaryrdfam=\tenstmaryrd \scriptfont\stmaryrdfam=\sevenstmaryrd
      \scriptscriptfont\stmaryrdfam=\fivestmaryrd

\def\eightpointlogic{%
   \textfont\eurmfam=\eighteurm \scriptfont\eurmfam=\sixeurm
      \scriptscriptfont\eurmfam=\fiveeurm
   \textfont\cmsmrfam=\eightsmr \scriptfont\cmsmrfam=\sixsmr
      \scriptscriptfont\cmsmrfam=\fivesmr
   \textfont\cmsmiufam=\eightsmiu \scriptfont\cmsmiufam=\sixsmiu
      \scriptscriptfont\cmsmiufam=\fivesmiu
   \textfont\eufmfam=\eighteufm \scriptfont\eufmfam=\sixeufm
      \scriptscriptfont\eufmfam=\fiveeufm
   \textfont\eusmfam=\eighteusm \scriptfont\eusmfam=\sixeusm
      \scriptscriptfont\eusmfam=\fiveeusm
   \textfont\msamfam=\eightmsam \scriptfont\msamfam=\sixmsam
      \scriptscriptfont\msamfam=\fivemsam
   \textfont\msbmfam=\eightmsbm \scriptfont\msbmfam=\sixmsbm
      \scriptscriptfont\msbmfam=\fivemsbm
   \ifscriptfamrsfs
      \textfont\scriptfam=\eightfs \scriptfont\scriptfam=\sixfs
         \scriptscriptfont\scriptfam=\fivefs
      \else
      \textfont\scriptfam=\eightmptwo \scriptfont\scriptfam=\sixmptwo
         \scriptscriptfont\scriptfam=\fivemptwo
      \fi
   \textfont\stmaryrdfam=\eightstmaryrd \scriptfont\stmaryrdfam=\sixstmaryrd
      \scriptscriptfont\stmaryrdfam=\fivestmaryrd}%
\let\oldeightpoint=\eightpoint
\def\eightpoint {\oldeightpoint\eightpointlogic}%

\def\ninepointlogic{%
   \textfont\eurmfam=\nineeurm \scriptfont\eurmfam=\sixeurm
      \scriptscriptfont\eurmfam=\fiveeurm
   \textfont\cmsmrfam=\ninesmr \scriptfont\cmsmrfam=\sixsmr
      \scriptscriptfont\cmsmrfam=\fivesmr
   \textfont\cmsmiufam=\ninesmiu \scriptfont\cmsmiufam=\sixsmiu
      \scriptscriptfont\cmsmiufam=\fivesmiu
   \textfont\eufmfam=\nineeufm \scriptfont\eufmfam=\sixeufm
      \scriptscriptfont\eufmfam=\fiveeufm
   \textfont\eusmfam=\nineeusm \scriptfont\eusmfam=\sixeusm
      \scriptscriptfont\eusmfam=\fiveeusm
   \textfont\msamfam=\ninemsam \scriptfont\msamfam=\sixmsam
      \scriptscriptfont\msamfam=\fivemsam
   \textfont\msbmfam=\ninemsbm \scriptfont\msbmfam=\sixmsbm
      \scriptscriptfont\msbmfam=\fivemsbm
   \ifscriptfamrsfs
      \textfont\scriptfam=\ninefs \scriptfont\scriptfam=\sixfs
         \scriptscriptfont\scriptfam=\fivefs
      \else
      \textfont\scriptfam=\ninemptwo \scriptfont\scriptfam=\sixmptwo
         \scriptscriptfont\scriptfam=\fivemptwo
      \fi
   \textfont\stmaryrdfam=\ninestmaryrd \scriptfont\stmaryrdfam=\sixstmaryrd
      \scriptscriptfont\stmaryrdfam=\fivestmaryrd}%
\let\oldninepoint=\ninepoint
\def\ninepoint {\oldninepoint\ninepointlogic}%

\def\twelvepointlogic{%
   \textfont\cmsmrfam=\twelvesmr \scriptfont\cmsmrfam=\eightsmr
      \scriptscriptfont\cmsmrfam=\sixsmr
      }%
\let\oldtwelvepoint=\twelvepoint
\def\twelvepoint {\oldtwelvepoint\twelvepointlogic}%

\def\script {\fam\scriptfam}%                        % Useless with zpmp2

\let\false=\Fss

\newbox\Nablassbox
\def\Nablass {{\mathchoice
      {\setbox\Nablassbox=\hbox{$\Deltass$}%
         \setbox\Nablassbox=\hbox{\rotu\Nablassbox}%
         \box\Nablassbox}%
      {\setbox\Nablassbox=\hbox{$\Deltass$}%
         \setbox\Nablassbox=\hbox{\rotu\Nablassbox}%
         \box\Nablassbox}%
      {\setbox\Nablassbox=\hbox{$\scriptstyle\Deltass$}%
         \setbox\Nablassbox=\hbox{\rotu\Nablassbox}%
         \box\Nablassbox}%
      {\setbox\Nablassbox=\hbox{$\scriptscriptstyle\Deltass$}%
         \setbox\Nablassbox=\hbox{\rotu\Nablassbox}%
         \box\Nablassbox}}}%

\let\true=\Tss
\mathchardef\weak="013C

\mathchardef\impl="221B
\catcode`\!=\active
   \edef!{\hexnumber\stmaryrdfam}%
   \mathchardef\squarebox="2!1F
   \mathchardef\lpar="2!4F
   \mathchardef\lplus="2!16
   \mathchardef\lmix="2!22
   \mathchardef\lprec="2!34
   \mathchardef\ltens="2!0F
   \mathchardef\lwith="2!4E
   \mathchardef\merge="2!05
   \edef!{\hexnumber\msamfam}%
   \mathchardef\limpalt="2!28
   \catcode`\!=12

\catcode`\!=\active
   \edef!{\hexnumber\msamfam}%
   \mathchardef\ge="3!3E
   \mathchardef\le="3!36
   \mathchardef\gex="3!3C
   \mathchardef\lex="3!34
   \edef!{\hexnumber\stmaryrdfam}%
   \mathchardef\bigcross="3!22
   \mathchardef\arrowequiv="3!2D
   \mathchardef\inplus="3!41
   \mathchardef\msin="3!41
   \mathchardef\subsetplus="3!44
   \mathchardef\supsetplus="3!45
   \mathchardef\subseteqplus="3!46
   \mathchardef\supseteqplus="3!47
   \catcode`\!=12

\let\turnstile=\vdash

\catcode`\!=\active
   \edef!{\hexnumber\stmaryrdfam}%
   \mathchardef\lbbrack="4!4A
   \mathchardef\rbbrack="5!4B
   \mathchardef\lbpar="4!4C
   \mathchardef\rbpar="5!4D
      \mathchardef\lstrange="4!2A
      \mathchardef\rstrange="5!2B
      \mathchardef\lstrange="4!48
      \mathchardef\rstrange="5!49
   \catcode`\!=12

\def\xyvdots {\raise6pt\hbox{$\vdots$}}%

\newdimen\dercldim                                % dcl
\newdimen\derccdim                                % dcc
\newdimen\dercrdim                                % dcr
\newdimen\derldim                                 % dl
\newdimen\dercdim                                 % dc
\newdimen\derrdim                                 % dr
\newdimen\derdim                                  % d
\newdimen\derdldim                                % ddl
\newdimen\derdrdim                                % ddr
\newbox\derboxone                                 % b1
\newbox\derboxtwo                                 % b2
\newbox\derboxthree                               % b3
\newbox\derboxfour                                % b4
\newdimen\derquad\derquad=\fontdimen6\textfont2
\newdimen\deropen\deropen=\fontdimen5\textfont2\divide\deropen by3
\def\leaf #1{\global\setbox\derboxone=\hbox{\strut$#1$}%
   \global\derldim=0pt                            % dl=0
   \global\dercdim=\wd\derboxone                  % dc=wd(b1)
   \global\derrdim=0pt                            % dr=0
   }%
\def\rootaux #1#2#3{\setbox\derboxtwo=\hbox{\unhbox\derboxone}%
   \setbox\derboxthree=\hbox 
      {$\smash{\lower\fontdimen22\textfont2\hbox{$#1$}}$}%
   \setbox\derboxfour=\hbox 
      {$\smash{\lower\fontdimen22\textfont2\hbox{$#2$}}$}%
   \leaf{#3}%                                     % dl=0, dc=wd(b1), dr=0
   \derdim=\dercdim\advance\derdim by-\derccdim\divide\derdim by2 
   \global\derldim=\dercldim\global\advance\derldim by-\derdim
   \global\derrdim=\dercrdim\global\advance\derrdim by-\derdim
   \deropen=\fontdimen5\textfont2\divide\deropen by3
   \setbox\derboxone=\hbox{\vbox{\offinterlineskip
         \hbox{\ifdim\derldim<0pt\kern-\derldim\fi
               \box\derboxtwo
               \ifdim\derrdim<0pt\kern-\derrdim\fi}%
         \kern\deropen
         \hbox{\ifdim\dercldim>\derldim
                  \ifdim\derldim>0pt\kern\derldim\fi
                  \else\kern\dercldim\fi
               \hbox to0pt{\hss\copy\derboxthree}%
               \vbox{\ifdim\derccdim>\dercdim\hsize=\derccdim
                                        \else\hsize=\dercdim \fi
                    \hrule height.2pt depth.2pt width\hsize}%
               \hbox to0pt{\copy\derboxfour\hss}%
               \ifdim\dercrdim>\derrdim
                  \ifdim\derrdim>0pt\kern\derrdim\fi
                  \else\kern\dercrdim\fi}%
         \kern\deropen
         \hbox{\ifdim\derldim>0pt\kern\derldim\fi
               \box\derboxone
               \ifdim\derrdim>0pt\kern\derrdim\fi}}}%
   \ifdim\derldim<0pt\global\derldim=0pt\fi       % dl=max(dl,0)
   \ifdim\derrdim<0pt\global\derrdim=0pt\fi       % dr=max(dr,0)
   \derdldim=\wd\derboxthree\advance\derdldim by-\dercldim
   \derdrdim=\wd\derboxfour \advance\derdrdim by-\dercrdim
   \ifdim\derdim<0pt
      \ifdim\derdldim<0pt
         \derdldim=0pt                            % d<0, ddl<0 -> ddl=0
      \fi
      \ifdim\derdrdim<0pt
         \derdrdim=0pt                            % d<0, ddr<0 -> ddr=0
      \fi
   \else
      \ifdim\derldim>0pt
         \ifdim\derdldim>-\derdim
            \advance\derdldim by\derdim           % d>=0, dl>0, ddl+d>0 -> 
         \else                                            %                    ddl=ddl+d
            \derdldim=0pt                         % d>=0, dl>0, ddl+d<=0 ->
         \fi                                      %                        ddl=0
      \else
         \advance\derdldim by\dercldim            % d>=0, dl=0 -> ddl=ddl+dcl
      \fi
      \ifdim\derrdim>0pt
         \ifdim\derdrdim>-\derdim
            \advance\derdrdim by\derdim           % d>=0, dr>0, ddr+d>0 -> 
         \else                                            %                    ddr=ddr+d
            \derdrdim=0pt                         % d>=0, dr>0, ddr+d<=0 ->
         \fi                                      %                        ddr=0
      \else
         \advance\derdrdim by\dercrdim            % d>=0, dr=0 -> ddr=ddr+dcr
      \fi
   \fi
   \global\setbox\derboxone=\hbox
      {\kern\derdldim\unhbox\derboxone\kern\derdrdim}%
   \global\advance\derldim by\derdldim            % dl=dl+ddl
   \global\advance\derrdim by\derdrdim            % dr=dr+ddr
   }%
\def\rootr #1#2#3#4{{#4}%
   \dercldim=\derldim
   \derccdim=\dercdim
   \dercrdim=\derrdim
   \rootaux{#1}{#2}{#3}}%
\def\rrootr #1#2#3#4#5{\derquad=\fontdimen6\textfont2
   {#4}%
           \dercldim  =\derldim
   \setbox\derboxtwo=\hbox{\unhbox\derboxone\kern\derquad}%
           \derccdim  =\dercdim
   \advance\derccdim by\derrdim
   \advance\derccdim by\derquad
   {#5}%
   \setbox\derboxone=\hbox{\unhbox\derboxtwo\unhbox\derboxone}%
   \advance\derccdim by\derldim
   \advance\derccdim by\dercdim
           \dercrdim  =\derrdim
   \rootaux{#1}{#2}{#3}}%
\def\rrrootr #1#2#3#4#5#6{\derquad=\fontdimen6\textfont2
   {#4}%
           \dercldim  =\derldim
   \setbox\derboxtwo=\hbox{\unhbox\derboxone\kern\derquad}%
           \derccdim  =\dercdim
   \advance\derccdim by\derrdim
   \advance\derccdim by\derquad
   {#5}%
   \setbox\derboxtwo=\hbox{\unhbox\derboxtwo\unhbox\derboxone\kern\derquad}%
   \advance\derccdim by\derldim
   \advance\derccdim by\dercdim
   \advance\derccdim by\derrdim
   \advance\derccdim by\derquad
   {#6}%
   \setbox\derboxone=\hbox{\unhbox\derboxtwo\unhbox\derboxone}%
   \advance\derccdim by\derldim
   \advance\derccdim by\dercdim
           \dercrdim  =\derrdim
   \rootaux{#1}{#2}{#3}}%
\def\root       #1#2#3{\rootr  {#1\;}{}{#2}{#3}}%
\def\rroot    #1#2#3#4{\rrootr {#1\;}{}{#2}{#3}{#4}}%
\def\deraux {\derldim=0pt\dercdim=0pt\derrdim=0pt}%
\def\der       #1#2#3{\deraux\root  {#1}{#2}{#3}        \box\derboxone}%
\def\dder    #1#2#3#4{\deraux\rroot {#1}{#2}{#3}{#4}    \box\derboxone}%
\def\dernote       #1#2#3#4{\deraux\rootr  {#1\;}{\;#2}{#3}{#4}\box\derboxone}%
\def\ddernote    #1#2#3#4#5{\deraux\rrootr {#1\;}{\;#2}{#3}{#4}{#5}\box
                                                                   \derboxone}%
\def\inf       #1#2#3{\der  {#1}{#2}{\leaf{#3}}}%
\def\iinf    #1#2#3#4{\dder {#1}{#2}{\leaf{#3}}{\leaf{#4}}}%
\newbox\derskelboxone
\newbox\derskelboxtwo
\newbox\derskelboxthree
\newbox\derskelboxfour
\newdimen\derskeldimenone
\newdimen\derskeldimentwo
\newdimen\derskeldimenthree
\newdimen\derskeldimenfour
\newdimen\derskeldimenfive
\newdimen\derskeldimensix
\newdimen\derskeldimenseven
\newdimen\derskeldimeneight
\def\derskel #1#2#3#4{%
   \setbox\derskelboxone=\hbox{$#1$\strut}%
   \derskeldimenone=\ht\derskelboxone
   \advance\derskeldimenone by\dp\derskelboxone
   \derskeldimentwo=\wd\derskelboxone
   \divide\derskeldimentwo by2
   \setbox\derskelboxone=\hbox to0pt{%
      \hss\raise\dp\derskelboxone\box\derskelboxone\hss}%
   \ht\derskelboxone=0pt
   \dp\derskelboxone=0pt
   \setbox\derskelboxtwo=\hbox{$#3$\strut}%
   \derskeldimenthree=\ht\derskelboxtwo
   \advance\derskeldimenthree by\dp\derskelboxtwo
   \derskeldimenfour=\wd\derskelboxtwo
   \divide\derskeldimenfour by2
   \setbox\derskelboxtwo=\hbox to0pt{%
      \hss\raise\dp\derskelboxtwo\box\derskelboxtwo\hss}%
   \ht\derskelboxtwo=0pt
   \dp\derskelboxtwo=0pt
   \ifdim\derskeldimenone>\derskeldimenthree
      \else\derskeldimenone=\derskeldimenthree\fi
   \setbox\derskelboxthree=\hbox{$#4$\strut}%
   \derskeldimenfive=\ht\derskelboxthree
   \advance\derskeldimenfive by\dp\derskelboxthree
   \derskeldimensix=\wd\derskelboxthree
   \divide\derskeldimensix by2
   \setbox\derskelboxthree=\hbox to0pt{%
      \hss\lower\ht\derskelboxthree\box\derskelboxthree\hss}%
   \ht\derskelboxthree=0pt
   \dp\derskelboxthree=0pt
   \setbox\derskelboxfour=\hbox{$#2$\strut}%
   \derskeldimenseven=\ht\derskelboxfour
   \advance\derskeldimenseven by\dp\derskelboxfour
   \derskeldimeneight=\wd\derskelboxfour
   \divide\derskeldimeneight by2
   \setbox\derskelboxfour=\hbox to0pt{%
      \hss\raise\dp\derskelboxfour\box\derskelboxfour\hss}%
   \ht\derskelboxfour=0pt
   \dp\derskelboxfour=0pt
   \ifdim\derskeldimenone>\derskeldimenseven
      \else\derskeldimenone=\derskeldimenseven\fi
   \derskeldimenthree=\derskeldimentwo
   \advance\derskeldimenthree by2\derskeldimeneight
   \advance\derskeldimenthree by\derskeldimenfour
   \advance\derskeldimenthree by2em
   \divide\derskeldimenthree by2
   \advance\derskeldimensix by-\derskeldimenthree
   \derskeldimenseven=\derskeldimensix
   \advance\derskeldimensix by-\derskeldimentwo
   \advance\derskeldimenseven by-\derskeldimenfour
   \ifdim\derskeldimensix>0pt
      \else\derskeldimensix=0pt\fi
   \ifdim\derskeldimenseven>0pt
      \else\derskeldimenseven=0pt\fi
   \vbox{\kern\derskeldimenone\hbox{\kern\derskeldimensix
         \kern\derskeldimentwo
         \xy
         <-\derskeldimenthree,\derskeldimenthree>="here"
            *{\box\derskelboxone}**\dir{-};
         "here"+<\derskeldimentwo,0pt>="here"**\dir{-};
         "here"+<1em,0pt>="here"**\dir{-};
         "here"+<\derskeldimeneight,0pt>="here"
            *{\box\derskelboxfour}**\dir{-};
         "here"+<\derskeldimeneight,0pt>="here"**\dir{-};
         "here"+<1em,0pt>="here"**\dir{-};
         "here"+<\derskeldimenfour,0pt>*{\box\derskelboxtwo}**\dir{-};
         0*{\box\derskelboxthree}**\dir{-};
         <-\derskeldimenthree,\derskeldimenthree>**\dir{-}
         \endxy
         \kern\derskeldimenfour\kern\derskeldimenseven}%
      \kern\derskeldimenfive}}%

\newbox\DerivOneBox
\newbox\DerivTwoBox
\newbox\DerivThreeBox
\newbox\DerivFourBox
\newdimen\DerivOneDimen
\newdimen\DerivTwoDimen
\newdimen\DerivThreeDimen
\newdimen\DerivFourDimen
\def\Derivationleaf #1#2#3#4#5{\global\setbox\derboxone=\hbox{\strut
                                    $\DerivationFactors{#1}{#2}{#3}{#4}{#5}11$}}%
\def\DerivationFactors #1#2#3#4#5#6#7{%
   \setbox\DerivOneBox=\hbox{$#1\strut$}%
      \DerivOneDimen=\wd\DerivOneBox\divide\DerivOneDimen by2
   \setbox\DerivThreeBox=\hbox{$#3\strut$}%
      \DerivThreeDimen=\wd\DerivThreeBox\divide\DerivThreeDimen by2
   \setbox\DerivTwoBox=\hbox{\box\DerivOneBox\hbox{$#2$}\box\DerivThreeBox}%
      \DerivTwoDimen=\wd\DerivTwoBox
   \setbox\DerivFourBox=\hbox{$#4\strut$}%
      \DerivFourDimen=\wd\DerivFourBox
   \ifdim\DerivFourDimen>\DerivTwoDimen
      \global\dercdim=\DerivFourDimen                % dc=wd(b4) see logicmac.tex
      \global\derldim=0pt                            % dl
      \global\derrdim=0pt                            % dr
      \advance\DerivFourDimen by-\DerivTwoDimen
      \divide \DerivFourDimen by2
      \advance\DerivTwoDimen  by-\DerivOneDimen
      \advance\DerivTwoDimen  by-\DerivThreeDimen
      \divide \DerivTwoDimen  by 2
   \else
      \global\dercdim=\DerivFourDimen                % dc=wd(b4) see logicmac.tex
      \DerivFourDimen=0pt
      \advance\DerivTwoDimen  by-\DerivOneDimen
      \advance\DerivTwoDimen  by-\DerivThreeDimen
      \global\derldim=\DerivTwoDimen
         \global\advance\derldim by-\dercdim
         \global\divide\derldim by2
         \global\advance\derldim by\DerivOneDimen    % dl
      \global\derrdim=\DerivTwoDimen
         \global\advance\derrdim by-\dercdim
         \global\divide\derrdim by2
         \global\advance\derrdim by\DerivThreeDimen  % dr
      \divide \DerivTwoDimen  by 2
   \fi
   \vbox{\offinterlineskip\hbox{\kern\DerivFourDimen\box\DerivTwoBox}%
         \hbox{\kern\DerivFourDimen\kern\DerivOneDimen
               \kern\DerivTwoDimen\kern-#6\DerivTwoDimen\hbox{$\xy
               0;<#6\DerivTwoDimen,0pt>:<0pt,#7\DerivTwoDimen>::
               (0,1);(2,1)**\crv{(1.25,1.1875)&(0.75,0.8125)};
               (1,0)**@{-};(0,1)**@{-};
               (1,0.625)*{\scriptstyle #5}
               \endxy$}}%
         \hbox{\kern\DerivFourDimen\kern\DerivOneDimen\kern\DerivTwoDimen
               \hbox to0pt{\hss\box\DerivFourBox\hss}%
               \kern\DerivFourDimen\kern\DerivOneDimen\kern\DerivTwoDimen}}}%
\newcommand{\rulebox}[1]{\parbox{7em}{$$#1$$}}

\def\at  {{\mathsf{at}}}
\def\occ  {{\mathsf{occ}}}

\def\commentthis #1{}%
\def\down #1#2#3{#1 \downarrow_{#3} #2}%

\def\S {{\mathsf{S}}}

\def\BV {\mathsf{BV}}

\def\NEL {\mathsf{NEL}}

\def\KSi  {\mathsf{KSi}}

\def\MLL {\mathsf{MLL}}

\def\ruleidown       {\mathsf{i}{\downarrow}}
\def\ruleaidown      {\mathsf{ai}{\downarrow}}
\def\ruleaiup        {\mathsf{ai}{\uparrow}}

\def\swir        {\mathsf{s}}

\def\promr       {\mathsf{p}{\downarrow}}

\def\weakr       {\mathsf{w}{\downarrow}}

\def\rulecdown   {\mathsf{c}{\downarrow}}

\def\rulewdown   {\mathsf{w}{\downarrow}}

\def\true {{\sf t}\!{\sf t}}%
\def\false {{\sf f}\!{\sf f}}%

\def\reslazyswir  {{\mathsf{lis}}}

\def\axiom {\mathsf{id}}

\def\lbot {\bot}

\def\lone{\mathsf{1}}

\def\ls #1{\underline{#1\phantom{\hbox to0pt{,\hss}}}{}\lower1ex\hbox
                                                   {$\mathsf{s}$}}%

\def\lv #1{\underline{#1\phantom{\hbox to0pt{,\hss}}}{}\lower1ex\hbox
                                                   {$\mathsf{v}$}}%

\def\lk #1{\underline{#1\phantom{\hbox to0pt{,\hss}}}{}\lower1ex\hbox
                                                   {$\mathsf{c}$}}%

\definecolor{rdxbackcolor}{gray}{0.91}
\def\rdx#1{\smash{\colorbox{rdxbackcolor}{\strut\smash{$#1$}}}}
\definecolor{Aquamarine}{rgb}{0.49,0.99,0.82}
\definecolor{MediumAquamarine}{rgb}{0.39,0.79,0.66}
\definecolor{Black}{rgb}{0.0,0.0,0.0}
\definecolor{Blue}{rgb}{0.0,0.0,0.99}
\definecolor{CadetBlue}{rgb}{0.36,0.61,0.62}
\definecolor{CornflowerBlue}{rgb}{0.38,0.58,0.92}
\definecolor{DarkSlateBlue}{rgb}{0.28,0.23,0.54}
\definecolor{LightBlue}{rgb}{0.67,0.84,0.89}
\definecolor{LightSteelBlue}{rgb}{0.68,0.76,0.86}
\definecolor{MediumBlue}{rgb}{0.0,0.0,0.79}
\definecolor{MediumSlateBlue}{rgb}{0.47,0.40,0.92}
\definecolor{MidnightBlue}{rgb}{0.09,0.09,0.43}
\definecolor{NavyBlue}{rgb}{0.0,0.0,0.49}
\definecolor{SkyBlue}{rgb}{0.52,0.80,0.91}
\definecolor{SlateBlue}{rgb}{0.41,0.35,0.79}
\definecolor{SteelBlue}{rgb}{0.27,0.50,0.70}
\definecolor{Coral}{rgb}{0.99,0.49,0.31}
\definecolor{Cyan}{rgb}{0.0,0.99,0.99}
\definecolor{Firebrick}{rgb}{0.69,0.13,0.13}
\definecolor{Brown}{rgb}{0.64,0.16,0.16}
\definecolor{Gold}{rgb}{0.99,0.83,0.0}
\definecolor{Goldenrod}{rgb}{0.84,0.64,0.12}
\definecolor{Green}{rgb}{0.0,0.99,0.0}
\definecolor{DarkGreen}{rgb}{0.0,0.38,0.0}
\definecolor{DarkOliveGreen}{rgb}{0.33,0.41,0.18}
\definecolor{ForestGreen}{rgb}{0.13,0.54,0.13}
\definecolor{LimeGreen}{rgb}{0.19,0.79,0.19}
\definecolor{MediumSeaGreen}{rgb}{0.23,0.69,0.44}
\definecolor{MediumSpringGreen}{rgb}{0.0,0.97,0.59}
\definecolor{PaleGreen}{rgb}{0.59,0.97,0.59}
\definecolor{SeaGreen}{rgb}{0.17,0.54,0.33}
\definecolor{SpringGreen}{rgb}{0.0,0.99,0.49}
\definecolor{YellowGreen}{rgb}{0.59,0.79,0.19}
\definecolor{DarkSlateGray}{rgb}{0.18,0.30,0.30}
\definecolor{DimGray}{rgb}{0.40,0.40,0.40}
\definecolor{LightGray}{rgb}{0.82,0.82,0.82}
\definecolor{Grey}{rgb}{0.73,0.73,0.73}
\definecolor{Khaki}{rgb}{0.93,0.89,0.54}
\definecolor{Magenta}{rgb}{0.99,0.0,0.99}
\definecolor{Maroon}{rgb}{0.68,0.18,0.37}
\definecolor{Orange}{rgb}{0.99,0.64,0.0}
\definecolor{Orchid}{rgb}{0.84,0.43,0.83}
\definecolor{DarkOrchid}{rgb}{0.59,0.19,0.79}
\definecolor{MediumOrchid}{rgb}{0.72,0.33,0.82}
\definecolor{Pink}{rgb}{0.99,0.74,0.79}
\definecolor{Plum}{rgb}{0.86,0.62,0.86}
\definecolor{Red}{rgb}{0.99,0.0,0.0}
\definecolor{IndianRed}{rgb}{0.79,0.35,0.35}
\definecolor{MediumVioletRed}{rgb}{0.77,0.08,0.51}
\definecolor{OrangeRed}{rgb}{0.99,0.26,0.0}
\definecolor{VioletRed}{rgb}{0.80,0.12,0.56}
\definecolor{Salmon}{rgb}{0.97,0.49,0.44}
\definecolor{Sienna}{rgb}{0.62,0.31,0.17}
\definecolor{Tan}{rgb}{0.81,0.70,0.54}
\definecolor{Thistle}{rgb}{0.84,0.74,0.84}
\definecolor{Turquoise}{rgb}{0.24,0.87,0.80}
\definecolor{DarkTurquoise}{rgb}{0.0,0.80,0.81}
\definecolor{MediumTurquoise}{rgb}{0.28,0.81,0.79}
\definecolor{Violet}{rgb}{0.92,0.50,0.92}
\definecolor{BlueViolet}{rgb}{0.53,0.16,0.88}
\definecolor{Wheat}{rgb}{0.95,0.86,0.69}
\definecolor{White}{rgb}{0.99,0.99,0.99}
\definecolor{Yellow}{rgb}{0.99,0.99,0.0}
\definecolor{GreenYellow}{rgb}{0.67,0.99,0.18}

\def\BV {{\mathsf{BV}}}
\def\NEL {{\mathsf{NEL}}}

\def\MLL {{\mathsf{MLL}}}

\def\SMLS {{\mathsf{SMLS}}}
\def\MLS {{\mathsf{MLS}}}
\def\MLSu {{\mathsf{MLSu}}}
\def\MLSd {{\mathsf{MLSd} }}
\def\MLSi {{\mathsf{MLSi} }}
\def\MLSl {{ \mathsf{MLSl} }}
\def\MLSli {{ \mathsf{MLSli} }}
\def\MLSdi {{\mathsf{MLSdi}}}
\def\MLSdli {{\mathsf{MLSdli}}}
\def\MLSdl {{\mathsf{MLSdl}}}

\def\KSli {{\mathsf{KSli}}}
\def\KSdl {{\mathsf{KSdl}}}
\def\KSdi {{\mathsf{KSdi}}}
\def\KSd {{\mathsf{KSd}}}
\def\KSi {{\mathsf{KSi}}}
\def\KSl {{\mathsf{KSl}}}
\def\KSu {{\mathsf{KSu}}}

\def\ruleaidown  {{\mathsf{ai}{\downarrow}}}
\def\swir  {\mathsf{s}}

\def\unitonedown {{\mathsf{u}_1{\downarrow}}}
\def\unittwodown {{\mathsf{u}_2{\downarrow}}}
\def\unitthreedown {{\mathsf{u}_3{\downarrow}}}
\def\unitfourdown {{\mathsf{u}_4{\downarrow}}}
\def\unitoneup {{\mathsf{u}_1{\uparrow}}}
\def\unittwoup {{\mathsf{u}_2{\uparrow}}}

\def\reslazyswir  {{\mathsf{lis}}}
\def\lazyintswir  {{\mathsf{lis}}}
\def\deeplazyswir  {{\mathsf{dls}}}
\def\deepreslazyswir  {{\mathsf{dlis}}}

\def\down #1#2#3{#1 \downarrow_{#3} #2}%

\def\occ  {{\mathsf{occ}}}
\def\at  {{\mathsf{at}}}

\def\deepswir  {{\mathsf{ds}}}

\def\lazyswir {{ \mathsf{ls}}}
\def\deepswir {{ \mathsf{ds}}}
\def\deeplazyintswir {{ \mathsf{dlis}}}
\def\intswir  {{ \mathsf{is}}}

\def\deepintswir {{ \mathsf{dis}}}

\def\ruleaiup  {{\mathsf{ai}{\uparrow}}}

\def\ruleidown  {{\mathsf{i}{\downarrow}}}

\def\KSdli {{\mathsf{KSdli}}}
\def\KSli {{\mathsf{KSli}}}
\def\KSdli {{\mathsf{KSdli}}}
\def\KSl {{\mathsf{KSl}}}
\def\KSi {{\mathsf{KSi}}}
\def\KSdi {{\mathsf{KSdi}}}
\def\KSd {{\mathsf{KSd}}}

\def\S {{\mathsf{S}}}

\theoremstyle{plain}

\begin{document}

\title[Interaction and Depth against Nondeterminism in Proof
  Search]{Interaction and Depth\\ against Nondeterminism in Proof
  Search}

\author[Ozan Kahramano\u gullar\i]{Ozan Kahramano\u gullar\i}	%required
\address{The Microsoft Research -- University of Trento  Centre for Computational and Systems Biology}	%required
\email{ozan@cosbi.eu}  %optional
\thanks{  }	%optional

%% required for running head on odd and even pages, use suitable
%% abbreviations in case of long titles and many authors:

%% mandatory lists of keywords and classifications:
\subjclass{Theory of Computation}
% \titlecomment{OPTIONAL comment concerning the title, , if a variant
% or an extended abstract of the paper has appeared elsewehere}
%%%%%%%%%%%%%%%%%%%%%%%%%%%%%%%%%%%%%%%%%%%%%%%%%%%%%%%%%%%%%%%%%%%%%%%%%%%

%% the abstract has to PRECEED the command \maketitle:
%% be sure not to issue the \maketitle command twice!

\begin{abstract}
  \noindent  Deep inference is a  proof theoretic methodology that generalizes 
the standard notion of inference of the sequent calculus, whereby 
inference rules become applicable at any depth inside logical expressions.
Deep inference provides more freedom in the design of deductive systems for 
different logics and a rich combinatoric analysis of proofs.  In particular, 
construction of exponentially shorter analytic proofs becomes possible, however 
with the cost of a greater nondeterminism than in the sequent calculus. 
In this paper, we show that the nondeterminism in proof search can be reduced 
without losing the shorter proofs and without sacrificing  proof theoretic cleanliness. 
For this, we exploit an interaction and depth scheme in the logical expressions.  
We demonstrate our method on deep inference systems for 
multiplicative linear logic and classical logic, discuss its proof 
complexity and its relation to focusing, and present implementations.
\end{abstract}

\maketitle

\section*{Introduction}

Proof search is of central importance in automated 
reasoning, especially for a broad  
range of applications in the fields of 
automated theorem proving, software verification and
artificial intelligence. 
The development of formalisms, techniques
and principles 
that allow the construction of shorter proofs 
in different logics
is a requirement for the applications in these fields.
This requirement gains an even greater emphasis 
for logics,  where highly optimized  techniques, 
such as  classical resolution,  
are not applicable.

Deep inference \cite{Gug02} is a proof theoretic 
methodology that generalizes the notion of inference
in formalisms such as the sequent calculus, natural 
deduction and analytic tableaux by introducing a top-down 
symmetry of the inference rules. This symmetry 
brings about  a combinatoric wealth that gives rise to properties
that are otherwise not observable, and  thereby results in 
a broad spectrum of theoretical and practical 
consequences.  
For example, at the theoretical front,
a manifestation of the top-down symmetry 
of the deep inference rules 
is observed  as the cut rule and the axiom 
become dual deep inference rules 
\cite{Gug02,Bru03a,Str03a}.\footnote{These are the 
rules $\ruleaiup$ 
and $\ruleaidown$ in Definition \ref{definition:SMLS}. }  
In the standard sequent calculus, this  duality 
remains implicit in the inference rules, and can only be  revealed explicitly, 
for example, by constructing a 
proof net or an atomic flow graph \cite{GG08,GGS10}, that is,
by removing the deductive information in the proofs that 
displays their step-wise logical construction. On the practical front, 
deep inference brings proof theory closer to computation by providing 
a meaningful deductive  interpretation of concepts 
such as  locality and sequentiality \cite{Gug02,Str02,Bru06,GS11,SG11}. 
The proof theoretic interpretation of these concepts make available 
potential applications in logic programming \cite{PaolaBVL02,Kah07b,Kah09}.

The top-down symmetry of the deep  inference rules has implications that also
bridge theoretical concepts with practical ones. This is because the symmetry of the 
inference rules
makes it possible to treat all parts of the proofs and
derivations at the same syntactic level as the logic. 
As a result, it becomes possible to
apply the inference rules at any depth inside logical expressions, similar to the
application of term rewriting rules \cite{KahTh,Kah07b}. This feature, delivering the name
deep inference, contrasts with the notion of inference of the sequent calculus,
which we call shallow inference. As an example, consider the following shallow
inference proof in multiplicative linear logic (Figure \ref{figure:MLL}) and 
a deep inference proof of the same formula, where the  rule instances
are marked:\\[-6 pt]
\[
\dernote{\; \lpar}{}{  \sqn{ \pars{  \aprs{\pars{a \lpar \neg{a}} \ltens \neg{b}}  \lpar b} }  }{
 {\rrootr{\ltens\;}{}{  \sqn{   \aprs{\pars{a \lpar \neg{a}} \ltens \neg{b}}  \; , \, b }  }
{\rootr{\lpar \;}{}{\sqn{a  \lpar  \neg a }\; }
{
                       \rootr{\mathsf{id}\,}{}{\sqn{a \; , \, \neg a} \;}{ 
                       \leaf{} }
                     }
} 
{\rootr{\mathsf{id}\,}{}{\sqn{b \; , \, \neg b} \;}{\leaf{}}}
}}
\qquad
\qquad
\dernote{\ruleaidown}{}{ \pars{  \aprs{\rdx{\pars{a \lpar \neg{a}}} \ltens \neg{b}}  \lpar b} }{
\rootr{\unittwodown\,}{}{  \pars{  \rdx{\aprs{\lone \ltens \neg{b}}}  \lpar b}  }{
\rootr{\ruleaidown\;}{}{  \rdx{\pars{  \neg{b}  \lpar b}}  }{
\leaf{\lone}}}}\\[6pt]
\]
Shallow inference proofs are constructed by breaking the logical expressions
into smaller components. These components are separated with meta-level operators:
the space between two branches in a proof tree denotes a meta-level
conjunction and the comma in a sequent denotes a meta-level disjunction. However
a separation as in the sequent calculus of meta-level and object-level causes
a mismatch for some logics \cite{Gug03} as the object level of the logic and the meta-level
correspond to different logical operators. For example, in linear logic, the logical
object-level employs multiplicative conjunctions as in the proof above as well as
additive conjunctions. However, in the sequent calculus the meta-level conjunction
implicitly plays a multiplicative role for some rules and an additive role for
others. Deep inference abolishes the meta-level in proofs as the meta-level is
integrated into the object-level.  Thereby, deep inference provides a solution to this 
mismatch as it clarifies the deductive reading of the proofs by keeping the inference steps
within the logic of the system. For example, in the proof above on the right, 
each proof step is a linear implication, and there is no meta-level information. 

Another consequence of the distinction between
meta-level and object-level is observed in term rewriting implementations of
shallow inference, as they require additional mechanisms, outside the logic, that
implement the meta-level \cite{MM96}. Absence of meta-level renders such an additional
mechanism unnecessary, and makes the term rewriting implementations much
simpler \cite{Kah07b}.

While indicating novel proof theoretic perspectives, 
the combinatoric wealth offered by deep inference  
 comes with a cost as the traditional  methods
do not meet the proof theoretic requirements.
This is because the long
established standard techniques of the sequent calculus 
for proving cut elimination and completeness are not applicable
in the deep inference setting. 
As a result of this,  the potential that came about due to deep inference required 
the development of proof theoretic methods for addressing the emerging technical challenges.
The consequent
efforts resulted in cut elimination results, which are very different from
the then-available methods for the sequent calculus \cite{Gug02,StraTh,BruTh,Bru06b,SG11}. 
Some of these cut-elimination results were obtained by fragmenting proofs into different 
phases, where in each phase distinct inference rules are used. 
The resulting \emph{decomposition} theorems  \cite{StraTh,BruTh,SG11} made available  
new ways for studying proofs, which are impossible in the sequent calculus.
These results can be seen as a consequence of the greater freedom in proof construction 
in deep inference that  provides many more proofs in comparison to the sequent calculus.

The concepts listed above qualify deep inference as a proof theoretic
methodology that is closer to computation. In fact, deep inference provides the
means to design deductive systems for different logics that reveal properties of
proofs of computational relevance and otherwise impossible. For example,  
the deep inference system $\BV$ \cite{Gug02,Kah08} gives a proof theoretic
interpretation of sequentiality as modeled in process algebra 
by extending multiplicative linear logic ($\MLL$) with the rules mix, 
nullary mix and a self-dual, non-commutative operator. $\BV$ 
and its Turing-complete extension \cite{GS11,SG11,StraUnd},
are provably not designable in the sequent calculus \cite{TiuTh01}. Moreover, the 
computer science notion of locality, i.e., an operation having a bounded computational 
cost, also finds a meaningful proof theoretic interpretation in the deep inference 
setting \cite{Bru06,Str02,TiuINT05}.

%WHY 

Deep inference deductive systems provide advantages also when they are
considered for \emph{computation as proof search}: the applicability
of the inference rules at arbitrary depths inside logical expressions makes
it possible to start the construction of the proofs from subformulae. This capability
\emph{provides many more different proofs, some of which are much shorter}
than those provided by shallow inference. For example, for a class of formulae
called Statman tautologies \cite{Sta78}, deep inference provides \emph{polynomially} 
growing  proofs in contrast to the \emph{exponentially} 
growing  sequent calculus proofs \cite{BG08}.
The reason for this drastic difference is because the sequent calculus can access the
subformulae only by opening up the formulae at the main connective, and proofs
are then constructed by spreading the context to the branches of the proof tree
as in the example above. In contrast, applying the inference rules without such
a restriction as in deep inference makes it possible to reduce the length of these
proofs to an extend that in some cases \emph{provides a polynomial length instead
of an exponential one}. In certain cases, the bounded depth formalism with alternative
compression mechanisms can be used for recuperating the polynomial
size proofs, as it is the case in \cite{Das11} for classical logic with a cubical transformation.
However, in non-deterministic proof search, where the search spaces grow
exponentially in the number of proof steps, even \emph{a linear speed up can extend
the margin of successful searches}.

Despite the existence of shorter proofs, a naive proof search approach with
deep inference results in transposing the problem. This is because deep inference
rules can be applied in many more ways, and this gives rise to a larger breadth
of the search space. Due to the resulting greater size of the search space, in
order to benefit from the shorter proofs provided by deep inference in proof
search, \emph{the non-determinism needs to be controlled}.

The trade off between shorter proofs and larger breadth of the search space
has been addressed by adapting the focusing technique \cite{And92,And01}
to deep inference systems \cite{CGS12}.
Although focusing provides a satisfactory means for reducing
nondeterminism in the sequent calculus, it has limitations in the deep
inference setting. This is because in deep inference, context management is
very different from the more rigid sequent calculus, which allows the application
of the inference rules only to the top-level formulae. Other techniques for
rendering sequent calculus based proof search more efficient, for example, for
linear logic \cite{CHP00}, include methods that use additional tagging mechanisms 
to account for formulas that are considered as consumed resources in proof search.

We present a technique for controlling nondeterminism in 
deep inference deductive systems.  Our technique, which is complementary to 
focusing and \emph{pure from a logical point of view},  provides a more immediate access 
to shorter proofs without sacrificing proof theoretic cleanliness. This is achieved 
by exploiting an interaction pattern together with depth within the rule instances 
during proof construction:  in deep inference deductive systems, the combinatoric 
behavior results from the instances of the switch rule, which is the rule responsible 
for context management. This rule is common to the deep inference systems for 
different fragments of  linear logic \cite{Str03a,Str02}, classical logic \cite{BruTh,Bru06}, 
modal logics \cite{GT06,Sto07}, intuitionistic logic \cite{TiuINT05}, the logic $\BV$ 
\cite{Gug02} and its extension with the exponentials of linear logic \cite{GS11,SG11}.

With linear logic operators, the switch rule is given as follows:
$$
\dernote{\swir}{}{S\pars{\aprs{R \ltens T} \lpar U}}{\leaf{S\aprs{\pars{R \lpar U} \ltens T}}}
$$
The intuition behind the role of the switch rule in proof construction can be
seen by using a communication metaphor \cite{Gug02}: when the formulae that are
connected with a disjunction are considered as communicating in a context
$S$, a bottom-up application of the switch rule can be perceived as breaking
the communication between the atoms in $T$ and $U$ and this way enhancing the
communication between the atoms in $R$ and $U$. In proof construction, this
mechanism is used to manage contexts to bring dual atoms closer as proofs
are constructed bottom-up by annihilating communicating dual atoms at the
instances of the atomic interaction rule, that is, 
the rule $\ruleaidown$ (see Definition \ref{definition:SMLS}),
as illustrated in the example above.

Our method is based on carrying the communication and interaction pattern 
used by the atomic interaction rule
to the switch rule. For this purpose, we redesign 
the switch rule to enhance the communication  of dual atoms 
that readily have an interaction potential.
This way, we manage the contexts during 
proof construction with respect to the interaction 
potentials of  the formula that are  processed at the 
rule instances.  The laziness and depth conditions 
that we impose make the processing of the formula 
more gradual, and reduce the size of the information 
being processed at every proof step, thereby resulting in a 
reduction in nondeterminism.  By combining 
the notions of laziness and depth, which are dual conditions 
together with the interaction condition, we 
introduce a rule that we call \emph{deep lazy 
interaction switch}. This rule results in a 
greater reduction in nondeterminism without losing 
the shorter proofs available due to deep inference,  
and reduces the cost of applying the interaction condition.

The  ideas on controlling nondeterminism in 
deep inference were initially introduced in \cite{Kah08}, 
where we  used the lazy interaction switch rule 
as a proof theoretic tool in proving NP-hardness of 
the logic $\BV$. As $\BV$ extends multiplicative linear logic
with the mix and nullary mix rules, the completeness argument 
in \cite{Kah08} benefits from the role played by 
the mix rule:  the presence of mix provides proofs for some formulae
that are not provable in $\MLL$. 
This phenomena, which results from the instances of the mix rule, 
 makes it possible to define a notion that we call \emph{independence} on 
the formulae, which, in \cite{Kah08}, simplifies the completeness argument. 
For example, 
the formulae $\pars{a \lpar \neg{a}}$ and $\pars{b \lpar \neg{b}}$
are independent, as the provability of one does not depend on the provability of the other,
and their disjunction  $\pars{a \lpar \neg{a} \lpar b \lpar \neg{b}}$ 
can be proved in $\BV$. 
However, the independence 
notion does not carry 
over to multiplicative linear logic directly, because, in contrast to $\BV$,  
the proofs of  two formulae do not deliver a proof of their disjunction.  
Moreover, the units $\lone$ and $\bot$ pose cases that require further attention.
We thus refine our technique and show that it does not rely on the mix rule.

In the following, 
we give a fragmented elaboration of different deep inference systems  
for multiplicative linear logic, where the nondeterminism is gradually  
reduced.  
For the presentation of the ideas, we use the standard notions and notations 
as presented in \cite{Gug02,StraTh,BruTh,Str03a,Str02,BG08,GS11,SG11}.
In the recent years, deep inference has grown to encompass
a family of proof theoretic formalisms \cite{GGP10}, and the calculus of structures 
remained as the simpler broadly-used deep inference framework. 
We therefore use the calculus of structures as the presentation framework.
We first show on a deep inference system for multiplicative linear logic 
that the nondeterminism due to the  units can be easily 
controlled by explicit inference rules. Following this,  within a lattice representation, 
we introduce the restrictions on the switch rule, which result in eight different systems.
We show the completeness of these systems  for multiplicative linear logic. 
We give a constructive cut elimination proof, 
which justifies that our technique is clean from a proof 
theoretic point of view. 
We show that our technique does not 
introduce any additional complexity in comparison to the sequent calculus
with respect to the size of the proofs. Following the approach in  
\cite{BG08,Das11,Das12},
 we show that the systems obtained with our technique polynomially simulate 
shallow inference systems. 
We then discuss the relation of our method with the focusing 
technique \cite{And92,And01,CGS12}.

Availability of a constructive cut-elimination 
procedure for the system with deep lazy interaction switch shows that this
system is independently meaningful from a proof theoretic point of view,  
in isolation from any other system for multiplicative linear logic. 
Moreover, the technique  and the rules that we use here are 
common to all the deep inference  systems, 
thus they should be applicable to other deep inference systems. As an 
evidence for this, 
we show that deep lazy interaction switch  can be replaced with 
the switch rule in a system for classical logic without breaking the
completeness of this system.

All the systems presented in this paper are implemented in Maude 
\cite{Kah07b}.\footnote{\texttt{http://sites.google.com/site/ozankahramanogullari/software/di\_in\_maude}}
An interactive prover, written in OCaml, is also available for 
download.\footnote{\texttt{http://sites.google.com/site/ozankahramanogullari/software/interana}}

%%%%%%%%%%%%%%%%%%%%%%%%%%%%%%%%%%%%%%%%%%%%%%%%%%%%
%%%%%%%%%%%%%%%%%%%%%%%%%%%%%%%%%%%%%%%%%%%%%%%%%%%%
%%%%%%%%%%%%%%%%%%%%%%%%%%%%%%%%%%%%%%%%%%%%%%%%%%%%
%%%%%%%%%%%%%%%%%%%%%%%%%%%%%%%%%%%%%%%%%%%%%%%%%%%%
%%%%%%%%%%%%%%%%%%%%%%%%%%%%%%%%%%%%%%%%%%%%%%%%%%%%
%%%%%%%%%%%%%%%%%%%%%%%%%%%%%%%%%%%%%%%%%%%%%%%%%%%%
%%%%%%%%%%%%%%%%%%%%%%%%%%%%%%%%%%%%%%%%%%%%%%%%%%%%
%%%%%%%%%%%%%%%%%%%%%%%%%%%%%%%%%%%%%%%%%%%%%%%%%%%%
%%%%%%%%%%%%%%%%%%%%%%%%%%%%%%%%%%%%%%%%%%%%%%%%%%%%
%%%%%%%%%%%%%%%%%%%%%%%%%%%%%%%%%%%%%%%%%%%%%%%%%%%%
%%%%%%%%%%%%%%%%%%%%%%%%%%%%%%%%%%%%%%%%%%%%%%%%%%%%
%%%%%%%%%%%%%%%%%%%%%%%%%%%%%%%%%%%%%%%%%%%%%%%%%%%%
%%%%%%%%%%%%%%%%%%%%%%%%%%%%%%%%%%%%%%%%%%%%%%%%%%%%
%%%%%%%%%%%%%%%%%%%%%%%%%%%%%%%%%%%%%%%%%%%%%%%%%%%%
%%%%%%%%%%%%%%%%%%%%%%%%%%%%%%%%%%%%%%%%%%%%%%%%%%%%
%%%%%%%%%%%%%%%%%%%%%%%%%%%%%%%%%%%%%%%%%%%%%%%%%%%%
%%%%%%%%%%%%%%%%%%%%%%%%%%%%%%%%%%%%%%%%%%%%%%%%%%%%
%%%%%%%%%%%%%%%%%%%%%%%%%%%%%%%%%%%%%%%%%%%%%%%%%%%%
%%%%%%%%%%%%%%%%%%%%%%%%%%%%%%%%%%%%%%%%%%%%%%%%%%%%
%%%%%%%%%%%%%%%%%%%%%%%%%%%%%%%%%%%%%%%%%%%%%%%%%%%%
%%%%%%%%%%%%%%%%%%%%%%%%%%%%%%%%%%%%%%%%%%%%%%%%%%%%
%%%%%%%%%%%%%%%%%%%%%%%%%%%%%%%%%%%%%%%%%%%%%%%%%%%%
%%%%%%%%%%%%%%%%%%%%%%%%%%%%%%%%%%%%%%%%%%%%%%%%%%%%
%%%%%%%%%%%%%%%%%%%%%%%%%%%%%%%%%%%%%%%%%%%%%%%%%%%%
%%%%%%%%%%%%%%%%%%%%%%%%%%%%%%%%%%%%%%%%%%%%%%%%%%%%
%%%%%%%%%%%%%%%%%%%%%%%%%%%%%%%%%%%%%%%%%%%%%%%%%%%%
%%%%%%%%%%%%%%%%%%%%%%%%%%%%%%%%%%%%%%%%%%%%%%%%%%%%
%%%%%%%%%%%%%%%%%%%%%%%%%%%%%%%%%%%%%%%%%%%%%%%%%%%%
%%%%%%%%%%%%%%%%%%%%%%%%%%%%%%%%%%%%%%%%%%%%%%%%%%%%
%%%%%%%%%%%%%%%%%%%%%%%%%%%%%%%%%%%%%%%%%%%%%%%%%%%%

\section{Proof Theory with Deep Inference} 
%%%%%%%%%%%%%%%%%%%%%%%%%%%%%%%%%%%%%%%%%%%%%%%%%%%%
%%%%%%%%%%%%%%%%%%%%%%%%%%%%%%%%%%%%%%%%%%%%%%%%%%%%
%%%%%%%%%%%%%%%%%%%%%%%%%%%%%%%%%%%%%%%%%%%%%%%%%%%%

In  deep inference deductive systems, the laws such as associativity 
and commutativity are  imposed on the logical expressions 
by means of an underlying equational system 
 \cite{Gug02,Str02,Str03a,Bru03a,TiuINT05,Bru06,GS11,SG11}. 
This is done at every inference step by working with congruence classes of 
formulae that are called structures: 
each structure, as a syntactic entity, denotes a set of formulae 
that are equivalent under a congruence relation, and inference rules are applied on 
these congruence classes. 
The notion of structure originates from graphs 
called relation webs \cite{Gug02}. 
The non-commutative multiplicative linear logic $\BV$, introduced in \cite{Gug02},
was obtained by preserving certain geometric 
properties of structures  at every  inference step. 
These inference rules gave rise to systems for various logics.

Structures, which we define 
for multiplicative linear logic,  
share properties with formulae and sequents, and they can be easily 
mapped to these entities.

\begin{defi}
\label{definition:formula}
There are countably many \emph{atoms}, denoted by $a,b,c,\ldots$  
The  \emph{structures}  $P$, $Q$, $R$, $S$\ldots \space  
of multiplicative  linear logic 
are generated by the grammar
$$
 R \grammareq a \mid
               \mathsf{1} \mid \bot \mid
  \pars{ R \lpar R } \mid
  \aprs{ R \ltens R } 
%  \mid \;  !R  \; \mid
%  \; ?R  \; 
\mid  \; \neg R  \; \quadcm
$$
where $a$ stands for any atom.  
$\lone$ and $\lbot$ 
are special atoms, called  \emph{one} and \emph{bottom}.
A structure
$\pars{R \lpar R}$ is a  \emph{par structure},
$\aprs{R \ltens  R}$ is  a \emph{copar (times) structure}
and $\neg R$ is the \emph{negation} of $R$. 
Structures are considered to be equivalent modulo the 
relation $\approx$, which is the smallest congruence relation 
induced by the equational system consisting of the equations
for 
\emph{associativity} and 
\emph{commutativity} for par and copar, and  
the \emph{equations} 
$$
 \overline{\strut\pars{R \lpar T}} \approx 
          \aprs{\overline{R} \ltens \overline{T}}, \quad
 \overline{\strut\aprs{R \ltens T}} \approx 
          \pars{\overline{R} \lpar \overline{T}}, \quad
 {\skew3\overline{\overline{R}} \approx  R}, \quad
\overline{\lbot}\approx  \lone, \quad
 \overline{\lone}\approx  \lbot
$$
\emph{for negation}. 
We denote the structures in 
the same equivalence class   
by picking a  structure from the equivalence class. 
If there is no ambiguity, 
we drop the superfluous brackets in the structures 
by resorting to the equations for associativity.
\end{defi}

\begin{rem} \label{remark:negation:equations}
Following \cite{Str03a}, in the definition of the structures above, we include the 
equations for negation  in the congruence relation imposed on structures.
However, in the systems discussed below, the inference rules do  not introduce 
any new negation symbols on the structures in proof construction. 
Because of this, we  consider the structures to be in negation normal form:
every structure can be considered in negation normal form by applying the equations 
for negation in Definition \ref{definition:formula} from left to right exhaustively 
as  term rewriting rules to push the negation symbol to the atoms. 
Given that a proof search is initiated with a structure in negation normal form,
no new negation symbol is introduced by the inference rules.
This allows us to disregard the equations for negation 
in the discussions below by assuming that proof construction
 is initiated with a structure in negation normal form. 
\end{rem}

\begin{nota}
In deep inference, the convention is using 
different types of parentheses to denote different logical 
operators by using the comma as an infix separator. 
For example, the expression $a \lpar b$ is often denoted 
by $\pars{a,b}$. 
In this paper, we adapt this notation to the usual 
notation of the logical operators by replacing the 
comma with the logical operators, 
for example,  as in $\pars{a \lpar b}$. 
\end{nota}

\begin{exa}
With respect to the congruence relation $\approx$ on the structures, 
we have  
$$
\pars{\aprs{\neg{b} \ltens \aprs{\neg{a} \ltens \lbot}} 
\lpar \pars{b \lpar \pars{a \lpar \lone}}} 
\approx 
\pars{ \pars{b \lpar \aprs{\aprs{\neg{a}   \ltens \lbot} \ltens \neg{b}}} 
\lpar \pars{\lone \lpar a}}
$$ 
and we can denote both of them  with 
$\pars{\aprs{\neg{a} \ltens \neg{b} \ltens \lbot} \lpar a \lpar b \lpar \lone}$.
\end{exa}

%%%%%%%%%%%%%%%%%%%%%%%%%%%%%%%%%%%%%%%
\begin{defi}
A \emph{structure context}, denoted as in  $S\cons{\;\;}$,
is a structure with a hole that does not appear in 
the scope of negation. The structure $R$ is 
a \emph{substructure}  of  $S\cons{R}$ and $S\cons{\;\;}$ 
is its \emph{context}.  
A context is empty if it is $\cons{\;\;}$. 
Context braces are omitted if no ambiguity is possible. That is,
whenever structural brackets of any kind surround the context of a hole, 
hole braces are omitted.
\end{defi}

\begin{exa}
Let $S\cons{\;\;}  = \pars{\cons{\;\;} 
\lpar a \lpar b \lpar \lone}$, $R = \neg{a}$ and  
$T = \aprs{\neg{b} \ltens \lbot}$. 
Then we have that
$$
S\aprs{R \ltens T}  = \pars{\aprs{\neg{a} \ltens \neg{b} \ltens \lbot} \lpar a \lpar b \lpar \lone} \; .
$$
\end{exa}

\begin{defi}
\label{definition:inference:rule}
An \emph{inference rule} is a scheme of the kind 
$$
\vcenter{\inf{\rho}{R}{\;T}} \; , 
$$ 
where
 $\rho$ is the \emph{name}  of the rule, 
$T$ is its \emph{premise}
 and $R$ is its \emph{conclusion}.
 A typical deep inference rule has the shape
$$
\vcenter{\inf{\rho}{S\cons{R}}{\;S\cons{T}}} \quad  ,
$$ 
which is 
determined by 
the implication 
$T \Rightarrow R$
inside a generic context $S\cons{\;\;}$. 
In the case of multiplicative linear logic, 
this  is  linear implication. 
In an instance of $\rho$, we say that $R$ is the \emph{redex} 
and $T$ is the \emph{contractum}.
A system $\sysS$ is a set of inference rules.
\end{defi}
 
An inference rule of the form in 
Definition \ref{definition:inference:rule}
specifies a step of rewriting. In this paper, 
these rewritings are those that rewrite 
the conclusion to the premise.
This is because we consider the inference rules 
for proof-search, thus we consider their  
bottom-up applications, which result in proofs 
that grow from the conclusion to the  premise.

\begin{rem}
In deductive systems with deep inference,
it is common to include a set of equations for the treatment of the
units of the logic in the
congruence relation 
in addition to those for associativity,
commutativity and negation.
Here, with a focus on proof search,
we choose a treatment of units by means of 
inference rules of the deductive system. This 
results in an equivalent proof theoretic consideration,
while providing more control over the units in proof construction
as we demonstrate below.
\end{rem}

\renewcommand{\rulebox}[1]{\mbox{$#1$}}
\begin{figure}[t]
%    \fbox{
%      \parbox{0.99\textwidth}{
%%%%%%%%%%%%%%%%%%%%%%%%%%%%%%%%%%%%%%%%%%%%%%
\begin{center}
{
\small
\framebox[5.2in]
{
\begin{minipage}[t]{5.2in}

\begin{center}
\vspace{2mm}
%%%%%%%%%%%%%%%%%%%%%%%%%%%%%%%%%%%%%%%%%%%%%%%
$$
\begin{array}{c}
       \vcenter{\dernote{\ruleaidown}{}
          {S\pars{a \lpar \neg a}}
          {\leaf{S\cons{1}}}}
\qquad \qquad
        \vcenter{\dernote{\swir}{}
        {S\pars{U \lpar \aprs{R \ltens T}   }}
        {\leaf{S\aprs{\pars{U \lpar  R} \ltens T}}}}
\qquad \qquad
      \vcenter{\dernote{\ruleaiup}{}{S\cons{R}}
                        {\leaf{S\pars{R \lpar \aprs{a \ltens \neg{a} }}}}}
%%%%%%%%%%%%%%%%%%
%%%%%%%%%%%%%%%%%%
\\[20pt]
%%%%%%%%%%%%%%%%%%
%%%%%%%%%%%%%%%%%%
\dernote{\unitonedown\!}{}{S\pars{R \lpar \lbot}}{\leaf{S\cons{R}}}
\qquad \quad
\dernote{\unitoneup\!}{}{S\cons{R}}{\leaf{S\aprs{R \ltens \lone}}}
\qquad \quad
\dernote{\unittwodown\!}{}{S\aprs{R \ltens \lone}}{\leaf{S\cons{R}}}
\qquad \quad 
\dernote{\unittwoup\!}{}{S\cons{R}}{\leaf{S\pars{R \lpar \lbot}}}
\end{array}
$$
%%%%%%%%%%%%%%%%%%%%%%%%%%%%%%%%%%%%%%%%%%%%%%%%%%
\vspace{2mm}

\end{center}
\end{minipage}
}  }

\end{center}
%%%%%%%%%%%%%%%%%%%%%%%%%%%%%%%%%%%%%%%%%%%%%%%%%%%
    \caption{Deep Inference System $\SMLS$}
    \label{figure:SMLS}
%\vspace{-4mm}
\end{figure}

\begin{defi} 
\label{definition:SMLS}
The rules of the 
$\mathsf{S}$ymmetric 
deep inference 
$\mathsf{M}$ultiplicative $\mathsf{L}$inear logic $\mathsf{S}$ystem 
or   $\SMLS$
are depicted in Figure \ref{figure:SMLS}. 
The rules are called 
\emph{atomic interaction} ($\ruleaidown$), 
$switch$ ($\swir$), 
\emph{atomic cut} ($\ruleaiup$), 
unit one down ($\unitonedown$),
unit one up ($\unitoneup$),
unit two down ($\unittwodown$) and
unit two up ($\unittwoup$),
respectively.
\end{defi}

\begin{defi}
The system $\MLS$ is obtained by removing the cut rule ($\ruleaiup$) from $\SMLS$.
\end{defi}

\begin{defi}
The system $\MLSu$ is obtained by removing 
the rules $\unitoneup$, $\unittwoup$ and 
the cut rule from $\SMLS$.
\end{defi}

%%%%%%%%%%%%%%%%%%%%%%%%%%%%%%%

\begin{rem}
In deep inference, deductive systems for different logics 
employ different treatments of the units in accordance with the 
underlying deductive setting.  
For example, in the logic $\BV$ \cite{Gug02,Kah08}, 
the units $\lbot$ and $1$ of multiplicative linear logic
merge into %%% a single 
one unit,  as they are logically 
equivalent in the presence of the mix rule ($\mathsf{mix}$) and the nullary mix ($\mathsf{mix0}$). 
In their sequent calculus
presentation, the rules $\mathsf{mix}$ and $\mathsf{mix0}$ 
are given as follows:
$$
\ddernote{\mathsf{mix}}{}{\sqn{ \Gamma, \Delta}}{\leaf{\sqn{\Gamma\;}}}{\leaf{\;\sqn{\Delta}}}
\qquad \qquad
\dernote{\mathsf{mix0}}{}{\;\sqn{\;}}{\leaf{}}
$$
When multiplicative linear logic in the sequent calculus (Definition \ref{definition:MLL}, Figure \ref{figure:MLL})
is extended with the rules $\mathsf{mix}$ and $\mathsf{mix0}$, 
it becomes possible to prove the logical equivalence of 
the units $\lbot$ and $\lone$, and as a result of this replace the two units 
with a single unit. In $\BV$ this equivalence
 is realized by such a single unit that replaces 
the units $\lbot$ and $\lone$ \cite{Gug02}. 
In fact, due to this equivalence, 
the unit can be removed completely from
the deductive system of the logic $\BV$ \cite{Kah04}. 
However, in multiplicative linear logic,
the units play a different role, similar to other atoms, 
which provides 
an expressive power for the unit-only fragment:  
as shown by 
Lincoln and Winkler \cite{LiWi94}, 
the fragment of $\MLL$ with only units and no other atoms 
is NP-complete. 
\end{rem}

\begin{defi}
A \emph{derivation} $\Delta$ is a finite chain
of instances of inference rules.
A derivation can consist of just one structure. The top-most
structure in a derivation  is called the
\emph{premise}, and the bottom-most structure is
called its \emph{conclusion}. 
A derivation $\Delta$ whose premise
is $T$, conclusion is $R$, and inference rules are in
 $\sysS$ will be written as
 $$
\vcenter{\xy\xygraph{[]!{0;<2.2pc,0pc>:}
      {\;T}-@{=}^<>(.5){\strut\sysS} _<>(.5){\strut\Delta}[d] {R}
    }\endxy}\, .
$$
A multiplicative linear logic \emph{proof} $\Pi$ in deep inference 
is either the unit $\lone$ or a finite  derivation whose premise is the unit $\mathsf{1}$.
The size $|\Pi|$ of a proof
$\Pi$ is the number of unit and atom occurrences appearing in it.
\end{defi}

In contrast to proofs and derivations in the sequent calculus, which are trees, 
the deep inference proofs are chains of instances of the inference rules. 
In deep inference,
the inference rules are not restricted to be applied at the main connective. 
That is,  when the the formula is seen as a term, 
they can be applied not only at the root position, 
but also at other positions of the term.
As it has been shown by Bruscoli and Guglielmi \cite{BG08}, 
in some cases this provides shorter proofs than those that are 
provided by the notion of inference of the sequent calculus that 
we call \emph{shallow inference}:  shallow inference 
permits the application of the inference rules only at the top level connective. 
However,   
applicability of the inference rules  at any depth inside a structure makes it 
possible to start the construction of a proof from the substructures. This brings 
about many more proofs, including those available with shallow inference. 
Some of these proofs are much shorter than the proofs in any bounded depth 
system \cite{Das11,Das12}: bounded depth systems are those that retain the top-down 
symmetry of deep inference, but can otherwise be designed at the same depth 
level of sequent calculus systems. 
In some cases, deep inference proofs are exponentially shorter than 
shallow inference proofs \cite{BG08}, 
which we discuss in  Section \ref{section:int:depth:KSg}.
Let us see the reason for this on an example.

\begin{exa}
\label{example:atomic:interaction}
Consider the two proofs below of the structure 
$ \pars{  \aprs{\pars{a \lpar \neg{a}} \ltens \neg{b}} \lpar b} $ 
in  $\MLS$. In the proof on the left, we use shallow inference rules, 
where the context $S\cons{\;\;}$ is $\cons{\;\;}$,
resembling  the application of the sequent calculus rules.
This requires prior rule instances at the top-level
in order to access other instances.
  In the proof on the right, 
we use deep inference rules. 
$$
\dernote{\swir}{}{  \rdx{\pars{  \aprs{\pars{a \lpar \neg{a}} \ltens \neg{b}}  \lpar b}  }}{
\rootr{\ruleaidown;\, \unittwodown\;}{}{  \aprs{  \pars{a \lpar \neg{a}} \ltens \pars{b \lpar  \neg{b}}}  }{
\rootr{\ruleaidown\; }{}{ \pars{b \lpar \neg{b}}  }{
\leaf{  \lone }}}}
\qquad \qquad 
\dernote{\ruleaidown;\, \unittwodown}{}{ 
\pars{  \aprs{\rdx{\pars{a \lpar \neg{a}}} \ltens \neg{b}}  \lpar b}  }{
\rootr{\ruleaidown\; }{}{ \pars{b \lpar \neg{b}} }{
\leaf{  \lone }}}\\[8pt]
$$
\end{exa}
%
%%%%%%%%%%%%%%%%%%%%%%%%%%%%%%%%%%%%%
%%%%%%%%%%%%%%%%%%%%%%%%%%%%%%%%%%%%%
%%%%%%%%%%%%%%%%%%%%%%%%%%%%%%%%%%%%%
%%%%%%%%%%%%%%%%%%%%%%%%%%%%%%%%%%%%%
%%%%%%%%%%%%%%%%%%%%%%%%%%%%%%%%%%%%%
%%%%%%%%%%%%%%%%%%%%%%%%%%%%%%%%%%%%%
Despite the availability of shorter proofs, 
deep inference causes a greater non-determinism in proof search: 
the inference rules can be applied at many more positions than in the sequent calculus.
This results in a much larger breadth of the search space in proof search. 
Some of this increased non-determinism is due to the treatment of units, 
which provides a rich proof theory, however it 
results in redundant rule instances.

\begin{exa}
\label{example:unit:redundant}
The following derivations involve rule  instances, which do not contribute to the construction of proofs. 
In particular, the two derivations on the left result in  
the premises and the conclusions  that are the same structures, whereas the 
derivation on the right transforms a provable structure in the 
conclusion to a  structure in the premise that is not provable.
These derivations  are instantiated 
with the application of  $\unitoneup$ and $\unittwoup$.

$$
\dernote{\unitoneup}{}{\pars{a \lpar\rdx{b}}}{
\rootr{\swir\;}{}{\pars{a \lpar \aprs{b \ltens \lone}}}{
\rootr{\unittwodown\;}{}{\aprs{\pars{a \lpar b}\ltens \lone}}{
\leaf{\pars{a \lpar b}}}}}
\qquad \qquad 
\dernote{\unittwoup}{}{\rdx{\aprs{a \ltens b}}}{
\rootr{\swir\;}{}{\pars{\lbot \lpar \aprs{a \ltens b}}}{
\rootr{\unitonedown\;}{}{\aprs{\pars{\lbot \lpar a}\ltens b}}{
\leaf{\aprs{a \ltens b}}}}}
\qquad \quad
\dernote{\unitoneup}{}{\pars{\rdx{a} \lpar \neg a}}{
\rootr{\swir\;}{}{\pars{\aprs{a \ltens \lone} \lpar \neg a}}{
\leaf{\aprs{a \ltens \pars{\lone \lpar \neg a}}}}}
\\[8pt]
$$
\end{exa}

The system $\MLS$ contains the rules $\unitoneup$ and $\unittwoup$, 
which are not included in $\MLSu$. 
In order to avoid derivations of the form in Example \ref{example:unit:redundant} 
in proof search, we thus remove the 
rules $\unitoneup$ and $\unittwoup$ from $\MLS$, and obtain the system $\MLSu$. 
The completeness of $\MLSu$ for multiplicative linear logic can be easily shown
 by inductively translating $\MLL$ proofs in the sequent calculus 
into $\MLSu$ proofs, or alternatively via cut elimination. 
We use the proof in \cite{Str03a}, where $\MLL$ proofs in the sequent calculus  
are transformed into proofs in  $\MLS$. Here the key argument is the fact that  
the rules $\unitoneup$ and $\unittwoup$ do not play a role in the translation.

\begin{defi}
\label{definition:MLL}
The Multiplicative Linear Logic system in the sequent calculus or $\MLL$ is depicted in Figure \ref{figure:MLL}. 
\end{defi}

\renewcommand{\rulebox}[1]{\mbox{$#1$}}
\begin{figure}[t]
%    \fbox{
%      \parbox{0.99\textwidth}{
%%%%%%%%%%%%%%%%%%%%%%%%%%%%%%%%%%%%%%%%%%%%%%
\begin{center}
{
\small
\framebox[5.2in]
{
\begin{minipage}[t]{5.2in}

\begin{center}
\vspace{2mm}
%%%%%%%%%%%%%%%%%%%%%%%%%%%%%%%%%%%%%%%%%%%%%%%
$$
\begin{array}{ll}
  \rulebox{\inf{\axiom}{\vdash A, \neg A} {}}
 \qquad
\rulebox{\iinf{\ltens} {\sqn{\aprs{A \ltens B}}, \Phi, \Psi}
  {\sqn{A,\Phi}}{\sqn{B,\Psi}}} 
\qquad
 \rulebox{\inf{\lpar}{\sqn{\pars{A \lpar B}, \Phi}}{\sqn{A,B,\Phi}} }
\qquad
 \rulebox{\inf{\lbot}{\sqn{\lbot, \Phi}}{\sqn{\Phi}} }
\qquad
 \rulebox{\inf{\lone}{\sqn{\lone}}{} }
\end{array}
$$
%%%%%%%%%%%%%%%%%%%%%%%%%%%%%%%%%%%%%%%%%%%%%%%%%%
\vspace{2mm}

\end{center}
\end{minipage}
}  }

\end{center}
%%%%%%%%%%%%%%%%%%%%%%%%%%%%%%%%%%%%%%%%%%%%%%%%%%%
%        }   }
 %   \vspace{-2mm}        
    \caption{$\MLL$ in the sequent calculus}
    \label{figure:MLL}
%\vspace{-4mm}
\end{figure}

The system $\MLL$ is defined on formulae, which are different from structures as the equalities for 
associativity and commutativity do not apply to formulae. Here, for simplicity 
we use the same notation for structures and formulae, and allow the equalities for associativity 
and commutativity to operate on structures. The following theorem, which makes use of this 
simplification, states that  $\MLS$ is complete for multiplicative linear logic. The proof of 
this statement can be found in \cite{Str03a}. 

\begin{thm}
\label{theorem:equivalent:MLL:MLS}
A structure $R$ has a proof in $\MLS$ if and only if the expression  obtained 
by translating $R$ into a formula has a proof in $\MLL$. 
\end{thm}

\begin{defi}
Two systems $\sysS$ and $\sysS'$ are 
\emph{equivalent} if they prove the same structures. 
A rule $\rho$ is \emph{admissible} for a system $\sysS$ if for every proof 
in $\sysS \cup \{ \rho \}$ there is a proof in $\sysS$ with the same conclusion. 
\end{defi}

\begin{thm}
\label{theorem:equivalent:MLS:MLSu}
The systems $\MLS$ and $\MLSu$ are equivalent.
\end{thm} 

\proof
Every proof in $\MLSu$ is a proof in $\MLS$.
For the proof of the other direction, by Theorem \ref{theorem:equivalent:MLL:MLS},
given an $\MLL$ proof $\nabla$, 
by structural induction we construct a proof $\Pi$ in $\MLSu$, omitting the obvious translation from 
the sequents to structures. The only difference  between this construction and  the proof in \cite{Str03a} 
 is in the use of the rules  $\unitonedown$ and $\unittwodown$
instead of the equalities for unit. 
\begin{itemize}
%%%%%%%%%%%%%%%%%%%%%%%
%%%%%%%%%%%%%%%%%%%%%%%
\item
If $\nabla$ is an instance of  the rule $\mathsf{id}$ in $\MLL$ then take the proof given with $\ruleidown$, 
which is the generic version of  $\ruleaidown$, and 
derivable for $\ruleaidown$, $\swir$ and $\unittwodown$. That is, for every instance of 
the rule $\mathsf{id}$, take an instance of the rule 
$$
\vcenter{\dernote{\ruleidown}{}{S\pars{R \lpar \neg R}}{\leaf{S\cons{\lone}}}}\, , 
\; \textrm{which we use as a short-hand for a derivation} 
\vcenter{\xy\xygraph{[]!{0;<2.4pc,0pc>:}
{
S\cons{\lone}
  }-@{=}^<>(.5){\{\, \ruleaidown,\, \swir,\, \unittwodown \,\}} _<>(.5){\Delta}[d] {
S\pars{R \lpar \neg R}
}
    }\endxy}.
$$
The derivation $\Delta$ is constructed by structural induction on $R$. The base case is given with an instance of 
$\ruleaidown$ or $\unittwodown$ for the cases where $R$ is an atom or a unit. For the inductive case, where 
$R = \pars{R_1 \lpar R_2}$ or $R = \aprs{R_1 \ltens R_2}$, apply the induction hypothesis to 
$$
\dernote{\swir}{\quad.}{S\pars{R_1 \lpar R_2 \lpar \aprs{\neg R_1 \ltens \neg R_2}}}{
\rootr{\ruleidown\;}{}{S\pars{ R_2 \lpar \aprs{\pars{ R_1 \lpar \neg R_1} \ltens \neg R_2}}}{
\rootr{\unittwodown\; }{}{S\pars{ R_2 \lpar \aprs{\lone \ltens \neg R_2}}}{
\rootr{\ruleidown\;}{}{S\pars{ R_2 \lpar \neg R_2}}{
\leaf{S\cons{\lone}}}}}}
\\[8pt]
$$

%%%%%%%%%%%%%%%%%%%%%%%
%%%%%%%%%%%%%%%%%%%%%%%
\item
If the rule $\ltens$ in $\MLL$ is the last rule applied in $\nabla$, then take the following  proof, 
where $\Pi_1$ and $\Pi_2$ are given by induction hypothesis.
$$
\vcenter{\xy\xygraph{[]!{0;<6.2pc,0pc>:}
{\lone
  }-@{=}^<>(.5){} _<>(.5){\Pi_2}[d] {
\vcenter{\xy\xygraph{[]!{0;<4.8pc,0pc>:}
{
\dernote{\unittwodown}{\quad}{ \aprs{\pars{A  \lpar \Phi} \ltens \lone}     }{\leaf{\pars{A  \lpar \Phi}}}
  }-@{=}^<>(.5){} _<>(.5){\Pi_1}[d] {
\dernote{\swir}{}{\pars{\aprs{A \ltens B} \lpar \Phi \lpar \Psi }}{
\rootr{\swir\;}{}{  \pars{\aprs{\pars{A  \lpar \Phi} \ltens B} \lpar \Psi } }{
\leaf{
\aprs{\pars{A  \lpar \Phi} \ltens \pars{B \lpar \Psi }} }}}
}
    }\endxy}
}
    }\endxy}
\\[8pt]
$$
%%%%%%%%%%%%%%%%%%%%%%%%
%%%%%%%%%%%%%%%%%%%%%%%%
\item
If the rule $\lpar$ in $\MLL$ is the last rule applied in $\nabla$, 
then the proof of the premise exists by induction hypothesis.\\

%%%%%%%%%%%%%%%%%%%%%%%%
%%%%%%%%%%%%%%%%%%%%%%%%
\item
If the rule $\lbot$ in $\MLL$ is the last rule applied in $\nabla$, then take the following  proof, 
where $\Pi_1$ is  given by induction hypothesis.
$$
\vcenter{\xy\xygraph{[]!{0;<3pc,0pc>:}
{
\lone
  }-@{=}^<>(.5){} _<>(.5){\Pi_1}[d] {
\dernote{\unitonedown}{\quad\:}{\pars{\Phi \lpar \lbot }}{
\leaf{ \Phi }}
}
    }\endxy}
\\[8pt]
$$
%%%%%%%%%%%%%%%%%%%%%%%
%%%%%%%%%%%%%%%%%%%%%%%
\item 
if $\nabla$ is an instance of the rule $\lone$ in $\MLL$ then take the proof given with $\lone$.
\end{itemize}
%%%%%%%%%%%%%%%%%%%%%%%
%%%%%%%%%%%%%%%%%%%%%%%
%%%%%%%%%%%%%%%%%%%%%%%
%%%%%%%%%%%%%%%%%%%%%%%
Alternatively, a completeness argument for $\MLSu$
that does not rely on the sequent calculus is given 
by removing the instances of the rules 
$\unitoneup$ and $\unittwoup$ from the proofs in $\MLS$ by using  
a cut elimination argument:
given a proof in $\MLS$, we replace each instance of 
$\unitoneup$ and $\unittwoup$ with the following derivations, 
which introduce the instances of the cut rule $\ruleaiup$.

$$
\vcenter{
\dernote{\ruleaiup}{}{S\cons{R}}{
\rootr{\swir\;}{}{S\pars{R \lpar \aprs{\lone \ltens \lbot }}}{
\rootr{\unitonedown\,}{}{S\aprs{\pars{\lbot \lpar R } \ltens \lone}}{
\leaf{S\aprs{R \ltens \lone}}}}}
}
\qquad
\qquad
\vcenter{
\dernote{\ruleaiup}{}{S\cons{R}}{
\rootr{\unittwodown\; }{}{S\pars{R \lpar  \aprs{\lone \ltens \lbot}}}{
\leaf{S\pars{R \lpar \lbot}}}}
}
\\[8pt]
$$
Then, to remove the instances of the cut rule, we apply a cut elimination procedure, 
which does not introduce any new instances of  $\unitoneup$ and $\unittwoup$. 
In Subsection \ref{subsection:cut:elimination}, we give such a 
cut elimination procedure for a more restricted system for multiplicative linear logic than $\MLSu$,
that is, $\MLSdli$. This proof  can be extended to $\MLSu$ by 
considering a similar case analysis, however without the need to use 
Lemma \ref{lemma:reslazyswitch:independence}.
\qed

The  treatment of units $\lone$ and $\lbot$ by inference rules as in $\MLSu$
becomes useful in controlling the non-determinism in proof search, in contrast to the 
case, where the units are treated with respect to a congruence relation.

\begin{exa}
\label{example:switch:instances}
Consider the structure $\pars{\aprs{\neg a \ltens \neg b} \lpar a \lpar b}$, 
which is provable in  $\MLSu$.
When the units are treated by the congruence 
relation that contains the equalities for associativity and commutativity as well as 
the equalities 
$$
\pars{R \lpar \lbot} \approx R \qquad \textrm{and}\qquad \aprs{R \ltens \lone} \approx R 
$$
for unit as in  \cite{Str03a}, the switch rule can be 
applied bottom-up to this structure in 28 different ways. 
However, only 2 of 
these instances can result in a proof. This can be checked by using the
 implementation of these systems, given in \cite{Kah07b}. Controlling the treatment of units 
with inference rules as in  $\MLSu$ reduces the number of instances of the switch rule 
from 28 to 6, listed below, where those resulting in a proof are  
the cases given with $( iii.)$ and $(vi.)$ below.

$$
\begin{array}{c}
(i.) \,
\vcenter{
\dernote{\swir}{}{\pars{\aprs{\neg a \ltens \neg b} \lpar a \lpar b}}{
\leaf{\aprs{ \neg b \ltens\pars{\neg a \lpar a \lpar b}}}}
}
\quad
(ii.) \;
\vcenter{
\dernote{\swir}{}{\pars{\aprs{\neg a \ltens \neg b} \lpar a \lpar b}}{
\leaf{\pars{\aprs{\pars{b \lpar \neg a} \ltens \neg b} \lpar a}}}
}
\quad
(iii.) \;
\vcenter{
\dernote{\swir}{}{\pars{\aprs{\neg a \ltens \neg b} \lpar a \lpar b}}{
\leaf{\pars{\aprs{\pars{a \lpar \neg a} \ltens \neg b} \lpar b}}}
}
%%%%%%%%%%%%%%%%%%
%%%%%%%%%%%%%%%%%%
\\[28pt]
%%%%%%%%%%%%%%%%%%
%%%%%%%%%%%%%%%%%%
(iv.)\;
\vcenter{
\dernote{\swir}{}{\pars{\aprs{\neg a \ltens \neg b} \lpar a \lpar b}}{
\leaf{\aprs{\neg a \ltens \pars{\neg b \lpar a \lpar b}}}}
}
\quad
(v.)\;
\vcenter{
\dernote{\swir}{}{\pars{\aprs{\neg a \ltens \neg b} \lpar a \lpar b}}{
\leaf{\pars{\aprs{\pars{a \lpar \neg b} \ltens \neg a} \lpar b}}}
}
\quad
(vi.)\;
\vcenter{
\dernote{\swir}{}{\pars{\aprs{\neg a \ltens \neg b} \lpar a \lpar b}}{
\leaf{\pars{\aprs{\pars{b \lpar \neg b} \ltens \neg a} \lpar a}}}
}
\\[8pt]
\end{array}
$$
\end{exa}

%%%%%%%%%%%%%%%%%%%%%%%%%
%%%%%%%%%%%%%%%%%%%%%%%%%
%%%%%%%%%%%%%%%%%%%%%%%%%
%%%%%%%%%%%%%%%%%%%%%%%%%
%%%%%%%%%%%%%%%%%%%%%%%%%
%%%%%%%%%%%%%%%%%%%%%%%%%
%%%%%%%%%%%%%%%%%%%%%%%%%
%%%%%%%%%%%%%%%%%%%%%%%%%
%%%%%%%%%%%%%%%%%%%%%%%%%
%%%%%%%%%%%%%%%%%%%%%%%%%
%%%%%%%%%%%%%%%%%%%%%%%%%
%%%%%%%%%%%%%%%%%%%%%%%%%

\section{Lazy and Deep Interaction}

%%%%%%%%%%%%%%%%%%%%%%%%%
%%%%%%%%%%%%%%%%%%%%%%%%%
%%%%%%%%%%%%%%%%%%%%%%%%%

A controlled treatment of units is useful for reducing the non-determinism in proof search.
However, an important part of the non-determinism in proof search with deep inference systems is due to 
the switch rule, which is the rule responsible for context management. 

\begin{exa}
\label{example:nested:structure}
Consider the structure 
$\pars{\aprs{ \pars{\aprs{  \neg c \ltens \neg d} \lpar c \lpar  d } \ltens  \neg a \ltens \neg b} \lpar a \lpar  b }$,
which is provable in $\MLSu$. This structure is obtained by nesting the structure in Example 
\ref{example:switch:instances} in itself. In the 
system, where the units are treated with the congruence relation, the switch rule can be 
applied bottom-up to this structure in 70 different ways. 
While $\MLSu$ prunes 46 of these instances, leaving 24 of them, only 4 of these instances 
provide proofs. The 2 of these instances at the deeper positions provide shorter proofs in comparison 
to the 2 others.
\end{exa}

%%%%%%%%%%%%%%%%%%%%%%%%%
%%%%%%%%%%%%%%%%%%%%%%%%%
%%%%%%%%%%%%%%%%%%%%%%%%%
%%%%%%%%%%%%%%%%%%%%%%%%%
%%%%%%%%%%%%%%%%%%%%%%%%%
%%%%%%%%%%%%%%%%%%%%%%%%%

\subsection{Controlling Context Management}

%%%%%%%%%%%%%%%%%%%%%%%%%
%%%%%%%%%%%%%%%%%%%%%%%%%

With the following definitions, 
we re-design the switch rule to control the non-determinism 
in proof search due to context management.

\begin{defi}
Given a structure $R$,  $\at \, R$ is the set of the atoms in $R$. 
\end{defi}

\begin{exa}  \label{example:at}
For $R = \pars{a \lpar \neg{a} \lpar b \lpar \aprs{\neg{a} \ltens \lbot} 
 \lpar \aprs{a \ltens \neg{b}}}$, 
we have $\at \, R = \{  a, \neg{a}, b,\neg{b}, \lbot \}$.
\end{exa}

\begin{defi}
\label{definition:deep:lazy:int:swir}
Consider the switch rule. %, given in Definition \ref{definition:SMLS}.

$$
        \vcenter{\dernote{\swir}{}
        {S\pars{U \lpar \aprs{R \ltens T}   }}
        {\leaf{S\aprs{\pars{U \lpar  R} \ltens T}}}}
\\[8pt]
$$
\begin{enumerate}[label=(\roman*)]
\item%[(i)]
We say that it is \emph{interaction switch ($\mathsf{is}$)} if  
$\at\, \overline{R} \, \cap \,   \at\, U \neq \emptyset$, that is, 
$R$ and $U$ contain complementary atoms. In this case, 
we say $R$ and $U$ can interact.\\
\item%[(ii)]
It is \emph{lazy switch ($\mathsf{ls}$)} if $U$ is a copar 
or an atom different from $\lbot$.\\
\item%[(iii)]
It is \emph{deep switch ($\mathsf{ds}$)}
if $R$ is a par or an atom different from $\lone$.\\
\end{enumerate} 
By combining these restrictions, we thus obtain 
$8$ different inference rules. For example, by imposing 
all these restrictions, we obtain the rule \emph{deep lazy 
interaction switch} ($\deepreslazyswir$).
\end{defi}

\begin{exa} \label{example:derivation}
Consider the structure 
$\pars{\aprs{\neg{a} \ltens \neg{b} \ltens \lbot} \lpar a \lpar b \lpar \lone}$
with the following instances of the switch rule where the 
$R$ and $U$ structures are shaded:

$$
\dernote{(i.)\,}{}{\pars{\aprs{\rdx{\neg{a}} \ltens \neg{b} \ltens \lbot} 
 \lpar a \lpar \rdx{b \lpar \lone}}}{
\leaf{\pars{\aprs{\pars{\neg{a} \lpar b \lpar \lone} \ltens \neg{b} \ltens \lbot} \lpar a}}}
\qquad \quad
\dernote{(ii.)\,}{}{\pars{\aprs{\rdx{\neg{a}} \ltens \neg{b} \ltens \lbot} 
 \lpar \rdx{a} \lpar b \lpar \lone}}{
\leaf{\pars{\aprs{\rdx{\pars{a \lpar \neg{a}}} \ltens \neg{b} \ltens \lbot} 
 \lpar b \lpar \lone}}}
\\[8pt]
$$
%(i) is an instance of $\lazyswir$, but not an instance of $\deepswir$ or $\intswir$.
(i) is an instance of $\deepswir$ but not an instance of $\lazyswir$ or $\intswir$.
(ii) is an instance of $\deeplazyintswir$ because it is  
an instance of the both $\lazyswir$ and $\deepswir$, 
and also    
$\at \, \neg{R} = \{ a \}$ and $\at \, U = \{ a \}$, and thus 
$\at \, \neg{R} \, \cap \, \at \, U = \{ a \} \neq \emptyset$.
\end{exa}

\begin{rem}
\label{remark:counter:example}
Replacing the interaction condition 
$\at \,\neg R \, \cap \, \at \, U \neq \emptyset$ in 
Definition \ref{definition:deep:lazy:int:swir} with 
$\at \,\neg U = \at \, R$
or 
$\at \,\neg U \subseteq \at \, R$
results in an incomplete system for multiplicative linear logic. 
To see this on an example, consider the structure:
$$
\pars{  \aprs{b \ltens \neg c} \lpar \aprs{a \ltens c}  \lpar 
     \aprs{ \pars{\neg a \lpar \neg b} \ltens \pars{\neg d \lpar  \neg e} }  \lpar 
          \aprs{ d \ltens f } \lpar \aprs{ \neg f \ltens e}  } 
$$
This structure is provable in $\MLSu$, 
however it cannot be proved in a system where the switch rule is
modified such that $\at \,\neg U = \at \, R$
or 
$\at \,\neg U \subseteq \at \, R$.
\end{rem}

The rule \emph{deep switch} was introduced in \cite{StraTh}
to simplify the cut elimination procedure by means of permutations of 
the inference rules. The intuition behind the naming of this rule 
is that it minimizes the extra depth introduced by the switch rule as 
a result of its bottom up application.  

\begin{figure}[t]
$$
\begin{array}{c}
 \psfrag{name1}{$\mathsf{dlis}$}
 \psfrag{name2}{$\mathsf{lis}$}
 \psfrag{name3}{$\mathsf{dls}$}
 \psfrag{name4}{$\mathsf{dis}$}
 \psfrag{name6}{$\mathsf{ls}$}
 \psfrag{name7}{$\mathsf{is}$}
 \psfrag{name8}{$\mathsf{ds}$}
 \psfrag{name9}{$\mathsf{s}$}
\includegraphics[width=4cm, height=3.2cm]{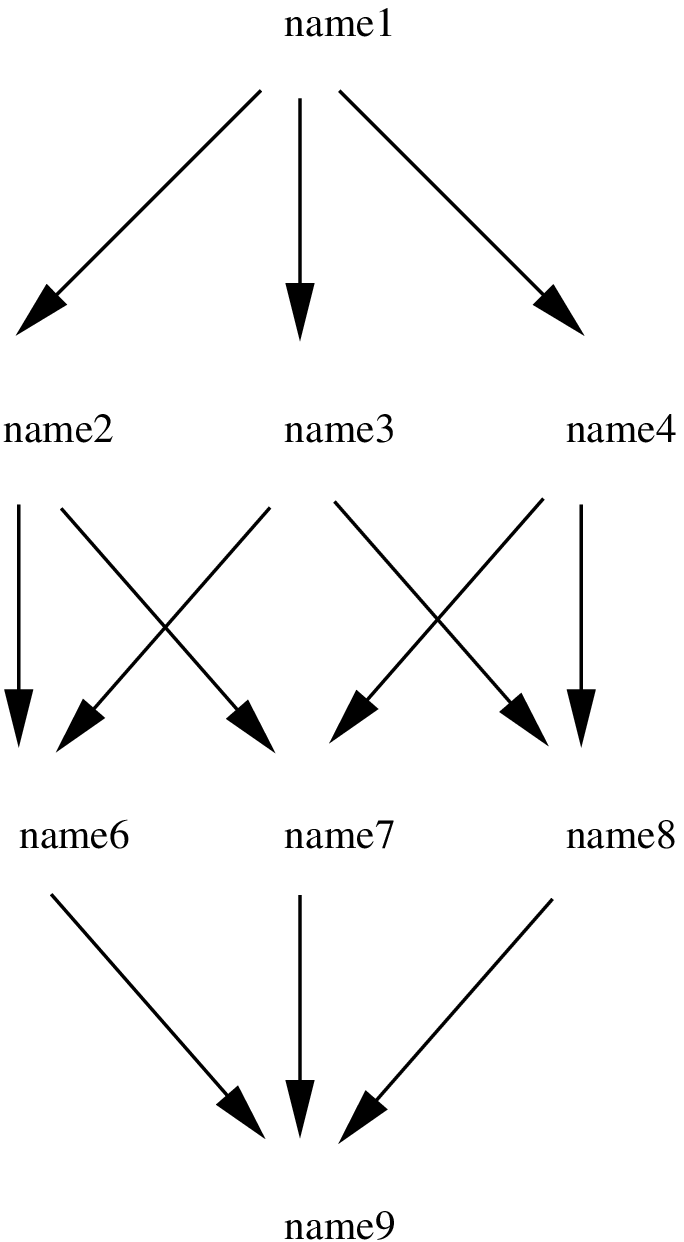}
\end{array}
\qquad \qquad
\begin{tabular}{|c||c|c|}
\cline{1-3}
{\footnotesize\textsf{MLSu}} & $\{ \, \mathsf{s} \, \}$ &\\
\cline{1-2}
{\footnotesize\textsf{MLSl}} & $\{ \, \mathsf{ls} \, \}$ & \\
\cline{1-2}
{\footnotesize\textsf{MLSd}}  & $\{ \, \mathsf{ds} \, \}$ & \\
\cline{1-2}
{\footnotesize\textsf{MLSi}}  & $\{ \, \mathsf{is} \, \} $ & \\
\cline{1-2}
{\footnotesize\textsf{MLSdl}} & $\{ \,\mathsf{dls} \, \}$ &   $\cup \; \{ \ruleaidown , \unitonedown ,\unittwodown \} $\\
\cline{1-2}
{\footnotesize\textsf{MLSli}} & $\{ \, \mathsf{lis} \, \}$ &  \\
\cline{1-2}
{\footnotesize\textsf{MLSdi}} & $\{ \,\mathsf{dis} \, \} $ &  \\
\cline{1-2}
{\footnotesize\textsf{MLSdli}} & $\{ \,\mathsf{dlis} \, \} $ & \\
\hline
\end{tabular}
$$ 
\caption{
\textbf{Left:} A partial order representation of the inclusion 
relation of the different switch rule instances, for example,
every instance of the rule $\deeplazyintswir$ is an instance of the 
rule $\lazyintswir$, which is an instance of the rule 
$\intswir$.
\textbf{Right: } The systems that are obtained by replacing the switch rule in $\MLSu$ 
with its restricted versions given in Definition \ref{definition:deep:lazy:int:swir}. 
}
    \label{figure:switch}
\end{figure}

The restrictions on lazy and deep switch are dual 
restrictions with respect to the structures that they are imposed on: 
in its instances, lazy switch imposes $U$ in a par 
context not to be a par structure, whereas deep switch imposes 
$R$ in a copar context not to be a copar structure.  
Figure \ref{figure:switch}
shows the relation between the 8 rules obtained, where the arrow denotes the 
inclusion relation of their instances,  for example, every instance of the 
rule $\deepintswir$ is an instance of $\deepswir$  and it is also  
an instance of $\intswir$.  

%%%%%%%%%%%%%%%%
%%%%%%%%%%%%%%%%
%%%%%%%%%%%%%%%%
%%%%%%%%%%%%%%%% 

\begin{defi} \label{definition:switch:systems}
\emph{System} $\MLSu$ \emph{with deep lazy interaction switch}, or 
$\MLSdli$ 
is the  system $\{ \, \ruleaidown\,  , \, \deeplazyintswir \, , 
\, \unitonedown \, , \unittwodown \}$. 
We define the other systems obtained by replacing in $\MLSu$ 
the switch rule with one of the rules given in 
Figure \ref{figure:switch} similarly by adding the prefix of the switch 
rule as a sufix to the name of the system. For example,
the \emph{system} $\MLSu$ \emph{with deep interaction switch}, or 
$\MLSdi$ 
is the  system $\{ \, \ruleaidown\,  , \, \deepintswir \, , 
\, \unitonedown \, , \unittwodown \, \}$. 
\end{defi}

In the sequent calculus, context management is performed by the inference rule 
$\mathsf{R}\ltens$, depicted below, which spreads the context of a copar formula 
at the top level to the branches of the proof. The switch rule simulates this 
rule as follows.

$$
\vcenter{\iinf{\ltens} {\sqn{ \phi, \aprs{R \ltens T}}, \psi}
  {\sqn{R,\phi}}{\sqn{T,\psi}}
}
\qquad
\leadsto
\qquad
\vcenter{
\dernote{\swir}{}{ \pars{\aprs{R \ltens T} \lpar \phi \lpar \psi}  }{
\rootr{\swir\; }{}{ \pars{\aprs{\pars{R \lpar \phi} \ltens T}  \lpar \psi}  }{
\leaf{\aprs{\pars{R \lpar \phi} \ltens \pars{T\lpar \psi}}   }}}
}
\\[8pt]
$$
Once the rule $\ltens$ is applied, the sequents 
in the two branches cannot exchange 
any formula, thus the communication between the formula 
at the two branches becomes impossible while going up in the proofs. 
However, in order for a proof to be constructed 
both of these two branches must contain dual 
atoms.

The rule interaction switch ($\intswir$)
exploits this observation while preserving the possibility to be applied at arbitrary
depths inside logical expressions. Thus, the rule $\intswir$ 
can simulate the sequent calculus proofs, but also provide 
shorter proofs due to deep instances.    

The rule $\deeplazyswir$ exploits the capability of the switch 
rule to manage contexts gradually. This is done by constraining the size 
of the formulae that are brought together for interaction, where interaction is 
eventually realized by the instances of the rule $\ruleaidown$ that 
correspond to the axiom of sequent calculus. 
In its shallow instances, the rule $\deeplazyswir$ is similar to constraining
the rule $\ltens$ such that
the sequent $\phi$ consists of a single non-par formula, 
and $R$ a non-copar formula. In fact, such 
a restriction in the sequent calculus would not affect 
the size of the proofs. However, as we show below, 
the system with the 
rule $\deeplazyintswir$ polynomially simulates the
sequent calculus proofs, while providing shorter 
proofs due to deep inference. 
In fact, in classical logic, the gain in proof length becomes 
exponential in certain cases.
Because the rule $\deeplazyintswir$ does not impose 
any restriction on the deep applicability of the 
inference rules, it does not cause a loss in shorter
proofs that are available due to deep inference.
However, restricting the sequent calculus in a similar manner 
would not reveal a complete system. 
For example, requiring in the instances of the 
sequent calculus context management rule $\ltens$
that $R$ and $\phi$ have common dual atoms as in the  
rule $\mathsf{dlis}$ would result in an incomplete system.
Thus, a straight-forward 
mapping from the sequent calculus to the system $\MLSdli$ 
cannot be used to prove the completeness of  $\MLSdli$.
The example below illustrates the reason for this.

\begin{exa}
\label{example:there:are:cases}
There are cases similar to the ones in the
proofs of the structure
$$
\pars{\aprs{\lbot \ltens \lbot} \lpar a \lpar \neg{a} \lpar b \lpar \neg{b}}
$$
below, which admit sequent calculus proofs that cannot be directly mapped to  
$\MLSdli$ without a transformation of these proofs. 
Both of the proofs below are in  $\MLSu$, however the proof on the 
left simulates the only shallow inference proof,
 whereas the proof in the right is  one of the many proofs in $\MLSdli$. 

$$
\dernote{\swir}{}{\pars{\aprs{\lbot \ltens \lbot} \lpar a \lpar \neg{a} \lpar b \lpar \neg{b}} }{
\rootr{\swir \;}{}{\pars{\aprs{\pars{\lbot \lpar a \lpar \neg{a}} \ltens \lbot}  \lpar b \lpar \neg{b}} }{
\rootr{\unitonedown\;}{}{\aprs{\pars{\lbot \lpar a \lpar \neg{a}} \ltens \pars{\lbot \lpar b \lpar \neg{b}}}   }{
\rootr{\unitonedown\;}{}{\aprs{\pars{ a \lpar \neg{a}} \ltens \pars{\lbot \lpar b \lpar \neg{b}}}   }{
\rootr{\ruleaidown;\, \unittwodown\;}{}{\aprs{\pars{ a \lpar \neg{a}} \ltens \pars{b \lpar \neg{b}}}   }{
\rootr{\ruleaidown\, }{}{\pars{ b \lpar \neg{b}}    }{
\leaf{\lone}}}}}}}
%%%%%%%%%%
%%%%%%%%%%
\qquad \qquad
%%%%%%%%%%
%%%%%%%%%%
\dernote{\ruleaidown}{}{\pars{\aprs{\lbot \ltens \lbot} \lpar a \lpar \neg{a} \lpar b \lpar \neg{b}} }{
\rootr{\ruleaidown\,}{}{\pars{\aprs{\lbot \ltens \lbot} \lpar \lone \lpar b \lpar \neg{b}} }{
\rootr{\deepreslazyswir\;}{}{\pars{\aprs{\lbot \ltens \lbot} \lpar \lone \lpar \lone} }{
\rootr{\ruleaidown;\, \unittwodown\;}{}{\pars{\aprs{\pars{\lbot\lpar \lone} \ltens \lbot}  \lpar \lone} }{
\rootr{\ruleaidown\,}{}{\pars{ \lbot  \lpar \lone} }{
\leaf{\lone}}}}}}
\\[8pt]
$$ 
In the proof on the right, we exploit the duality between $\lone$ and $\lbot$, which remains 
hidden in the rigidity of the sequent-calculus-like shallow inference
proof on the left, where the instances of the rule $\ruleaidown$  simulate the sequent calculus 
axioms at the leafs of the proof tree. 
\end{exa}

\begin{figure}[t]
$$
{\footnotesize
\begin{tabular}{|c||c|c|c|c|c|c|c|c|}
\cline{1-9}
    & 1. & 2. & 3. &  4.  &  5. & 6. & 7. & 8.\\
\hline
\textsf{MLS} & 22 & 28 & 70 & 114 & 112  & 112 & 112 & 118 \\
\hline
\textsf{MLSu} & 6 & 8 & 24 & 42 & 42 & 42 & 42 & 42 \\
\hline
\textsf{MLSl} & 2 & 5 & 16 & 18 & 28 & 28 & 28 & 18 \\
\hline
\textsf{MLSd}  & 3 & 8 & 15 & 21 & 24 & 24 & 24 & 42 \\
\hline
\textsf{MLSi}  & 6 & 6 & 15  & 30 & 26 & 31 & 36 & 26 \\
\hline
\textsf{MLSdl} & 1 & 5 & 10 & 9  & 16  & 16 & 16 &  18 \\
\hline
\textsf{MLSli} & 2 & 3 & 8 & 9 & 14 & 18 & 22 & 7 \\
\hline
\textsf{MLSdi} & 3 & 6 & 8  & 12 & 12 & 15 & 18 & 26 \\
\hline
\textsf{MLSdli} & 1 & 3 & 4 & 3 & 6 & 8 & 10  & 7 \\
\hline
\end{tabular}
%\end{tabular}
}
$$
\caption{The number of immediate redexes at the 
beginning of proof search for the systems 
$\MLS$, $\MLSu$, $\MLSl$, $\MLSi$, $\MLSd$, 
$\MLSli$, $\MLSdl$, $\MLSdi$ and $\MLSdli$
for the structures given
in Example \ref{example:structures:improvement}.
Unlike other systems, $\MLS$ is implemented with equalities for unit,
shown in Example \ref{example:switch:instances}, 
as in \cite{Str03a}.
}
    \label{figure:comparison}
\end{figure}

\begin{exa}
\label{example:structures:improvement}
The following structures are provable in  $\MLSu$.
$$
{\footnotesize
\begin{array}{l}
\\
1. \;
% [ [ [ 1 , 1 ] , 1 ] , { bot, { bot , bot } } ]
% [  1 | 1  | 1 | ( bot * bot * bot ) ]
\pars{  \lone \lpar \lone  \lpar \lone \lpar \aprs{ \bot \ltens \bot \ltens \bot } }
\\[4pt]
%%%%%%%%%%%%%%%%%%%%%%%
2. \;
% [ [ { a , b } , {  a , b } ] , { [- a , - b ] , [ - a , - b ] } ]
% [  ( a * b ) | ( a * b )  | ( [ -a | -b ]  * [ -a | -b ] ) ]
\pars{  \aprs{ a \ltens b } \lpar \aprs{ a \ltens b }  \lpar  \aprs{ \pars{ \neg a \lpar \neg b }  \ltens \pars{ \neg a \lpar \neg b } } }
\\[4pt]
%%%%%%%%%%%%%%%%%%%%%%%
3. \;
% [ - a , [ - b , { a , { b , [ - c , [ - d , { c , d } ] ] } } ] ]
% [ -a |  -b | ( a *  b * [ -c | -d | ( c * d )  ]  )  ]
\pars{   \neg a \lpar  \neg b \lpar \aprs{ a \ltens  b \ltens \pars{  \neg c \lpar \neg d \lpar \aprs{ c \ltens d }  }  }  }
\\[4pt]
%%%%%%%%%%%%%%%%%%%%%%%
4. \;
% [ [ [ a , b ] , c ] , { - a , { - b , - c } } ]
% [  a | b  | c  | ( -a  *  -b  * -c )  ]
\pars{   a \lpar b  \lpar c  \lpar \aprs{ \neg a  \ltens  \neg b  \ltens \neg c }  }
\\[4pt]
%%%%%%%%%%%%%%%%%%%%%%%
5. \;
% [ - a , [ - b , { a , { b , [ - c , [ - d , { c , { d , [ - e , [ - f , { e , f } ] ] } } ] ] } } ] ]
% [ -a |  -b | ( a  *  b  * [ -c  | -d  | ( c * d * [ -e  | -f  | ( e * f ) ] ) ] ) ]
\pars{  \neg a \lpar  \neg b \lpar \aprs{ a  \ltens  b  \ltens \pars{  \neg c  \lpar \neg d  \lpar \aprs{ c \ltens d \ltens \pars{  \neg e  \lpar \neg f  \lpar \aprs{ e \ltens f } } } } } }
\\[4pt]
%%%%%%%%%%%%%%%%%%%%%%%
6. \;
% [ - a , [ - b , { a , { b , [ - c , [ - d , { c , { d , [ - a , [ - b , { a , b } ] ] } } ] ] } } ] ]
% [ -a |  -b | ( a * b * [ -c  |  -d  |  ( c  *  d  * [ -a  |  -b  | ( a * b ) ] ) ] ) ]
\pars{  \neg a \lpar  \neg b \lpar \aprs{ a \ltens b \ltens \pars{  \neg  c  \lpar  \neg  d  \lpar  \aprs{ c  \ltens  d  \ltens \pars{  \neg  a  \lpar  \neg  b  \lpar \aprs{ a \ltens b } } } } } }
\\[4pt]
%%%%%%%%%%%%%%%%%%%%%%%
7. \;
% [ - a , [ - b , { a , { b , [ - a , [ - b , { a , { b , [ - a , [ - b , { a , b } ] ] } } ] ] } } ] ] 
% [ -a | -b | ( a *  b * [ -a |  -b | ( a  *  b  * [ -a |  -b | ( a * b ) ] ) ] ) ] 
\pars{  \neg a \lpar \neg b \lpar \aprs{ a \ltens  b \ltens \pars{  \neg a \lpar  \neg b \lpar \aprs{ a  \ltens  b  \ltens \pars{  \neg a \lpar  \neg b \lpar \aprs{ a \ltens b } } } } } } 
\\[4pt]
%%%%%%%%%%%%%%%%%%%%%%%
8. \;
% [ [ - c , { - d , - e} ] , [ { e , - a } , { c , [ a , d ] } ] ]
% [  -c | ( -d  * - e )  |  ( e * -a )  | ( c * [ a | d ] )  ]
\pars{   \neg c \lpar \aprs{ \neg d  \ltens \neg e }  \lpar  \aprs{ e \ltens \neg a }  \lpar \aprs{ c \ltens \pars{  a \lpar d } }  }
\\[8pt]
\end{array}
}
$$
%%%%%%%%%%%%%%%%%%%%%%%%%%
%%%%%%%%%%%%%%%%%%%%%%%%%%
%%%%%%%%%%%%%%%%%%%%%%%%%%
Figure \ref{figure:comparison} depicts the number of all the immediate 
rule instances on these structures in the systems $\MLS$, $\MLSu$, 
$\MLSl$, $\MLSi$, $\MLSd$, $\MLSli$, $\MLSdl$, $\MLSdi$ and $\MLSdli$.
While controlling the nondeterminism in context management, the rule
$\deeplazyintswir$ provides a more immediate access to the shorter 
proofs which are available in nested structures as in $(3.)$, $(5.)$, $(6.)$ and $(7.)$ 
above. This is because the proof can be constructed by starting from the 
substructures as illustrated below on structure $(3.)$.

$$
\dernote{\deeplazyintswir}{}{
\pars{\neg{a} \lpar \neg{b} \lpar  \aprs{a \ltens b \ltens  
            \pars{\neg{c} \lpar \rdx{\neg{d} \lpar \aprs{c \ltens d} }}}} }{
\rootr{\ruleaidown\, ; \,\unittwodown\; }{}{\pars{\neg{a} \lpar \neg{b} \lpar  \aprs{a \ltens b \ltens  
            \pars{\neg{c} \lpar  \aprs{c \ltens \rdx{\pars{d \lpar \neg{d}}}} }}}}{            
\rootr{\ruleaidown\, ; \,\unittwodown\; }{}{\pars{\neg{a} \lpar \neg{b} \lpar  \aprs{a \ltens b \ltens  
            \rdx{\pars{\neg{c} \lpar  c  }}}}}{            
\rootr{\deeplazyintswir\; }{}{\pars{\neg{a} \lpar \rdx{\neg{b} \lpar  \aprs{a \ltens b }}}}{            
\rootr{\ruleaidown\, ; \,\unittwodown\; }{}{\pars{\neg{a}\lpar  \aprs{a \ltens \rdx{\pars{b \lpar \neg{b} }} }}}{            
\rootr{\ruleaidown\, }{}{\rdx{\pars{a\lpar \neg{a}}}}{            
\leaf{\lone}}}}}}}
\\[10pt]
$$
An alternative proof of this structure in $\MLSdli$ is the following one.
This proof simulates a sequent calculus proof,  where the 
inference rules are applied only at the top level connectives.

$$
\dernote{\deeplazyintswir}{}{
\pars{\neg{a} \lpar \neg{b} \lpar  \aprs{a \ltens b \ltens  
            \pars{\neg{c} \lpar \neg{d} \lpar \aprs{c \ltens d} }}} }{
\rootr{\deeplazyintswir \;}{}{
\pars{ \neg{a} \lpar  \aprs{\pars{b  \lpar \neg{b}} \ltens  a \ltens  
            \pars{\neg{c} \lpar \neg{d} \lpar \aprs{c \ltens d} }}}  }{ 
\rootr{\deeplazyintswir \;}{}{
  \aprs{\pars{b  \lpar \neg{b}} \ltens  \pars{a \lpar \neg a} \ltens  
            \pars{\neg{c} \lpar \neg{d} \lpar \aprs{c \ltens d} }}  }{ 
\rootr{\deeplazyintswir \;}{}{
  \aprs{\pars{b  \lpar \neg{b}} \ltens  \pars{a \lpar \neg a} \ltens  
            \pars{\neg{c}  \lpar \aprs{c \ltens \pars{d \lpar \neg{d}}} }}  }{ 
\rootr{\ruleaidown\, ; \,\unittwodown\; }{}{
  \aprs{\pars{b  \lpar \neg{b}} \ltens  \pars{a \lpar \neg a} \ltens  
            \pars{c  \lpar \neg c} \ltens \pars{d \lpar \neg{d}} }  }{ 
\rootr{\ruleaidown\, ; \,\unittwodown\; }{}{
  \aprs{  \pars{a \lpar \neg a} \ltens  
            \pars{c  \lpar \neg c} \ltens \pars{d \lpar \neg{d}} }  }{ 
\rootr{   \ruleaidown\, ; \,\unittwodown\; }{}{
  \aprs{\pars{c  \lpar \neg c} \ltens \pars{d \lpar \neg{d}} }  }{ 
\rootr{  \ruleaidown\, }{}{ \pars{d \lpar \neg{d}}   }{ 
\leaf{\lone}}}}}}}}}
$$
\end{exa}

%%%%%%%%%%%%%%%%%%%%%%%%%
%%%%%%%%%%%%%%%%%%%%%%%%%
%%%%%%%%%%%%%%%%%%%%%%%%%
%%%%%%%%%%%%%%%%%%%%%%%%%
%%%%%%%%%%%%%%%%%%%%%%%%%
%%%%%%%%%%%%%%%%%%%%%%%%%

\subsection{Correctness: Exploring the Switch Lattice}

%%%%%%%%%%%%%%%%%%%%%%%%%
%%%%%%%%%%%%%%%%%%%%%%%%%

The restrictions on the switch rule, given in Definition \ref{definition:deep:lazy:int:swir},
control the nondeterminism in context management  during proof construction.  
By exploring the lattice in Figure \ref{figure:switch}, we now show that these
different versions of the switch rule preserve completeness. 
Showing the completeness  of the systems 
 in the direction of the arrows in Figure \ref{figure:switch}  is straightforward. 
This is because all instances 
of a system at the source of an arrow are also instances of the system at the target 
of that arrow. For example, every instance of the rule $\deeplazyintswir$
is an instance of the rules  
$\mathsf{lis}$, 
$\mathsf{dls}$ or
 $\mathsf{dis}$. 
Similarly, every instance of  $\mathsf{dls}$ is 
an instance of  $\mathsf{ds}$ or
$\mathsf{ls}$, which are all instances of the switch rule. 
However, exploring this lattice from the bottom to the top requires a 
global analysis of proofs. This is because the deep applicability 
of the inference rules in these systems  
brings about the possibility for the substructures to interact with 
their contexts in nested structures.
The example below  illustrates this idea. 

\begin{exa}
\label{example:global:view}
Consider the following proof on the left in  $\MLSu$ and a proof 
of the same structure in $\MLSdli$ on the right. In the proof transformation below,
we remove the bottom-most instance of the switch rule in the proof on the left, and thereby
replace this proof with the proof on the right, where all the instances of  switch are
instances of the rule $\deeplazyintswir$. 

$$
\vcenter{
\dernote{\swir}{}{\pars{\rdx{a \lpar \aprs{b \ltens c}} \lpar \neg{b} \lpar \aprs{\neg{a} \ltens \neg{c}}} }{
\rootr{\deeplazyintswir\;}{}{ 
\pars{\aprs{b \ltens \pars{a \lpar c}} \lpar  \neg{b} \lpar \aprs{\neg{a} \ltens \neg{c}}} }{
\rootr{\ruleaidown; \,\unittwodown\;}{}{ 
\pars{\aprs{\pars{b \lpar \neg{b}} \ltens \pars{a \lpar c}} \lpar \aprs{\neg{a} \ltens \neg{c}}} }{
\rootr{\deeplazyintswir\;}{}{ 
\pars{a \lpar c \lpar \aprs{\neg{a} \ltens \neg{c}}} }{
\rootr{\ruleaidown;\,\unittwodown\;}{}{ 
\pars{c \lpar \aprs{\pars{a \lpar \neg{a}} \ltens \neg{c}}} }{
\rootr{\ruleaidown\; }{}{ \pars{c \lpar \neg{c}} }{
\leaf{\lone}}}}}}}
}
\qquad \;
\leadsto
\qquad
\vcenter{
\dernote{\deeplazyintswir}{}{ 
   \pars{a \lpar \aprs{b \ltens c} \lpar \neg{b} \lpar \aprs{\neg{a} \ltens \neg{c}}}  }{
\rootr{\ruleaidown;\,\unittwodown\;}{}{ 
   \pars{ \aprs{b \ltens c} \lpar \neg{b} \lpar \aprs{\pars{a \lpar \neg{a}} \ltens \neg{c}}} }{
\rootr{\deeplazyintswir\; }{}{
  \pars{ \aprs{b \ltens c} \lpar \neg{b} \lpar \neg{c}} }{
\rootr{\ruleaidown;\, \unittwodown\;}{}{
  \pars{\aprs{\pars{b \lpar \neg{b}} \ltens c}  \lpar \neg{c}}  }{
\rootr{\ruleaidown\;}{}{
  \pars{c  \lpar \neg{c}}  }{
\leaf{\lone}}}}}}
}
\\[8pt]
$$
\end{exa}

We explore the lattice in Figure \ref{figure:switch} step by step 
to consider all the systems obtained by replacing the switch rule with 
its restricted versions given in Definition \ref{definition:deep:lazy:int:swir}.
Thereby, we show that proofs in  $\MLSu$
can be transformed into proofs in $\MLSdli$ and other intermediate systems, and 
show that these systems are complete for multiplicative linear logic.

Let us first give some definitions, which we use in the proofs below.

\begin{defi}
\cite{Gug02}
Given a structure $R$, its \emph{atom occurrences} are obtained by
considering all the atoms appearing in $R$ as distinct
(for example, by indexing them so that two atoms, which are 
equal, get different indices). 
Then, $\occ \,R$ is the set of all the atom 
occurrences 
appearing in $R$. 
The \emph{size} of $R$ is the cardinality of the set
$\occ \, R$.    
\end{defi}

\begin{exa}
For $R$ in Example \ref{example:at}, 
$\occ \, R 
= \{a_1, \neg{a}_1, b,\neg{b}, a_2,\neg{a}_2 , \lbot_1 \}$.
\end{exa}

\begin{defi}
\label{definition:down:arrow}
\cite{Gug02}
Let $R$ be a structure. The   
\emph{structural relation} 
$\down{}{}{R}$ is the minimal set  such that 
$\down{}{}{R} \; \subset \, (\occ \,R)^2$
and,
for every $S\cons{\;\;}$, $U$ and $V$ and for 
every $a$ in $U$ and $b$ in $V$, the following 
holds: if $R = S\pars{U \lpar V}$ then  $\down{a}{b}{R}$.
The notation $| \down{}{}{R} |$ denotes the 
cardinality of the set $\down{}{}{R}$.
\end{defi}

\begin{exa}
For $R$ in Example \ref{example:at}, 
in $\downarrow_R$ we have   
$(a_1 , \neg{a}_1)$, 
$(a_1 , b)$, 
$(a_1 , \neg{a}_2)$,
$(a_1 , \lbot_1)$, 
$(a_1 , a_2)$, 
$(a_1 , \neg{b})$, 
$(\neg{a}_1, b)$, 
$(\neg{a}_1 , \neg{a}_2)$, 
$(\neg{a}_1 , \lbot_1)$, 
$(\neg{a}_1, a_2)$, 
$(\neg{a}_1, \neg{b})$,
$(b, \neg{a}_2)$, 
$(b, \lbot_1)$, 
$(b, a_2)$, 
$(b, \neg{b})$,
$ (\neg{a}_2, a_2)$, 
$(\neg{a}_2, \neg{b})$,
$(\lbot_1,a_2)$, 
$(\lbot_1,\neg b)$
and their symmetric instances.
\end{exa}

When considered from the point of view of  
Definition \ref{definition:down:arrow}, 
the role of the inference rules on a structure $R$ can be seen as 
modifying the relation $\down{}{}{R}$. The Figure \ref{figure:BV:relationweb}
displays the inference rules of  $\MLSu$ from such a point of view. The 
following remarks explicitly identify these modifications,
and demonstrate how the 
relation $\downarrow$ is used as an 
induction measure.

%%%%%%%%%%%%%%%%%%%%%%%%%%%%%%%%
%%%%%%%%%%%%%%%%%%%%%%%%%%%%%%%%
%%%%%%%%%%%%%%%%%%%%%%%%%%%%%%%%
%%%%%%%%%%%%%%%%%%%%%%%%%%%%%%%%
%%%%%%%%%%%%%%%%%%%%%%%%%%%%%%%%
%%%%%%%%%%%%%%%%%%%%%%%%%%%%%%%%
%%%%%%%%%%%%%%%%%%%%%%%%%%%%%%%%

\begin{figure}[!t]
%%%%%%%%%%%%%%%%%%%%%%%%%%%%%%%%%%%%%%%%%%%%%%
\begin{center}
{
\small
\framebox[5.4in]
{
\begin{minipage}[t]{5.4in}

\begin{center}
\vspace{4mm}
%%%%%%%%%%%%%%%%%%%%%%%%%%%%%%%%%%%%%%%%%%%%%%%
$$
\begin{array}{c}
\dernote{\ruleaidown}{}
        {
\, S\left[
\begin{array}{c} 
\xymatrix{ a \ar@{.}[r]   & \neg{a} }
\end{array}
\right] \,
} 
        {\leaf{S\cons{\, \lone\, }}} 
\end{array}
%%%%%%%%%%%%%%%%%
%%%%%%%%%%%%%%%%%
\qquad  %%%%%%%%%%%%%
%%%%%%%%%%%%%%%%%
%%%%%%%%%%%%%%%%%
\begin{array}{c}
\dernote{\swir}{}
        {
\, S\left[
\begin{array}{c} 
 \xymatrix{ R \ar@{.}[r]  & T   \\  
                                      & U\ar@{.}[ul]  } 
\end{array}
\right] \,
} 
        {\leaf{
\, S\left[
\begin{array}{c} 
\xymatrix{ R   & T    \\  
                    & U\ar@{.}[ul]   }
\end{array}
\right] \,
}}
\end{array} 
%%%%%%%%%%%%%%%%%
%%%%%%%%%%%%%%%%%
\qquad  %%%%%%%%%%%%%
%%%%%%%%%%%%%%%%%
%%%%%%%%%%%%%%%%%
\begin{array}{c}
\begin{array}{c}
\dernote{\unitonedown}{}
        {
\, S\left[
\begin{array}{c} 
\xymatrix{ R \ar@{.}[r]   & \lbot }
\end{array}
\right] \,
} 
        {\leaf{S\cons{\, R \, }}} 
\end{array}
%%%%
%%%%
%%%%
\\[34pt]
%%%%
%%%%
%%%%
\begin{array}{c}
\dernote{\unittwodown}{}
        {
\, S\left(
\begin{array}{c} 
\xymatrix{ R    & \lone }
\end{array}
\right) \,
} 
        {\leaf{S\cons{\, R\, }}} 
\end{array}
\end{array}
$$
%%%%%%%%%%%%%%%%%%%%%%%%%%%%%%%%%%%%%%%%%%%%%%%%%%
\vspace{4mm}

\end{center}
\end{minipage}
}  }

\end{center}
%%%%%%%%%%%%%%%%%%%%%%%%%%%%%%%%%%%%%%%%%%%%%%%%%%%
\caption{The effect of the inference rules of  $\MLSu$ 
on structures with respect to Definition 
\ref{definition:down:arrow}, 
where the dashed lines indicate the 
the relation $\down{}{}{}$.}
\label{figure:BV:relationweb}
\end{figure}

%%%%%%%%%%%%%%%%%%%%
%%%%%%%%%%%%%%%%%%%%
%%%%%%%%%%%%%%%%%%%%
%%%%%%%%%%%%%%%%%%%%
%%%%%%%%%%%%%%%%%%%%
%%%%%%%%%%%%%%%%%%%%
%%%%%%%%%%%%%%%%%%%%
%%%%%%%%%%%%%%%%%%%%
%%%%%%%%%%%%%%%%%%%%
%%%%%%%%%%%%%%%%%%%%
%%%%%%%%%%%%%%%%%%%%
%%%%%%%%%%%%%%%%%%%%
%%%%%%%%%%%%%%%%%%%%
%%%%%%%%%%%%%%%%%%%%
%%%%%%%%%%%%%%%%%%%%

\renewcommand{\rulebox}[1]{\mbox{$#1$}}
\begin{figure}[!t]
%%%%%%%%%%%%%%%%%%%%%%%%%%%%%%%%%%%%%%%%%%%%%%
\begin{center}
{
\small
\framebox[5.4in]
{
\begin{minipage}[t]{5.4in}

\begin{center}
\vspace{2mm}
%%%%%%%%%%%%%%%%%%%%%%%%%%%%%%%%%%%%%%%%%%%%%%%

$$
\begin{array}{c}
\vcenter{\dernote {\swir}
       {}{
\xymatrix{             
                a\ar@{.}[rr] 
                  \ar@{.}[ddrr]
                  \ar@{.}[dd]
                             &   &  \neg{a}   & \\   
                                &     &             &  \pars{\aprs{\neg{a} \ltens \neg{b}} \lpar a \lpar b} \; \,\\   
                 b \ar@{.}[rr]
                      \ar@{.}[uurr]     
                                &    &  \neg{b}     &         
 }         
}{
 \root{\ruleaidown}{
\;\;
\xymatrix{   a\ar@{.}[rr] 
                  \ar@{.}[dd]
                               &  &  \neg{a}   & \\
                                &  &                  & \pars{\aprs{\pars{ a \lpar\neg{a}} \ltens \neg{b}} \lpar b}  \\
                 b \ar@{.}[rr]
                      \ar@{.}[uurr]     
                                &    & \neg{b} & 
 }         
}{
\root{\unittwodown}{\;
\xymatrix{  \lone \ar@{.}[ddrr]
                           \ar@{.}[dd]
                                &  &     & \\
                                &    &   &    \pars{\aprs{\lone \ltens \neg{b}} \lpar b} \quad \quad \\
                b \ar@{.}[rr]
                                 &     &     \neg{b} &               
 }         
}{
\root{\ruleaidown}{\;
\xymatrix{  
       &                        &                      & \\          
                 b \ar@{.}[rr]
                                &    &    \neg{b}     &       \pars{\neg{b} \lpar b} \qquad \qquad\\   
           &                        &        & } 
}{
       \leaf{\xymatrix{   \quad \\ \lone \quad \qquad \qquad \qquad \lone \quad \\ \quad } }}}}}}
\end{array}
$$
%%%%%%%%%%%%%%%%%%%%%%%%%%%%%%%%%%%%%%%%%%%%%%%%%%%
%%%%%%%%%%%%%%%%%%%%%%%%%%%%%%%%%%%%%%%%%%%%%%%%%%
\vspace{2mm}

\end{center}
\end{minipage}
}  }

\end{center}
%%%%%%%%%%%%%%%%%%%%%%%%%%%%%%%%%%%%%%%%%%%%%%%%%%%
\caption{A proof of the structure 
 $\pars{\aprs{\neg{a} \ltens \neg{b}} \lpar a \lpar b}$
and the corresponding  relation $\down{}{}{}$
as in Definition 
\ref{definition:down:arrow}, where the 
dashed lines denote the relation $\down{}{}{}$.}
\label{figure:BV:relationweb:example}
\end{figure}

%%%%%%%%%%%%%%%%%%%%%%%%%%
%%%%%%%%%%%%%%%%%%%%%%%%%%
%%%%%%%%%%%%%%%%%%%%%%%%%%
%%%%%%%%%%%%%%%%%%%%%%%%%%
%%%%%%%%%%%%%%%%%%%%%%%%%%

%%%%%%%%%%%%%%%%%%%%%%%%%%%%%%%%
%%%%%%%%%%%%%%%%%%%%%%%%%%%%%%%%
%%%%%%%%%%%%%%%%%%%%%%%%%%%%%%%%
%%%%%%%%%%%%%%%%%%%%%%%%%%%%%%%%
%%%%%%%%%%%%%%%%%%%%%%%%%%%%%%%%
%%%%%%%%%%%%%%%%%%%%%%%%%%%%%%%%
%%%%%%%%%%%%%%%%%%%%%%%%%%%%%%%%

\begin{rem} \label{remark:relweb0a}
Let $R = S\pars{P \lpar \lbot}$ and  $R' = S\cons{P}$
be structures such that $R$
consists of pairwise distinct atoms and $\lbot$ is uniquely marked.
If
$$      
\vcenter{\dernote{\unitonedown}{\,}
 {R}{\leaf{R'}}} 
\quad
then 
\quad
  \down{}{}{R'} \; =  \; \down{}{}{R} \setminus \; \,
  \{ \, ( x , y) \;  | \; (x,y) \, \in  \, \down{}{}{R} \land \; 
( \, x = \lbot \, \lor \, y = \lbot \, ) \, \}\; .
$$
\end{rem}

\begin{rem} \label{remark:relweb0b}
Let $R = S\aprs{P \ltens \lone}$ and  $R' = S\cons{P}$
be structures such that $R$
consists of pairwise distinct atoms and $\lone$ is uniquely marked.
If 
$$      
\vcenter{\dernote{\unittwodown}{\,}
 {R}{\leaf{R'}}}
\quad
then 
\quad 
 \down{}{}{R'} \; =  \; \down{}{}{R} \setminus \; \,
  \{ \, ( x , y) \;  | \; (x,y) \, \in  \, \down{}{}{R} \land \; 
( \, x = \lone \, \lor \, y = \lone \, ) \, \} \; .
$$
\end{rem}

\begin{rem} \label{remark:relweb1}
Let $R = S\pars{a \lpar \neg{a}}$ and  $R' = S\cons{\lone}$
be  structures such that $R$
consists of pairwise distinct atoms and
the occurrence of $\lone$ is uniquely marked.
Let $\Phi = \occ \,R  \setminus \{ a , \neg a \}$. 
If 
$$      
\vcenter{\dernote{\ruleaidown}{\,}
 {R}{\leaf{R'}}} \; ,
$$
given that $\occ \,R' = \Phi \cup \{ \lone \}$, then
$$
 \begin{array}{ll}
\down{}{}{R'} \; =  &  (\down{}{}{R} \; \cup \;  
\{ \, ( \lone , x), (x , \lone) \; | \;  x  \in \Phi \, \} \, )
\\[2pt]
& 
\qquad \setminus \; 
  \{ \, ( x , y) \; | 
\; (x,y) \, \in  \, \down{}{}{R} \land \; 
( \, x = a \, \lor \, y = a \, \lor \, x = \neg{a} \, \lor \, y = \neg{a} \, ) \, \} \, .\\[8pt]
\end{array}
$$
\end{rem}

\begin{rem}   \label{remark:relweb2}
Let $R = S\pars{\aprs{P \ltens T} \lpar U}$ and  
   $R' = S\aprs{\pars{P \lpar U} \ltens T}$
be structures 
such that $R$
consists of pairwise distinct atoms.
If 
$$      
\vcenter{\dernote{\swir}{\,}
 {R}{\leaf{R'}}}
\quad
then 
\quad  
\down{}{}{R'} \; = \; 
\down{}{}{R}  \setminus \; 
     \{ \, ( x , y) , \, (y , x) \, | \; x \, \in \occ \, T   
                     \,  \land \, y \in \occ \, U \, \} \; .
 $$
\end{rem}

With these observations of the remarks above, 
we can now state the proposition below.

\begin{prop}
\label{proposition:induction:measure}
For any rule instance 
$      
\vcenter{\dernote{\rho}{\,}
 {R}{\leaf{R'}}}
$
of any rule 
$\rho \in \{  \ruleaidown , \swir, \unitonedown \}$ 
of  $\MLSu$, we have that 
$|\down{}{}{R'}| \, < \, | \down{}{}{R}|$, and 
for $\rho = \unittwodown$ 
 we have that 
$|\down{}{}{R'}| \, \leq \, | \down{}{}{R}|$.
\end{prop}

\proof
As depicted in Figure \ref{figure:BV:relationweb}, 
it immediately follows from  
Remark \ref{remark:relweb0a},
Remark \ref{remark:relweb0b},
Remark \ref{remark:relweb1}
and Remark \ref{remark:relweb2} that
in any instance of the 
rules $\unitonedown$, $\ruleaidown$ and $\swir$
the cardinality of the relation $\downarrow$ is smaller in 
the premise than in the conclusion, and in an instance 
of the rule $\unittwodown$ it is either smaller 
or remains unchanged.
\qed

%%%%%%%%%%%%%%%%%%%%
%%%%%%%%%%%%%%%%%%%%
%%%%%%%%%%%%%%%%%%%%
%%%%%%%%%%%%%%%%%%%%
%%%%%%%%%%%%%%%%%%%%
%%%%%%%%%%%%%%%%%%%%
%%%%%%%%%%%%%%%%%%%%
%%%%%%%%%%%%%%%%%%%%
%%%%%%%%%%%%%%%%%%%%
%%%%%%%%%%%%%%%%%%%%
%%%%%%%%%%%%%%%%%%%%
%%%%%%%%%%%%%%%%%%%%
%%%%%%%%%%%%%%%%%%%%
%%%%%%%%%%%%%%%%%%%%
%%%%%%%%%%%%%%%%%%%%

\begin{prop} \label{proposition:MLL:inNP}
The length of any proof of any structure $R$ in $\MLSu$   
is bounded by $\mathcal{O} ({| \occ \, R |}^2 )$.
\end{prop}
\proof
Follows immediately from  Proposition \ref{proposition:induction:measure}. 
\qed

\begin{exa}
%%%%%%%%%%%%%%%
%%%%%%%%%%%%%%%
Let $R$ be the structure 
$\pars{\aprs{\neg{a} \ltens \neg{b} \ltens \lbot} \lpar a \lpar b \lpar \lone}$.
We consider the rule instance (ii) in 
Example \ref{example:derivation}. If  $R'$
denotes the premise of this rule instance, we have
$\downarrow_{R'} \: = \: \downarrow_{R} \setminus \,
\{ (a,\neg{b}) , (a,\lbot) , (\neg{b},a) , (\lbot,a) \}.
$  
\end{exa}

\begin{exa}
Figure \ref{figure:BV:relationweb:example} 
displays an example proof of the structure 
$\pars{\aprs{\neg a \ltens \neg b} \lpar a \lpar b}$ 
and the corresponding  modifications on 
relation $\down{}{}{}$.
\end{exa}

%%%%%%%%%%%%%%%%%%%%%%%%%%%%%%%%
%%%%%%%%%%%%%%%%%%%%%%%%%%%%%%%%
%%%%%%%%%%%%%%%%%%%%%%%%%%%%%%%%
%%%%%%%%%%%%%%%%%%%%%%%%%%%%%%%%
%%%%%%%%%%%%%%%%%%%%%%%%%%%%%%%%
%%%%%%%%%%%%%%%%%%%%%%%%%%%%%%%%
%%%%%%%%%%%%%%%%%%%%%%%%%%%%%%%%

%########################
%########################
%########################
%########################
%########################

\begin{prop}
\label{proposition:unit:invertible}
For any multiplicative linear logic structure $R$ and context $S$,
 in $\MLSu$, $S\cons{R}$ has a proof if and only if\\
\begin{enumerate}[label=(\roman*)]
\item%[$(i.$)]
 $S\pars{R \lpar \lbot}$ has a proof;\\
\item%[(ii)]
 $S\aprs{R \ltens \lone}$ has a proof.\\
\end{enumerate}
\end{prop}

\proof
For the if direction, the required proofs are constructed in  $\MLSu$ 
by applying the rules  $\unitonedown$ and  $\unittwodown$. For the only if direction, we  
construct the required proof  first in  $\MLS$ by applying the 
rules $\unitoneup$ and  $\unittwoup$, then transform 
these proofs into proofs in $\MLSu$ by applying 
Theorem \ref{theorem:equivalent:MLS:MLSu}.
\qed

\begin{prop}
\label{proposition:lazy:switch}
The systems $\MLSu$ and $\MLSl$ are equivalent.
\end{prop}

\proof
Every proof in $\MLSl$ is a proof in $\MLSu$ as every instance 
of $\lazyswir$ is an instance of $\swir$. 
For the other direction, we transform  proofs in $\MLSu$ into proofs in 
$\MLSl$ as follows: let us call redundant each occurrence  of $\lbot$ 
in a structure $Q$ if for some structure $U$ and context $S$, 
we have that $Q = S\pars{U \lpar \lbot}$.

Let $P$ be a structure with a proof $\Pi$ in $\MLSu$.
If $P$ has any redundant $\lbot$ than we apply to $P$
the rule $\unitonedown$ exhaustively to obtain $P'$ as depicted below 
such that $P'$ does not have any redundant $\lbot$.
By Proposition \ref{proposition:unit:invertible}, if $P$
has a proof $\Pi$ then $P'$ has a proof $\Pi'$ in $\MLSu$. 

$$
\vcenter{\xy\xygraph{[]!{0;<2.4pc,0pc>:}
{
1
  }-@{=}^<>(.5){\MLSu} _<>(.5){\Pi}[d] {
P
}
    }\endxy}
%%%
%%%%%%%%%%%%%%%%%%%%%
%%%%%%%%%%%%%%%%%%%%%
%%%
\qquad  
\leadsto 
\qquad
%%%
%%%%%%%%%%%%%%%%%%%%%
%%%%%%%%%%%%%%%%%%%%%
%%%
\vcenter{\xy\xygraph{[]!{0;<1.3pc,0pc>:}
         {\lone}-@{=}^<>(.5){     
                   \MLSu  
                                  }  _<>(.5){
                                            \Pi'
                } [dd] {   P'  }
           -@{=}^<>(.5){\{ \, \unitonedown \, \}
                                  } _<>(.5){ \Delta_1
               } [dd] {  P }
       }\endxy}  
\quad
.
\\[8pt]
$$
In $\Pi'$  we single out the bottom-most rule instance that is   
either (i)  an instance of $\mathsf{ls}$ that introduces a redundant $\lbot$
or (ii) an instance of $\unittwodown$ that introduces a redundant $\lbot$
or (iii) an instance of $\swir$ that is not an instance of $\lazyswir$. (If there is no such instance then we are done.)
We depict case (i) and (iii) below as case (ii) is similar to case (i).

$$
%%%%%%%%%%%%%%%%%%%%%
%%%%%%%%%%%%%%%%%%%%%
%%%
\vcenter{\xy\xygraph{[]!{0;<6pc,0pc>:}
{
%%%%%%%%%%%%
%%%%%%%%%%%%
\vcenter{\xy\xygraph{[]!{0;<1.7pc,0pc>:}
         {\lone  
                                  }-@{=}^<>(.5){     
                   \MLSu  
                                  }  _<>(.5){
                                            \Pi_1'
                } [dd] {   \dernote{\mathsf{ls}}{\;}{S\pars{\aprs{\lbot \ltens T} \lpar U}}{
                                       \leaf{  S\aprs{ \pars{ \lbot \lpar U} \ltens T} }}
                        }-@{=}^<>(.5){     
                   \MLSl  
                                  }  _<>(.5){
                                            \Delta_2
                } [dd] {  P' }
       }\endxy}  
%%%%%%%%%%%%%%%%%%%%%
%%%%%%%%%%%%%%%%%%%%%
  }-@{=}^<>(.5){\{ \unitonedown \} } _<>(.5){\Delta_1}[d] {
P
}
    }\endxy}
%%%
%%%%%%%%%%%%%%%%%%%%%
%%%%%%%%%%%%%%%%%%%%%
%%%
\quad  
\stackrel{(i.)}{\rotatebox[origin=c]{180}{$\leadsto$}}
\qquad
%%%
%%%%%%%%%%%%%%%%%%%%%
%%%%%%%%%%%%%%%%%%%%%
%%%
\vcenter{\xy\xygraph{[]!{0;<1.3pc,0pc>:}
         {\lone}-@{=}^<>(.5){     
                   \MLSu  
                                  }  _<>(.5){
                                            \Pi'
                } [dd] {   P'  }
           -@{=}^<>(.5){\{ \, \unitonedown \, \}
                                  } _<>(.5){ \Delta_1
               } [dd] {  P }
       }\endxy}  
%%%
%%%%%%%%%%%%%%%%%%%%%
%%%%%%%%%%%%%%%%%%%%%
%%%
\qquad  
\stackrel{(iii.)}{ \leadsto}
\quad
%%%
%%%%%%%%%%%%%%%%%%%%%
%%%%%%%%%%%%%%%%%%%%%
%%%
\vcenter{\xy\xygraph{[]!{0;<6pc,0pc>:}
{
%%%%%%%%%%%%
%%%%%%%%%%%%
\vcenter{\xy\xygraph{[]!{0;<1.7pc,0pc>:}
         {\lone  
                                  }-@{=}^<>(.5){     
                   \MLSu  
                                  }  _<>(.5){
                                            \Pi_3'
                } [dd] {   \dernote{\swir}{\;}{S\pars{\aprs{R \ltens T} \lpar U_1 \lpar \ldots \lpar U_n}}{
                                       \leaf{  S\aprs{ \pars{ R \lpar U_1 \lpar \ldots \lpar U_n } \ltens T} }}
                        }-@{=}^<>(.5){     
                   \MLSl  
                                  }  _<>(.5){
                                            \Delta_2
                } [dd] {  P' }
       }\endxy}  
%%%%%%%%%%%%%%%%%%%%%
%%%%%%%%%%%%%%%%%%%%%
  }-@{=}^<>(.5){\{ \unitonedown \} } _<>(.5){\Delta_1}[d] {
P
}
    }\endxy}
.
\\[8pt]
$$  
In these three cases we proceed as follows:
\begin{itemize}
%%%%%%%%%%%%%%%%%%%%%%%%%%%
%%%%%%%%%%%%%%%%%%%%%%%%%%%
\item[(i)]
To the premise $S\aprs{\pars{\lbot \lpar U} \ltens T}$ of the instance of $\mathsf{ls}$, we apply the rule $\unitonedown$. 
By Proposition  \ref{proposition:unit:invertible}, we obtain the proof $\Pi''_1$ below, which we replace with 
the proof $\Pi'_1$.
%%%%%%%%%
%%%%%%%%%

$$
\vcenter{\xy\xygraph{[]!{0;<4pc,0pc>:}
{
1
  }-@{=}^<>(.5){\MLSu} _<>(.5){\Pi''_1}[d] {
\dernote{\mathsf{ls}}{\quad}{S\pars{\aprs{\lbot \ltens T} \lpar U}}{
                  \rootr{\unitonedown}{}{ S\aprs{ \pars{ \lbot \lpar U} \ltens T} }{
                                       \leaf{  S\aprs{ U \ltens T} }}}
}
    }\endxy}
\\[8pt]
$$
%%%%%%%%%%%%%%%%
%%%%%%%%%%%%%%%%
\item[(ii)]
To the premise $S\pars{\lbot \lpar U}$ of the instance of $\unittwodown$, we apply the rule $\unitonedown$. 
By Proposition  \ref{proposition:unit:invertible}, we obtain the proof $\Pi''_2$ below, which we replace with 
the proof $\Pi'_2$.

$$
\vcenter{\xy\xygraph{[]!{0;<4pc,0pc>:}
{
1
  }-@{=}^<>(.5){\MLSu} _<>(.5){\Pi''_2}[d] {
\dernote{\mathsf{\unittwodown}}{\quad}{S\pars{\aprs{\lbot \ltens \lone} \lpar U}}{
                  \rootr{\unitonedown}{}{ S\pars{ \lbot \lpar U} }{
                                       \leaf{  S\cons{ U} }}}
}
    }\endxy}
\\[8pt]
$$
%%%%%%%%%%%%%%%
%%%%%%%%%%%%%%%
\item[(iii)]
Because none of the $U_1, \ldots, U_n$ is $\lbot$, we can replace the instance of the switch with a 
derivation that consists of $n$ instances of the rule $\mathsf{ls}$.

$$
\vcenter{
\dernote{\swir}{}{S\pars{\aprs{R \ltens T} \lpar  U_1 \lpar\ldots \lpar  U_n}}{
\leaf{S\aprs{\pars{R \lpar U_1 \lpar \ldots \lpar U_n} \ltens T}}}}
\quad
\leadsto
\quad
\vcenter{
\dernote{\lazyswir}{}{S\pars{\aprs{R \ltens T} \lpar  U_1 \lpar U_2 \lpar  \ldots \lpar U_n}}{
\rootr{\lazyswir\; }{}{S\pars{\aprs{\pars{R \lpar U_1} \ltens T} \lpar  U_2 \ldots \lpar U_n }}{
\rootr{\lazyswir\; }{}{ \vdots }{
\leaf{S\aprs{\pars{R \lpar U_1 \lpar \ldots \lpar U_n} \ltens T}}}}}}
\\[8pt]
$$
If $R = \bot$,  as in case (i), we apply the rule $\unitonedown$ to the
premise of the derivation on the right above, 
and obtain a proof $\Pi''_3$ in $\MLSu$, which we replace with $\Pi'_3$. 
If $R$ is different from  $\lbot$, we proceed with  $\Pi'_3$. 
%%%%%%%%%%%%%%%%%%%
%%%%%%%%%%%%%%%%%%%
\end{itemize}

By Proposition \ref{proposition:induction:measure},  
for the case (i) we have that  $|\down{}{}{S\aprs{U \ltens T}}| < |\down{}{}{P}|$ 
and for the case (ii) we have that 
$|\down{}{}{S\aprs{\pars{R \lpar U_1 \lpar \ldots \lpar U_n} \ltens T}}| < |\down{}{}{P}|$ 
and
$|\down{}{}{S\aprs{\pars{U_1 \lpar \ldots \lpar U_n} \ltens T}}| < |\down{}{}{P}|$. 
We thus repeat the procedure above inductively until all the instances of the rule $\swir$
that are not instances of the rule $\mathsf{ls}$ are removed.
\qed

%########################
%########################
%########################
%########################
%########################

The transformation in the proof of Proposition \ref{proposition:lazy:switch}  
maps instances of the switch rule to derivations that 
consist of sequences of the lazy switch rule instances. 
This transformation is linear in the number of the structures connected by par in $U$.  
Although lazy instances of the switch rule appear to increase the length of the proofs in comparison 
to the instances of the switch rule, because the rule $\ruleaidown$
is applied to pairs of dual atoms, this restriction becomes beneficial in non-deterministic proof search.  
This is because the laziness condition reduces the size of the information processed at every proof step.

The restrictions of the rule $\lazyswir$ and $\mathsf{ds}$
are dual restrictions.  
In \cite{StraTh},
 Strassburger proves this proposition 
by permuting the instances of the switch rule that are not instances 
of the rule $\mathsf{ds}$ up in the proofs.  Here, we give a similar proof.

\begin{prop}
\label{proposition:deep:switch}
The systems 
$\MLSu$, $\MLSl$, $\MLSd$ and $\MLSdl$ are equivalent.
\end{prop}

\proof 
By Proposition \ref{proposition:lazy:switch}, we have that
$\MLSu$ and $\MLSl$ are equivalent. 
Every proof in $\MLSdl$ is a proof in $\MLSd$ and $\MLSl$, 
 and every proof in $\MLSd$ is a proof in  $\MLSu$.
For the proof of the other direction, 
given Proposition \ref{proposition:lazy:switch}, we transform proofs in $\MLSl$  
into proofs in $\MLSdl$, as this suffices to make the equality commute for all four systems. 
(Note that the procedure in the proof of Proposition \ref{proposition:lazy:switch}    
can also be used to transform proofs in $\MLSd$ into proofs in $\MLSdl$.)

Given a proof $\Pi$ of $\MLSl$, 
by induction 
we  transform it into a proof in $\MLSdl$.
We associate with each proof $\Pi$ a pair 
$\#\Pi=(n,r),$
where $n$ is the number of inferences of the switch rule in $\Pi$ that are not deep, 
and $r$ is the number of inferences above the top-most non-deep instance of 
the switch rule in $\Pi$ (it is zero if there is no such instance).
We prove by induction on $\#\Pi$ under the usual lexicographical order.  
In the base case, where $n = 0$, we are done as all the instances of the 
switch rule must be the instances of deep switch.
For the inductive cases, we single out the top-most non-deep instance of switch in $\Pi$.
This cannot be the top-most rule instance of $\Pi$, that is, $r > 0$.  
There is thus an instance of a rule $\rho$ and a structure $Q$ such that 

$$
   \quad \vcenter{\xy \xygraph{[]!{0;<3.8pc,0pc>:}
 {\lone}*=<14pt>{}:@{=}^<>(.5){} _<>(.5){} [d] {
\;\;\,
\dernote{\mathsf{ls}}{\quad.}{
S\pars{\aprs{R \ltens T} \lpar U}
}{
\rootr{\rho \; }{}{S\aprs{\pars{R \lpar U} \ltens T}}{
\leaf{
   Q
}}}
    }} \endxy}
\\[8pt]
$$
If the instance of $\rho$ is inside any of $S$, $U$, $R$, or $T$, 
we can exchange the instances of $\rho$ and the instance of switch, and obtain $\Pi'$ 
such that $\#\Pi'=(n,r-1)$, where we can apply the induction hypothesis.  
Otherwise, $\rho$ cannot be $\ruleaidown$ as this would 
contradict with the instance of switch being non-deep.
Thus, there are three cases, where $\rho = \unitonedown$, 
$\rho = \unittwodown$ or 
 $\rho = \mathsf{dls}$. 

If $\rho = \unitonedown$, the only case to consider is the one where $U = \lbot$, 
which is impossible if the switch instance is an instance of lazy switch. 

If $\rho = \unittwodown$, the only case to consider is the one where $T = \lone$. In this case, 
we permute trivially by first applying the rule $\unittwodown$ and then the switch.

 If $\rho = \mathsf{dls}$, because the instance of $\mathsf{ls}$ below it is not an instance 
of deep switch, we have either $R = \lone$ or $R = \aprs{R_1 \ltens R_2}$.
We discard the case $R = \lone$  by resorting to 
Proposition \ref{proposition:unit:invertible} and assuming that 
the  rule $\unittwodown$  is applied exhaustively in proof construction,
and prove the latter case.   We have two possibilities: \\
%%%%%%%%%%%%%
%%%%%%%%%%%%%

$\bullet$
If we have the situation on the left below, 
we can then replace the two instances of the switch 
rule with a single instance:

$$
\vcenter{
\dernote{\mathsf{ls}}{}{
S\pars{\aprs{R_1 \ltens R_2  \ltens T} \lpar U}
}{
\rootr{\mathsf{dls} \; }{}{S\aprs{\pars{\aprs{R_1 \ltens R_2} \lpar U} \ltens T}}{
\leaf{
   S\aprs{{\pars{U \lpar R_1} \ltens R_2} \ltens T}
}}}}
\qquad
\leadsto
\qquad
\vcenter{
\dernote{\mathsf{dls}}{}{
S\pars{\aprs{R_1 \ltens R_2  \ltens T} \lpar U}
}{
\leaf{
   S\aprs{{\pars{U \lpar R_1} \ltens R_2} \ltens T}
}}}\quad ,
\\[8pt]
$$
where $R_1$ is not a copar.  We obtain a proof $\Pi'$ such that 
$\#\Pi'=(n-1,r')$ to  which we can then apply the induction hypothesis, 
independent from the value of $r'$.  \\

%%%%%%%%%%%%%%%%%%
%%%%%%%%%%%%%%%%%%

$\bullet$
If we have the situation on the left below, 
we can permute the non-deep instance of switch up in the derivation 
as follows:

$$
\vcenter{
\dernote{\mathsf{ls}}{}{
S\pars{\aprs{R_1 \ltens R_2  \ltens T} \lpar \aprs{U_1 \ltens U_2}}
}{
\rootr{\mathsf{dls} \; }{}{S\aprs{\pars{\aprs{R_1 \ltens R_2} \lpar \aprs{U_1 \ltens U_2}} \ltens T}}{
\leaf{
   S\aprs{\pars{\aprs{R_1 \ltens R_2} \lpar U_1 } \ltens U_2 \ltens T}
}}}}
\qquad
\leadsto
\qquad
\vcenter{
\dernote{\mathsf{dls}}{}{
S\pars{\aprs{R_1 \ltens R_2  \ltens T} \lpar\aprs{U_1 \ltens U_2}}
}{
\rootr{\mathsf{ls} \; }{}{S\aprs{\pars{\aprs{R_1 \ltens R_2  \ltens T} \lpar U_1}  \ltens U_2}}{
\leaf{
S\aprs{\pars{\aprs{R_1 \ltens R_2  \ltens T} \lpar U_1}  \ltens U_2}
}}}}\quad .
\\[8pt]
$$
  We obtain a proof $\Pi'$ such that 
$\#\Pi'=(n,r-1)$ to  which we can then apply the induction hypothesis.  
%%%%%%%%%%%%%
%%%%%%%%%%%%%
\qed

As  Example \ref{example:global:view} demonstrates, 
a constructive transformation that replaces all the instances of the switch rule 
that are not interaction switch rule  instances requires an 
observation mechanism on the context of the switch rule instances to be removed. 
We use below a splitting theorem \cite{Gug02},
which provides such a global view of the rule instances. 
This way, we show that $\MLSdli$ is complete for 
multiplicative linear logic 
by exploiting the relation between completeness 
and cut-elimination.  We prove this result on $\MLSdli$, because the 
other systems with interaction switch follow simpler versions of the 
same argument.

%%%%%%%%%%%%%%%%%%%%%%%%%%%%%%%%%%%%%
%%%%%%%%%%%%%%%%%%%%%%%%%%%%%%%%%%%%%
%%%%%%%%%%%%%%%%%%%%%%%%%%%%%%%%%%%%%
%%%%%%%%%%%%%%%%%%%%%%%%%%%%%%%%%%%%%
%%%%%%%%%%%%%%%%%%%%%%%%%%%%%%%%%%%%%
%%%%%%%%%%%%%%%%%%%%%%%%%%%%%%%%%%%%%
%%%%%%%%%%%%%%%%%%%%%%%%%%%%%%%%%%%%%
%%%%%%%%%%%%%%%%%%%%%%%%%%%%%%%%%%%%%
%%%%%%%%%%%%%%%%%%%%%%%%%%%%%%%%%%%%%
%%%%%%%%%%%%%%%%%%%%%%%%%%%%%%%%%%%%%
%%%%%%%%%%%%%%%%%%%%%%%%%%%%%%%%%%%%%
%%%%%%%%%%%%%%%%%%%%%%%%%%%%%%%%%%%%%
%%%%%%%%%%%%%%%%%%%%%%%%%%%%%%%%%%%%%
%%%%%%%%%%%%%%%%%%%%%%%%%%%%%%%%%%%%%
%%%%%%%%%%%%%%%%%%%%%%%%%%%%%%%%%%%%%
%%%%%%%%%%%%%%%%%%%%%%%%%%%%%%%%%%%%%
%%%%%%%%%%%%%%%%%%%%%%%%%%%%%%%%%%%%%
%%%%%%%%%%%%%%%%%%%%%%%%%%%%%%%%%%%%%
%%%%%%%%%%%%%%%%%%%%%%%%%%%%%%%%%%%%%
%%%%%%%%%%%%%%%%%%%%%%%%%%%%%%%%%%%%%
%%%%%%%%%%%%%%%%%%%%%%%%%%%%%%%%%%%%%

\begin{lem}  \label{lemma:reslazyswitch:independence} 
For any structures $P$, $U$ and $R$, if 
$\pars{P \lpar U}$ has a proof in  $\MLSdli$, 
then there is a derivation
$
\vcenter{\xy\xygraph{[]!{0;<2pc,0pc>:}
{
R
  }-@{=}^<>(.5){\MLSdli} _<>(.5){}[d] {
\pars{\aprs{R \ltens P} \lpar U}
}
    }\endxy}
\;\; .
$
\end{lem}

%%%%%%%%%%%%%%%%%%%%%%%%%%%%%%%%%%%%%
%%%%%%%%%%%%%%%%%%%%%%%%%%%%%%%%%%%%%
%%%%%%%%%%%%%%%%%%%%%%%%%%%%%%%%%%%%%

\proof
Let $\Pi$ be the proof of the structure 
$\pars{ P \lpar  U}$ in $\MLSdli$. 
We associate with every structure $Q$
a pair $\# Q  = (n,r)$, where $n$ is
the cardinality of $\down{}{}{Q}$
and $r$ is the cardinality of $\occ\, Q$.
We prove by induction   on $\# \pars{P \lpar U}$ by 
Proposition \ref{proposition:induction:measure}.
We single out the bottom-most rule instance in $\Pi$ such that for some structure $Q$ 
and rule $\rho \in \{ \, \ruleaidown, \mathsf{dlis}, \unitonedown, \unittwodown \, \}$

$$
\vcenter{\xy\xygraph{[]!{0;<3.2pc,0pc>:}
{
R
  }-@{=}^<>(.5){\MLSdli} _<>(.5){}[d] {
\dernote{\rho}{\;\,}{\pars{\aprs{R \ltens P} \lpar U}}{\leaf{Q}}
}
    }\endxy}
\;\; .
\\[8pt]
$$

The base case is given with one of the following: 
(i) $P = \lbot$ and $U = \lone$; 
(ii)  $P = \lone$ and $U = \lbot$;
and 
(iii) $P = a$ and $U = \neg a$ or vice versa,
which are proved by replacing the derivation above with the following derivations.

$$
\begin{array}{c}
\dernote{\deeplazyintswir}{}{\pars{\aprs{R \ltens \lbot} \lpar \lone}}{
\rootr{\ruleaidown\,}{}{\aprs{R \ltens \pars{\lbot \lpar \lone}}}{
\rootr{\unittwodown\,}{}{\aprs{R \ltens \lone}}{
\leaf{R}}}}
\end{array}
\qquad \quad
\begin{array}{c}
\dernote{\unitonedown}{}{\pars{\aprs{R \ltens \lone} \lpar \lbot}}{
\rootr{\unittwodown\,}{}{\aprs{R \ltens \lone}}{
\leaf{R}}}
\end{array}
\qquad \quad
\begin{array}{c}
\dernote{\deeplazyintswir}{}{\pars{\aprs{R \ltens a} \lpar \neg a}}{
\rootr{\ruleaidown\,}{}{\aprs{R \ltens \pars{a \lpar \neg a}}}{
\rootr{\unittwodown\,}{}{\aprs{R \ltens \lone}}{
\leaf{R}}}}
\end{array}
\\[8pt]
$$
For the inductive cases,
let us reason on the position of the instance of the rule $\rho$ 
in $\pars{R \lpar P}$. There are the following
possibilities:\\

%%%%%%%%%%
\begin{enumerate}
%%%%%%%%%%
%%%%%%%%%%%%%%%%%%%%%%%%
 \item  %%
   $\rho = \ruleaidown\; :$ 
  There are the following cases:\\

   \begin{enumerate}
    \item
    If $\ruleaidown$ is applied inside $P$

$$    
\textrm{such that}
  \vcenter{\xy \xygraph{[]!{0;<3pc,0pc>:}
 {\lone}*=<14pt>{}:@{=}^<>(.5){} _<>(.5){} [d] {
\qquad \dernote{\ruleaidown}{\; ,\;\; \textrm{then}}
        {\pars{P \lpar U}}
        {\leaf{\pars{P' \lpar U }}}
    }} \endxy} 
\quad 
 \vcenter{\xy \xygraph{[]!{0;<3pc,0pc>:}
 {R}*=<14pt>{}:@{=}^<>(.5){} _<>(.5){\Delta} [d] {
       \dernote{\ruleaidown}{\; ,}
        {\pars{\aprs{R \ltens P} \lpar U }}
        {\leaf{ \pars{\aprs{R \ltens P'} \lpar U } }}
    }} \endxy} 
\\[8pt]
$$
where $\Delta$ is delivered by the induction hypothesis.\\
%%%%%%%%%%%%%%%%%%%%%%
\item
  The $\ruleaidown$ is applied inside $U$: similar to the previous case.\\
\item 
If $P = \pars{P' \lpar a}$, $U = \pars{\neg a \lpar U'  }$, and 
$$
  \vcenter{\xy \xygraph{[]!{0;<3pc,0pc>:}
 {\lone}*=<14pt>{}:@{=}^<>(.5){} _<>(.5){\Pi'} [d] {
   \qquad       \dernote{\ruleaidown}{\; , \; \; \textrm{then}}
        {\pars{P'  \lpar a \lpar \neg a \lpar U'}}
        {\leaf{\pars{P'  \lpar \lone \lpar U'}}}
    }} \endxy} 
\quad 
 \vcenter{\xy \xygraph{[]!{0;<3.8pc,0pc>:}
 {R}*=<14pt>{}:@{=}^<>(.5){} _<>(.5){\Delta} [d] {
         \dernote{\deeplazyintswir}{\; ,}
        {\pars{\aprs{R \ltens \pars{P' \lpar a}} \lpar \neg a \lpar U' }}{
        \rootr{\ruleaidown}{}{\pars{\aprs{R \ltens \pars{P' \lpar a \lpar \neg a}} \lpar U' }}{  
        \leaf{ \pars{\aprs{R \ltens \pars{P' \lpar \lone}} \lpar U' } }}}
    }} \endxy} 
\\[8pt]
$$
where $\Delta$ is obtained by applying the induction hypothesis to $\Pi'$.\\
%%%%%%%%%
\end{enumerate}
%%%%%%%%%
%
%%%%%%%%%%%%%%%%%%
\item
%%%%%%%%%%%%%%%%%%
%%%%%%%%%%%%%%%%%%
%%%%%%%%%%%%%%%%%%%%
  $\rho = \unitonedown$:  
if the rule $\unitonedown$  is applied inside  $P$ or $U$, then we have cases
that are similar to the case $(a)$  of $\rho = \ruleaidown$. Otherwise either it is the 
case that $P \approx \lbot$ or $U \approx \lbot$. For these cases, we prove as follows.\\
\begin{itemize}
\item
If
$$
  \vcenter{\xy \xygraph{[]!{0;<3pc,0pc>:}
 {\lone}*=<14pt>{}:@{=}^<>(.5){} _<>(.5){\Pi'} [d] {
     \dernote{\unitonedown}{\quad}{\pars{\bot \lpar U}}{\leaf{U}}   
 }} \endxy}
\qquad 
\textrm{ then }
\qquad
%%%%%%
%%%%%%
%%%%%%
\vcenter{\xy\xygraph{[]!{0;<2.6pc,0pc>:}
 {                             
\dernote{\deeplazyintswir}{\quad;}{\pars{\aprs{R \ltens \bot} \lpar \lone} }{
       \rootr{\unitonedown\;}{}{  
                  \aprs{R \ltens \pars{\bot \lpar \lone}}}{
      \rootr{\unittwodown\;}{}{  
                  \aprs{R \ltens \lone}}   
      {\leaf{R}}}}
                                 }
         -@{=}^<>(.5){ 
             }_<>(.5){\Pi'} [dd] {
 \pars{\aprs{R \ltens \bot} \lpar U} 
}
       }\endxy}
\\[8pt]
$$ 
%%%%%%%%
%%%%%%%%
%%%%%%%%
\item
and if
$$
  \vcenter{\xy \xygraph{[]!{0;<3pc,0pc>:}
 {\lone}*=<14pt>{}:@{=}^<>(.5){} _<>(.5){\Pi'} [d] {
     \dernote{\unitonedown}{\quad\:}{\pars{P \lpar \bot}}{\leaf{P}}   
 }} \endxy}
\qquad 
\textrm{ then }
\qquad
%%%%%%
%%%%%%
%%%%%%
\vcenter{\xy\xygraph{[]!{0;<2.2pc,0pc>:}
 {      \dernote{\unitonedown}{\qquad}{\aprs{R \ltens \lone} }{
             \leaf{R}}
                     }
         -@{=}^<>(.5){ 
             }_<>(.5){\Pi'} [dd] {
\dernote{\unitonedown}{\qquad}{\pars{\aprs{R \ltens P} \lpar \bot} }{
             \leaf{\aprs{R \ltens P} }}
}
       }\endxy} .
\\[8pt]
$$
\end{itemize}
%%%%%%%%%%%%%%%%%%
%
%
%%%%%%%%%%%%%%%%%%
%%%%%%%%%%%%%%%%%%
%%%%%%%%%%%%%%%%%%
\item
%%%%%%%%%%%%%%%%%%%%
  $\rho = \unittwodown$:  
the only possible cases are those that are applied inside  $P$ or $U$,
which are similar to the case $(a)$  of $\rho = \ruleaidown$.
In this case, if   $n$ in $\#(n,r)$ does not 
decrease, $r$ always decreases by $1$, thus induction 
hypothesis can be applied.\\
%%%%%%%%%%%%%%%%%%
%
\item
%%%%%%%%%%%%%%%%%%%%
  $\rho = \deeplazyintswir$ :  
If the rule  is inside $P$ or $U$,
similar to the case (a) of $\rho = \ruleaidown$.
Otherwise there are the following cases:\\
\begin{enumerate}
\item 
If $P = \pars{P' \lpar \aprs{ Q \ltens T }}$, $U = \pars{W \lpar U' }$, and 

$$
  \vcenter{\xy \xygraph{[]!{0;<3pc,0pc>:}
 {\lone}*=<14pt>{}:@{=}^<>(.5){} _<>(.5){\Pi'} [d] {
 \qquad \dernote{\deeplazyintswir}{\qquad}
        {\pars{P' \lpar U' \lpar \aprs{ Q \ltens T } \lpar W}}
        {\leaf{  \pars{P' \lpar U' \lpar \aprs{ \pars{Q \lpar W} \ltens T } } }}
    }} \endxy} 
\\[8pt]
$$
then we construct the derivation

$$
 \vcenter{\xy \xygraph{[]!{0;<3.8pc,0pc>:}
 {R}*=<14pt>{}:@{=}^<>(.5){} _<>(.5){\Delta} [d] {
 \; \dernote{\deeplazyintswir}{\; ,}
        {\pars{\aprs{R \ltens \pars{P' \lpar  \aprs{ Q \ltens T }}}  \lpar W \lpar U'}}{
        \rootr{\deeplazyintswir}{}{\pars{\aprs{R \ltens \pars{P' \lpar \aprs{ Q \ltens T } \lpar W}} \lpar U' }}{  
        \leaf{ \pars{\aprs{R \ltens \pars{P' \lpar \aprs{ \pars{Q \lpar W} \ltens T } }} \lpar U' } }}}
    }} \endxy} 
\\[8pt]
$$
where $\Delta$ is obtained by applying the induction hypothesis to $\Pi'$.\\
%%%%%%%%%
\item 
The case where $P = \pars{P' \lpar W}$, $U = \pars{ \aprs{ Q \ltens T } \lpar U' }$, and
the rule $\deeplazyintswir$ is applied as in case (a) is similar.\qed  
\end{enumerate}
%
%%%%%%%%%%%%%%%%%%
%%%%%%%%%%%%%%%%%%
%
\end{enumerate}

%%%%%%%%%%%%%%%%%%%%%%%%%%%%%%%%%%%%%
%%%%%%%%%%%%%%%%%%%%%%%%%%%%%%%%%%%%%
%%%%%%%%%%%%%%%%%%%%%%%%%%%%%%%%%%%%%
%%%%%%%%%%%%%%%%%%%%%%%%%%%%%%%%%%%%%
%%%%%%%%%%%%%%%%%%%%%%%%%%%%%%%%%%%%%
%%%%%%%%%%%%%%%%%%%%%%%%%%%%%%%%%%%%%
%%%%%%%%%%%%%%%%%%%%%%%%%%%%%%%%%%%%%
%%%%%%%%%%%%%%%%%%%%%%%%%%%%%%%%%%%%%
%%%%%%%%%%%%%%%%%%%%%%%%%%%%%%%%%%%%%
%%%%%%%%%%%%%%%%%%%%%%%%%%%%%%%%%%%%%
%%%%%%%%%%%%%%%%%%%%%%%%%%%%%%%%%%%%%
%%%%%%%%%%%%%%%%%%%%%%%%%%%%%%%%%%%%%
%%%%%%%%%%%%%%%%%%%%%%%%%%%%%%%%%%%%%
%%%%%%%%%%%%%%%%%%%%%%%%%%%%%%%%%%%%%
%%%%%%%%%%%%%%%%%%%%%%%%%%%%%%%%%%%%%
%%%%%%%%%%%%%%%%%%%%%%%%%%%%%%%%%%%%%
%%%%%%%%%%%%%%%%%%%%%%%%%%%%%%%%%%%%%

\begin{prop}
\label{proposition:multiplicative:conjunction}
In $\MLSdli$, a structure $\aprs{R \ltens T}$ has a proof  if and only if 
$R$ and $T$ have proofs.
\end{prop}

\proof
If direction being trivial, for the only if direction 
construct the proofs of $R$ and $T$ by induction on the length of the 
proof of $\aprs{R \ltens T}$. 
\qed

%%%%%%%%%%%%%%%%%%%%%%%%%%%%%%%%
%%%%%%%%%%%%%%%%%%%%%%%%%%%%%%%%
%%%%%%%%%%%%%%%%%%%%%%%%%%%%%%%%
%%%%%%%%%%%%%%%%%%%%%%%%%%%%%%%%
%%%%%%%%%%%%%%%%%%%%%%%%%%%%%%%%
%%%%%%%%%%%%%%%%%%%%%%%%%%%%%%%%
%%%%%%%%%%%%%%%%%%%%%%%%%%%%%%%%

In the theorem below,  we use Lemma \ref{lemma:reslazyswitch:independence}
together with an argument that takes into consideration the role of the units $\lone$
and $\lbot$ in $\MLSdli$ proofs.  The aim of this theorem is enabling the breaking 
of a formula into smaller pieces, while preserving provability. This way, the obtained pieces
can be used to build a constructive induction argument for proving properties of
$\MLSdli$ such as completeness and cut-elimination.

%%%%%%%%%%%%%%%%%%%%%%%%%%%%%%%%%
%%%%%%%%%%%%%%%%%%%%%%%%%%%%%%%%%
%%%%%%%%%%%%%%%%%%%%%%%%%%%%%%%%%
%%%%%%%%%%%%%%%%%%%%%%%%%%%%%%%%%
%%%%%%%%%%%%%%%%%%%%%%%%%%%%%%%%%
%%%%%%%%%%%%%%%%%%%%%%%%%%%%%%%%%
%%%%%%%%%%%%%%%%%%%%%%%%%%%%%%%%%
%%%%%%%%%%%%%%%%%%%%%%%%%%%%%%%%%

\begin{thm}[Shallow Splitting for $\MLSdli$] 
\label{theorem:split:MLSdi}
For all structures $R$, $T$ and $P$,
 if $\pars{\aprs{R \ltens T} \lpar  P}$ is provable in $\MLSdli$ 
then  
\begin{enumerate}
\item[(i)]
either $\pars{R \lpar P}$ and $T$ 
are provable in $\MLSdli$;\\
\item[(ii)]
or $R$ and $\pars{P \lpar T}$ 
are provable in $\MLSdli$;\\
\item[(iii)]
or there exist $P_1$, $P_2$ and 
 $
\vcenter{\xy\xygraph{[]!{0;<2.2pc,0pc>:}
{\pars{P_1 \lpar  P_2}
  }-@{=}^<>(.5){\MLSdli} _<>(.5){}[d] {P}
    }\endxy}\, $
such that $\pars{R \lpar P_1}$ and $\pars{T \lpar P_2}$ 
are provable in $\MLSdli$.\\
\end{enumerate}
\end{thm}

\proof
We associate with every structure $Q$
a pair $\# Q  = (n,r)$, where $n$ is
the cardinality of $\down{}{}{Q}$
and $r$ is the cardinality of $\occ\, Q$.
We prove by induction   on $\# \pars{\aprs{R \ltens T} \lpar  P}$ by 
Proposition \ref{proposition:induction:measure}.
The base case is given with the structures 
$\pars{\aprs{\lone \ltens \lone} \lpar \lbot}$ 
and $\pars{\aprs{\lbot  \ltens \lone} \lpar \lone}$, 
which are trivially covered by the cases (i) or (ii). 

For the inductive cases, we single out the 
bottom most rule application $\rho$ in the 
proof of $\pars{\aprs{R \ltens T} \lpar  P}$. 
We reason on the position where $\rho$ is 
applied in $\pars{\aprs{R \ltens T} \lpar P}$. 
There are the following possibilities:
\begin{enumerate}
%%%%%%%%%%%%%%%%%%%%%%%%
%%%%%%%%%%%%%%%%%%%%%%%%
 \item
   $\rho = \ruleaidown\; :$ 
  There are the following cases:\\
   \begin{enumerate}
    \item
    If $\ruleaidown$ is applied inside $R$ such that

$$    
\qquad 
  \vcenter{\xy \xygraph{[]!{0;<3.2pc,0pc>:}
 {\lone}*=<14pt>{}:@{=}^<>(.5){} _<>(.5){} [d] {
\dernote{\ruleaidown}{}
        {\pars{\aprs{R \ltens T} \lpar P}}
        {\leaf{\pars{\aprs{R' \ltens T} \lpar P}}}
    }} \endxy} 
%%%%%
\quad 
\textrm{then}
\qquad
%%%%%
 \vcenter{\xy \xygraph{[]!{0;<3.2pc,0pc>:}
 {\lone}*=<14pt>{}:@{=}^<>(.5){} _<>(.5){\Pi_1} [d] {
 \dernote{\ruleaidown}{ \quad}
        {\pars{R \lpar P}}
        {\leaf{\pars{R' \lpar P}}}
    }} \endxy} 
%%%%%
\;
\textrm{,}
\qquad
%%%%%
 \vcenter{\xy \xygraph{[]!{0;<3.2pc,0pc>:}
 {\lone}*=<14pt>{}:@{=}^<>(.5){} _<>(.5){\Pi_2} [d] {
\dernote{\ruleaidown}{\quad}
        {R}
        {\leaf{R' }}
    }} \endxy} 
%%%%%%
\;
 \textrm{and}
\quad
%%%%%%
 \vcenter{\xy \xygraph{[]!{0;<3.2pc,0pc>:}
 {\lone}*=<14pt>{}:@{=}^<>(.5){} _<>(.5){\Pi_3} [d] {
 \dernote{\ruleaidown}{\;\; ,}
        {\pars{R \lpar P_1}}
        {\leaf{\pars{R' \lpar P_1}}}
    }} \endxy} 
\\[8pt]
$$
where $\Pi_1$, $\Pi_2$ and $\Pi_3$, respectively, 
are delivered by induction hypothesis for the cases (i), (ii) and (iii).
\\
%%%%%%%%%%%%%%%%%%%%%%
\item
  If the rule $\ruleaidown$ is applied inside $T$, we have a case similar to the previous one.
\\
\item
If the rule $\ruleaidown$ is applied inside $P$, such that 

$$ 
\qquad \qquad   
  \vcenter{\xy \xygraph{[]!{0;<3pc,0pc>:}
 {\lone}*=<14pt>{}:@{=}^<>(.5){} _<>(.5){} [d] {
\; \; \, \dernote{\ruleaidown}{\quad }
        {\pars{\aprs{R \ltens T} \lpar P}}
        {\leaf{\pars{\aprs{R \ltens T} \lpar P'}}}
    }} \endxy} 
\textrm{ then }
\qquad 
\vcenter{\xy\xygraph{[]!{0;<3pc,0pc>:}
{\lone
  }-@{=}^<>(.5){} _<>(.5){\Pi_1}[d] {
 \dernote{\ruleaidown}{\quad}
        {\pars{R \lpar P}}
        {\leaf{\pars{R \lpar P'}}}
}
    }\endxy}
\quad
\textrm{and }
\qquad 
\vcenter{\xy\xygraph{[]!{0;<3pc,0pc>:}
{\lone
  }-@{=}^<>(.5){} _<>(.5){\Pi_2}[d] {
 \dernote{\ruleaidown}{\quad ,}
        {\pars{P \lpar T}}
        {\leaf{\pars{P' \lpar T}}}
}
    }\endxy}
\\[8pt]
$$
for (i) and (ii), where $\Pi_1$  and $\Pi_2$ are delivered by induction hypothesis.
For (iii), we have

$$
%%%%%%%%%%%%%%%%
%%%%%%%%%%%%%%%%
\vcenter{\xy\xygraph{[]!{0;<3.2pc,0pc>:}
{\pars{P_1 \lpar  P_2}
  }-@{=}^<>(.5){} _<>(.5){\Delta}[d] {
 \dernote{\ruleaidown}{\quad}
        {P}
        {\leaf{P'}}
}
    }\endxy}
\\[8pt]
$$
where $\Delta$ is delivered by induction hypothesis.\\
\end{enumerate}
%
%%%%%%%%%%%%%%%%%%
%%%%%%%%%%%%%%%%%%%%%%%%%%%%%%%%%%%%%%%%%%%%%%%%
%%%%%%%%%%%%%%%%%%%%%%%%%%%%%%%%%%%%%%%%%%%%%%%%
%%%%%%%%%%%%%%%%%%%%%%%%%%%%%%%%%%%%%%%%%%%%%%%%
%%%%%%%%%%%%%%%%%%
%
\item
%%%%%%%%%%%%%%%%%%%%
$\rho = \unitonedown$:
if the rule is applied inside $R$, $T$ or $P$,
the proof is similar to the cases for $\rho = \ruleaidown$.
Otherwise, it must be that $P = \lbot$ such that

$$    
  \vcenter{\xy \xygraph{[]!{0;<3pc,0pc>:}
 {\lone}*=<14pt>{}:@{=}^<>(.5){} _<>(.5){} [d] {
             \dernote{\unitonedown}{\quad \;}
               {\pars{ \aprs{R \ltens T} \lpar \lbot}}
        {\leaf{ \aprs{R \ltens T}}}
    }} \endxy}  
\textrm{then}
\qquad \; 
\vcenter{\xy\xygraph{[]!{0;<3pc,0pc>:}
{\lone
  }-@{=}^<>(.5){} _<>(.5){}[d] {
 \dernote{\unitonedown}{\quad\;\; }
        {\pars{R \lpar \lbot}}
        {\leaf{ R }}
}
    }\endxy}
\\[8pt]
$$
as $R$ and $T$ must have proofs by 
Proposition \ref{proposition:multiplicative:conjunction}. \\
%
%%%%%%%%%%%%%%%%%%
%%%%%%%%%%%%%%%%%%%%%%%%%%%%%%%%%%%%%%%%%%%%%%%%
%%%%%%%%%%%%%%%%%%%%%%%%%%%%%%%%%%%%%%%%%%%%%%%%
%%%%%%%%%%%%%%%%%%%%%%%%%%%%%%%%%%%%%%%%%%%%%%%%
%%%%%%%%%%%%%%%%%%
%
\item
$\rho = \unittwodown$:  
%%%%%%%%%%%%%%%%%%%%
if the rule is applied inside $R$, $T$ or $P$,
the proof is similar to the cases for $\rho = \ruleaidown$.
Otherwise it must be that $T = \lone$ such that 

$$    
  \vcenter{\xy \xygraph{[]!{0;<3pc,0pc>:}
 {\lone}*=<14pt>{}:@{=}^<>(.5){} _<>(.5){\Pi} [d] {
\quad         \dernote{\unittwodown}{\qquad .}
        {\pars{ \aprs{R \ltens \lone} \lpar P}}
        {\leaf{ \pars{R \lpar P}}}
    }} \endxy} 
\\[8pt]
$$
Then $\pars{R \lpar  P}$ has the proof $\Pi$. \\
%%%%%%%%%%%%%%%%%%%%%%%%%%%%%%%%%%%%%%%%%%%%%%%%
%%%%%%%%%%%%%%%%%%%%%%%%%%%%%%%%%%%%%%%%%%%%%%%%
%%%%%%%%%%%%%%%%%%%%%%%%%%%%%%%%%%%%%%%%%%%%%%%%
%%%%%%%%%%%%%%%%%%%%%%%%%%%%%%%%%%%%%%%%%%%%%%%%
%%%%%%%%%%%%%%%%%%%%%%%%%%%%%%%%%%%%%%%%%%%%%%%%
%%%%%%%%%%%%%%%%%%%%%%%%%%%%%%%%%%%%%%%%%%%%%%%%
\item
  $\rho = \deeplazyintswir$ :  
if the rule is applied inside $R$, $T$ or $P$,
the proof is similar to the cases for $\rho = \ruleaidown$.
Otherwise, there are the following possibilities:
(Note that we dismiss the possibility that $T \approx \aprs{T' \ltens T''}$ 
as this would not introduce a new case for the instances of the rule $\deeplazyintswir$.)\\
\begin{enumerate}
 \item
   $R \approx \aprs{R' \ltens  R''}$ 
  and
   $P \approx \pars{P' \lpar  P''}$ 
   such that the bottom most rule instance is of the 
   following form:

 $$
\dernote{\deeplazyintswir}{\quad ,}
    {\pars{\aprs{R' \ltens R'' \ltens T} \lpar  P' \lpar P''}}
    {\leaf{ \pars{\aprs{\pars{R'  \lpar  P'} \ltens R'' \ltens T} \lpar P''}}}
\\[8pt]
$$
where    
the structure $R'$ is not a copar and $P'$ is not a par: 
we apply the induction hypothesis to the premise, 
by Proposition \ref{proposition:induction:measure},
and get\\
\begin{enumerate}
%%%%%%%%%%%
%%%%%%%%%%%
\item[(i)] either

$$
 \vcenter{\xy \xygraph{[]!{0;<2.2pc,0pc>:}
 {\lone}*=<14pt>{}:@{=}^<>(.5){} _<>(.5){\Pi_1} [d] {
       \pars{R'  \lpar P' \lpar P''}     }}
    \endxy}
\qquad
\textrm{and}
\qquad
 \vcenter{\xy \xygraph{[]!{0;<2.2pc,0pc>:}
 {\lone}*=<14pt>{}:@{=}^<>(.5){} _<>(.5){\Pi_2} [d] {
       \aprs{R'' \ltens T} } }
    \endxy}
\quad; 
\\[8pt]
$$
%%%%%%%%%%%
%%%%%%%%%%%
\item[(ii)]
or 

$$
\vcenter{\xy \xygraph{[]!{0;<2.2pc,0pc>:}
 {\lone}*=<14pt>{}:@{=}^<>(.5){} _<>(.5){\Pi_3} [d] {
       \pars{\aprs{R'' \ltens T}  \lpar P''}} }
    \endxy}
\qquad
\textrm{and}
\qquad
\vcenter{\xy \xygraph{[]!{0;<2.2pc,0pc>:}
 {\lone}*=<14pt>{}:@{=}^<>(.5){} _<>(.5){\Pi_4} [d] {
       \pars{R'  \lpar P'}     }}
    \endxy}
 \quad; 
\\[8pt]
$$
%%%%%%%%%%%
%%%%%%%%%%%

\item[(iii)]
or

$$
\qquad \qquad
\vcenter{\xy\xygraph{[]!{0;<2.2pc,0pc>:}
{\pars{P'_1 \lpar  P'_2}
  }-@{=}^<>(.5){} _<>(.5){\Delta_1}[d] {
P''
}
    }\endxy}
\qquad 
,
\qquad
 \vcenter{\xy \xygraph{[]!{0;<2.2pc,0pc>:}
 {\lone}*=<14pt>{}:@{=}^<>(.5){} _<>(.5){\Pi_5} [d] {
       \pars{R'  \lpar P' \lpar P'_1}     }}
    \endxy}
\qquad
\textrm{and}
\qquad
 \vcenter{\xy \xygraph{[]!{0;<2.2pc,0pc>:}
 {\lone}*=<14pt>{}:@{=}^<>(.5){} _<>(.5){\Pi_6} [d] {
       \pars{ \aprs{R'' \ltens T} \lpar  P'_2 }} }
    \endxy}
\quad .
\\[8pt]
$$
%%%%%%%%%%%
%%%%%%%%%%%
\end{enumerate}
%%%%%%%%%%%
%%%%%%%%%%%
%%%%%%%%%%%
%%%%%%%%%%%
%%%%%%%%%%%
%%%%%%%%%%%

\begin{itemize}
\item
If (i) is the case, we have all the required derivations to complete the proof.
%%%%%%%%%%%%%%%%%%%%%%%%%%
%%%%%%%%%%%%%%%%%%%%%%%%%%
%%%%%%%%%%%%%%%%%%%%%%%%%%
By Proposition \ref{proposition:multiplicative:conjunction},
from $\Pi_2$ we can construct the proof $\Pi'$  
of $R''$ and the proof $\Pi''$ of $T$. With $\Pi'$ and $\Pi_1$, 
we then construct

$$
\qquad \quad
\vcenter{\xy\xygraph{[]!{0;<5pc,0pc>:}
{\lone
  }-@{=}^<>(.5){} _<>(.5){\Pi_1}[d] {
\vcenter{\xy\xygraph{[]!{0;<3.4pc,0pc>:}
{
\dernote{\unittwodown}{\quad\,}{\pars{\aprs{R' \ltens \lone} \lpar P' \lpar P''}}{\leaf{\pars{R' \lpar P' \lpar P''}}}
  }-@{=}^<>(.5){} _<>(.5){\Pi'}[d] {
\pars{\aprs{R' \ltens R''} \lpar P' \lpar P''}
}
    }\endxy}
}
    }\endxy}\; .
\\[8pt]
$$

\item
If (ii) is the case, we can apply the induction hypothesis to $\Pi_3$, 
%$$
%\textrm{ because we have that } | \down{}{}{\pars{\aprs{R' \ltens R'' \ltens T}% \lpar  P' \lpar  P''}}| >
%| \down{}{}{\pars{\aprs{R'' \ltens T} \lpar  P'' }}|.
%\\[8pt]
%$$
 because we have that  $| \down{}{}{\pars{\aprs{R' \ltens R'' \ltens T} \lpar  P' \lpar  P''}}| >
| \down{}{}{\pars{\aprs{R'' \ltens T} \lpar  P'' }}|.$
\medskip

This way, from $\Pi_3$ we obtain, by Proposition \ref{proposition:induction:measure}, 
the following three cases:\\
%%%%%%%%%%%%%%%%%%%%%%%%%%
%%%%%%%%%%%%%%%%%%%%%%%%%%
%%%%%%%%%%%%%%%%%%%%%%%%%%
\begin{enumerate}[label=(ii.\roman*)]
\item%[($ii.i.$)]
We have the proof $\Pi_7$ of $\pars{R'' \lpar P''}$ and the proof $\Pi_8$ of $T$:
 together with $\Pi_{8}$,
 we construct 

$$
\qquad \quad
\vcenter{\xy\xygraph{[]!{0;<5pc,0pc>:}
{\lone
  }-@{=}^<>(.5){} _<>(.5){\Pi_7}[d] {
\vcenter{\xy\xygraph{[]!{0;<3.5pc,0pc>:}
{
\dernote{\unittwodown}{\quad\,.}{\pars{\aprs{R'' \ltens \lone} \lpar P''}}{\leaf{\pars{R''  \lpar P''}}}
  }-@{=}^<>(.5){} _<>(.5){\Delta'}[d] {
\pars{\aprs{R' \ltens R''} \lpar P' \lpar P'' }
}
    }\endxy}
}
    }\endxy}
\\[8pt]
$$
where $\Delta'$ is delivered by Lemma \ref{lemma:reslazyswitch:independence} 
with proof $\Pi_4$.\\

%%%%%%%%%%%%%%%%%%%%%%%%%%
%%%%%%%%%%%%%%%%%%%%%%%%%%
\item%[($ii.ii.$)]
We have the proof $\Pi_9$ of $R''$ and the proof $\Pi_{10}$ of $\pars{T \lpar  P''}$:
together with $\Pi_{10}$,
 we then construct 

$$
\qquad \quad
\vcenter{\xy\xygraph{[]!{0;<5pc,0pc>:}
{\lone
  }-@{=}^<>(.5){} _<>(.5){\Pi_4}[d] {
\vcenter{\xy\xygraph{[]!{0;<3.5pc,0pc>:}
{
\dernote{\unittwodown}{\quad.}{\pars{\aprs{R' \ltens \lone} \lpar P'}}{\leaf{\pars{R'  \lpar P'}}}
  }-@{=}^<>(.5){} _<>(.5){\Pi_9}[d] {
\pars{\aprs{R' \ltens R''} \lpar P' }
}
    }\endxy}
}
    }\endxy}
\\[8pt]
$$

%%%%%%%%%%%%%%%%%%%%%%%%%%
%%%%%%%%%%%%%%%%%%%%%%%%%%
\item%[($ii.iii.$)]
We have: 

$$
\qquad 
\vcenter{\xy\xygraph{[]!{0;<2.2pc,0pc>:}
{\pars{P''_1 \lpar  P''_2}
  }-@{=}^<>(.5){} _<>(.5){\Delta_2}[d] {
P''
}
    }\endxy}
\qquad 
,
\qquad
 \vcenter{\xy \xygraph{[]!{0;<2.2pc,0pc>:}
 {\lone}*=<14pt>{}:@{=}^<>(.5){} _<>(.5){\Pi_{11}} [d] {
       \pars{R''  \lpar P''_1}     }}
    \endxy}
\qquad
\textrm{and}
\qquad
 \vcenter{\xy \xygraph{[]!{0;<2.2pc,0pc>:}
 {\lone}*=<14pt>{}:@{=}^<>(.5){} _<>(.5){\Pi_{12}} [d] {
       \pars{ T \lpar  P''_2 }} }
    \endxy}
\quad .
\\[8pt]
$$
We then construct the derivation and the proofs

$$
\qquad \qquad
\vcenter{\xy\xygraph{[]!{0;<2.4pc,0pc>:}
{
\pars{P' \lpar P''_1 \lpar P''_2}
  }-@{=}^<>(.5){} _<>(.5){\Delta_2}[d] {
\pars{P' \lpar P''}
}
    }\endxy}
\qquad
,
\qquad
\vcenter{\xy \xygraph{[]!{0;<5pc,0pc>:}
 {\lone}*=<14pt>{}:@{=}^<>(.5){} _<>(.5){\Pi_4} [d] {
\vcenter{\xy\xygraph{[]!{0;<3.5pc,0pc>:}
{  \dernote{\unittwodown}{\quad\,}{\pars{\aprs{R' \ltens \lone} \lpar P' }}{\leaf{\pars{R' \lpar P' }}}
  }-@{=}^<>(.5){} _<>(.5){\Delta'}[d] {
\pars{\aprs{R' \ltens R''} \lpar P'  \lpar P''_1}
}
    }\endxy}
}}
    \endxy}
\quad
\textrm{ and }
\;
\qquad
%%%%%%%%%
\vcenter{\xy\xygraph{[]!{0;<2.2pc,0pc>:}
{\lone
  }-@{=}^<>(.5){} _<>(.5){\Pi_{12}}[d] {
\pars{T \lpar P''_2}
}
    }\endxy},
\\[8pt]
$$
where 
$\Delta'$ is  delivered by 
Lemma \ref{lemma:reslazyswitch:independence}  
with proof $\Pi_{11}$.\\
\end{enumerate}
%%%%%%%%%%%%%%%%%%%%%%%%%%
%%%%%%%%%%%%%%%%%%%%%%%%%%
%%%%%%%%%%%%%%%%%%%%%%%%%%

\item
If (iii) is the case, we can apply the induction hypothesis to $\Pi_6$, 
%$$
%\textrm{ because we have that } | \down{}{}{\pars{\aprs{R' \ltens R'' \ltens T} \lpar  P' \lpar  P''}}| >
%| \down{}{}{\pars{\aprs{R'' \ltens T} \lpar  P'_2 }}|.
%\\[8pt]
%$$
because we have that $| \down{}{}{\pars{\aprs{R' \ltens R'' \ltens T} \lpar  P' \lpar  P''}}| >
| \down{}{}{\pars{\aprs{R'' \ltens T} \lpar  P'_2 }}|.$
\medskip

This way, from $\Pi_6$ we obtain, by Proposition \ref{proposition:induction:measure}, the following three cases:\\
%%%%%%%%%%%%%%%%%%%%%%%%%%
%%%%%%%%%%%%%%%%%%%%%%%%%%
%%%%%%%%%%%%%%%%%%%%%%%%%%
\begin{enumerate}[label=(iii.\roman*)]
\item%[($iii.i.$)]
We have the proof $\Pi'_7$ of $\pars{R'' \lpar P'_2}$ and the proof $\Pi'_8$ of $T$:
we then construct the derivation and the proofs

 $$
\qquad \qquad 
\vcenter{\xy\xygraph{[]!{0;<2.4pc,0pc>:}
{
\pars{P' \lpar  P'_1 \lpar  P'_2}
  }-@{=}^<>(.5){} _<>(.5){\Delta_1}[d] {
\pars{P' \lpar P''}
}
    }\endxy}
%%%%
\qquad , 
\qquad
%%%%
\vcenter{\xy \xygraph{[]!{0;<5pc,0pc>:}
 {\lone}*=<14pt>{}:@{=}^<>(.5){} _<>(.5){\Pi_5} [d] {
\vcenter{\xy\xygraph{[]!{0;<3.5pc,0pc>:}
{ \dernote{\unittwodown}{\quad\,}{\pars{\aprs{R' \ltens \lone}\lpar P' \lpar P'_1}}{\leaf{
\pars{R' \lpar P' \lpar P'_1}}}
  }-@{=}^<>(.5){} _<>(.5){\Delta'}[d] {
\pars{\aprs{R' \ltens R''} \lpar P'  \lpar P'_1 \lpar P'_2}
}
    }\endxy}
}}
    \endxy}
\quad
\textrm{ and }
\qquad \;
%%%%%%%%%
\vcenter{\xy\xygraph{[]!{0;<2.2pc,0pc>:}
{\lone
  }-@{=}^<>(.5){} _<>(.5){\Pi'_{8}}[d] {
T
}
    }\endxy} \quad ,
\\[8pt]
$$
where 
$\Delta'$ is delivered by 
Lemma \ref{lemma:reslazyswitch:independence}  
with proof $\Pi'_{7}$.\\
%%%%%%%%%%%%%%%%%%%%%%%%%%
%%%%%%%%%%%%%%%%%%%%%%%%%%
\item%[($iii.ii.$)]
We have the proof $\Pi'_9$ of $R''$ and the proof $\Pi'_{10}$ of $\pars{T \lpar  P'_2}$:
We then construct the derivation and the proofs
 
$$
\qquad \qquad
\vcenter{\xy\xygraph{[]!{0;<2.4pc,0pc>:}
{
\pars{P' \lpar  P'_1 \lpar  P'_2}
  }-@{=}^<>(.5){} _<>(.5){\Delta_1}[d] {
\pars{P' \lpar P''}
}
    }\endxy}
%%%%
\quad , 
\qquad
%%%%
\vcenter{\xy \xygraph{[]!{0;<5pc,0pc>:}
 {\lone}*=<14pt>{}:@{=}^<>(.5){} _<>(.5){\Pi_5} [d] {
\vcenter{\xy\xygraph{[]!{0;<3.5pc,0pc>:}
{
\dernote{\unittwodown}{\quad\,}{\pars{\aprs{R' \ltens \lone} \lpar P'  \lpar P'_1}}{\leaf{\pars{R' \lpar P'  \lpar P'_1}}}
  }-@{=}^<>(.5){} _<>(.5){\Pi'_9}[d] {
\pars{\aprs{R' \ltens R''} \lpar P'  \lpar P'_1}
}
    }\endxy}
}}
    \endxy}
\quad
\textrm{ and }
\qquad
%%%%%%%%%
\vcenter{\xy\xygraph{[]!{0;<2.2pc,0pc>:}
{\lone
  }-@{=}^<>(.5){} _<>(.5){\Pi'_{10}}[d] {
\pars{T \lpar P'_2}
}
    }\endxy}.
\\[8pt]
$$

%%%%%%%%%%%%%%%%%%%%%%%%%%
%%%%%%%%%%%%%%%%%%%%%%%%%%
\item%[($iii.iii.$)]
We have:

$$
\qquad \quad
\vcenter{\xy\xygraph{[]!{0;<2pc,0pc>:}
{\pars{P''_1 \lpar  P''_2}
  }-@{=}^<>(.5){} _<>(.5){\Delta'_2}[d] {
P'_2
}
    }\endxy}
\qquad 
,
\qquad
 \vcenter{\xy \xygraph{[]!{0;<2.2pc,0pc>:}
 {\lone}*=<14pt>{}:@{=}^<>(.5){} _<>(.5){\Pi'_{11}} [d] {
       \pars{R''  \lpar P''_1}     }}
    \endxy}
\qquad
\textrm{and}
\qquad
 \vcenter{\xy \xygraph{[]!{0;<2.2pc,0pc>:}
 {\lone}*=<14pt>{}:@{=}^<>(.5){} _<>(.5){\Pi'_{12}} [d] {
       \pars{ T \lpar  P''_2 }} }
    \endxy}
\quad .
\\[8pt]
$$
We construct the derivation and the proofs

$$
\qquad \quad
\vcenter{\xy\xygraph{[]!{0;<4.6pc,0pc>:}
{\pars{P' \lpar  P'_1 \lpar  P''_1 \lpar  P''_2}
  }-@{=}^<>(.5){} _<>(.5){\Delta'_2}[d] {
\vcenter{\xy\xygraph{[]!{0;<2.6pc,0pc>:}
{
\pars{P' \lpar  P'_1 \lpar  P'_2}
  }-@{=}^<>(.5){} _<>(.5){\Delta_1}[d] {
\pars{P' \lpar P''}
}
    }\endxy}
}
    }\endxy}
\quad
,
\qquad
\vcenter{\xy \xygraph{[]!{0;<4.8pc,0pc>:}
 {\lone}*=<14pt>{}:@{=}^<>(.5){} _<>(.5){\Pi_5} [d] {
\vcenter{\xy\xygraph{[]!{0;<3.2pc,0pc>:}
{  \dernote{\unittwodown}{\quad}{\pars{\aprs{R' \ltens \lone} \lpar P'  \lpar P'_1}}
{\leaf{\pars{R' \lpar P' \lpar P'_1}}}
  }-@{=}^<>(.5){} _<>(.5){\Delta'}[d] {
\pars{\aprs{R' \ltens R''} \lpar P'  \lpar P'_1 \lpar P''_1}
}
    }\endxy}
}}
    \endxy}
\;\;
\textrm{ and }
\qquad
%%%%%%%%%
\vcenter{\xy\xygraph{[]!{0;<2.2pc,0pc>:}
{\lone
  }-@{=}^<>(.5){} _<>(.5){\Pi'_{12}}[d] {
\pars{T \lpar P''_2}
}
    }\endxy},
\\[8pt]
$$
%%%%%%%%%%%%%%%%%%%%%%%%%%%
%%%%%%%%%%%%%%%%%%%%%%%%%%%
where 
$\Delta'$ is delivered by 
Lemma \ref{lemma:reslazyswitch:independence}  
with proof $\Pi'_{11}$.\\
\end{enumerate}
\end{itemize}

%%%%%%%%%%%%%%%%%%%%%%%%%%%%
%%%%%%%%%%%%%%%%%%%%%%%%%%%%
%%%%%%%%%%%%%%%%%%%%%%%%%%%%
 \item
   $P \approx \pars{P' \lpar  P''}$ 
   such that the bottom most rule instance is given with:

 $$
\dernote{\deeplazyintswir}{\quad .}
    {\pars{\aprs{R \ltens T} \lpar  P' \lpar P''}}
    {\leaf{ \pars{\aprs{\pars{R  \lpar  P'} \ltens T} \lpar P''}}}
\\[8pt]
$$
We prove as  in case ($a$) where we replace $R'$ with $R$ 
and $R''$ with $\lone$. \\

%%%%%%%%%%%%%%%%%%%%%%%%%%%%
%%%%%%%%%%%%%%%%%%%%%%%%%%%%
%%%%%%%%%%%%%%%%%%%%%%%%%%%%

%
 \item
   $R \approx \aprs{R' \ltens  R''}$ 
   such that the bottom-most rule instance is  given with:

 $$
\dernote{\deeplazyintswir}{\quad .}
    {\pars{\aprs{R' \ltens R'' \ltens T} \lpar  P}}
    {\leaf{ \aprs{\pars{R'  \lpar  P} \ltens R'' \ltens T}}}
\\[8pt]
$$

By Proposition \ref{proposition:multiplicative:conjunction}, 
because the structure 
$\aprs{\pars{R'  \lpar  P} \ltens R'' \ltens T}$
has a proof, we have that the structures 
$\pars{R'  \lpar  P}$,  $R''$  and $T$ have proofs, with 
which we can easily construct a proof of
$\pars{\aprs{R' \ltens R''}  \lpar  P}$.\\
%%%%%%%%%%%%%%%%%%%%%%%%%%%%
%%%%%%%%%%%%%%%%%%%%%%%%%%%%
%%%%%%%%%%%%%%%%%%%%%%%%%%%%
 \item
   The bottom-most rule instance is given with:

 $$
\dernote{\deeplazyintswir}{\quad .}
    {\pars{\aprs{R \ltens T} \lpar  P}}
    {\leaf{ \aprs{\pars{R  \lpar  P} \ltens T}}}
\\[8pt]
$$
By Proposition \ref{proposition:multiplicative:conjunction}, 
we have that  $\pars{R  \lpar  P}$ and $T$ have proofs.\\

%%%%%%%%%%%%%%%%%%%%%%%%%%%%
%%%%%%%%%%%%%%%%%%%%%%%%%%%%
%%%%%%%%%%%%%%%%%%%%%%%%%%%%
%%%%%%%%%%%%%%%%%%%%%%%%%%%%
%%%%%%%%%%%%%%%%%%%%%%%%%%%%
%%%%%%%%%%%%%%%%%%%%%%%%%%%%

\item
   $P \approx \pars{\aprs{P' \ltens P''} \lpar  U }$  
   such that the bottom-most rule instance is

$$
\dernote{\deeplazyintswir}{\quad .}
    {\pars{\aprs{R \ltens T} \lpar  \aprs{P' \ltens P''} \lpar  U}}
    {\leaf{ \pars{\aprs{\pars{\aprs{R \ltens T} \lpar P'} \ltens P''} \lpar U}
}}
\\[8pt]
$$

%%%%%%%%%%%%%%%%%%%%%%%%%%%%%
We apply the induction hypothesis, by Proposition \ref{proposition:induction:measure}, and get \\
\begin{enumerate}
%%%%%%%%%%%%%%%%%%%%%%%%%%%%%%%%%%%%%%%
%%%%%%%%%%%%%%%%%%%%%%%%%%%%%%%%%%%%%%%
%%%%%%%%%%%%%%%%%%%%%%%%%%%%%%%%%%%%%%%
\item
either
%%%%%%%%%%%%%%%%%%%%%%%%%%%%%%%%%%%%%%%
%%%%%%%%%%%%%%%%%%%%%%%%%%%%%%%%%%%%%%%
%%%%%%%%%%%%%%%%%%%%%%%%%%%%%%%%%%%%%%%

$$
\qquad \qquad
 \vcenter{\xy \xygraph{[]!{0;<2.2pc,0pc>:}
 {\lone}*=<14pt>{}:@{=}^<>(.5){} _<>(.5){\Pi_1} [d] {
       \pars{\aprs{R \ltens T} \lpar P' \lpar U} }}
    \endxy}
\qquad
\textrm{and}
\qquad
 \vcenter{\xy \xygraph{[]!{0;<2.2pc,0pc>:}
 {\lone}*=<14pt>{}:@{=}^<>(.5){} _<>(.5){\Pi_2} [d] {
        P''} }
    \endxy}
\quad ;
\\[8pt]
$$
%%%%%%%%%%%%%%%%%%%%%%%%%%%%%%%%%%%%%%%
%%%%%%%%%%%%%%%%%%%%%%%%%%%%%%%%%%%%%%%
%%%%%%%%%%%%%%%%%%%%%%%%%%%%%%%%%%%%%%%
\item
%%%%%%%%%%%%%%%%%%%%%%%%%%%%%%%%%%%%%%%
%%%%%%%%%%%%%%%%%%%%%%%%%%%%%%%%%%%%%%%
%%%%%%%%%%%%%%%%%%%%%%%%%%%%%%%%%%%%%%%
or

$$
\qquad \qquad
 \vcenter{\xy \xygraph{[]!{0;<2.2pc,0pc>:}
 {\lone}*=<14pt>{}:@{=}^<>(.5){} _<>(.5){\Pi_3} [d] {
       \pars{\aprs{R \ltens T} \lpar P'} }}
    \endxy}
\qquad
\textrm{and}
\qquad
 \vcenter{\xy \xygraph{[]!{0;<2.2pc,0pc>:}
 {\lone}*=<14pt>{}:@{=}^<>(.5){} _<>(.5){\Pi_4} [d] {
       \pars{ P'' \lpar  U }} }
    \endxy}
\quad ;
\\[8pt]
$$
%%%%%%%%%%%%%%%%%%%%%%%%%%%%%%%%%%%%%%%
%%%%%%%%%%%%%%%%%%%%%%%%%%%%%%%%%%%%%%%
%%%%%%%%%%%%%%%%%%%%%%%%%%%%%%%%%%%%%%%
\item
%%%%%%%%%%%%%%%%%%%%%%%%%%%%%%%%%%%%%%%
%%%%%%%%%%%%%%%%%%%%%%%%%%%%%%%%%%%%%%%
%%%%%%%%%%%%%%%%%%%%%%%%%%%%%%%%%%%%%%%
or

$$
\qquad \qquad
\vcenter{\xy\xygraph{[]!{0;<2.2pc,0pc>:}
{\pars{U_1 \lpar  U_2}
  }-@{=}^<>(.5){} _<>(.5){\Delta_1}[d] {
U
}
    }\endxy}
\qquad 
,
\qquad
 \vcenter{\xy \xygraph{[]!{0;<2.2pc,0pc>:}
 {\lone}*=<14pt>{}:@{=}^<>(.5){} _<>(.5){\Pi_5} [d] {
       \pars{\aprs{R \ltens T} \lpar P' \lpar U_1} }}
    \endxy}
\qquad
\textrm{and}
\qquad
 \vcenter{\xy \xygraph{[]!{0;<2.2pc,0pc>:}
 {\lone}*=<14pt>{}:@{=}^<>(.5){} _<>(.5){\Pi_6} [d] {
       \pars{ P'' \lpar  U_2 }} }
    \endxy}
\quad .
\\[8pt]
$$
\end{enumerate}
Other cases being similar, we consider (iii):
by Proposition \ref{proposition:induction:measure},   given

$$
|\down{}{}{\pars{\aprs{ R \ltens T} \lpar  \aprs{P' \ltens  P''} \lpar  U}} |
>
|\down{}{}{\pars{ \aprs{ R \ltens T} \lpar P' \lpar U_1}} | 
\\[8pt]
$$
we can apply the induction hypothesis on $\Pi_5$,
and get \\
%%%%%%%%%%%%%%%%%%%%%%%%%%%%%%
\begin{enumerate}[label=(iii.\roman*)]
%%%%%%%%%%%%%%%%%%%%%%%%%%%%%%%%%%%%%%%
%%%%%%%%%%%%%%%%%%%%%%%%%%%%%%%%%%%%%%%
%%%%%%%%%%%%%%%%%%%%%%%%%%%%%%%%%%%%%%%
\item%[($iii.i.$)]
%%%%%%%%%%%%%%%%%%%%%%%%%%%%%%%%%%%%%%%
%%%%%%%%%%%%%%%%%%%%%%%%%%%%%%%%%%%%%%%
%%%%%%%%%%%%%%%%%%%%%%%%%%%%%%%%%%%%%%%
either

$$
\qquad \qquad
 \vcenter{\xy \xygraph{[]!{0;<2.2pc,0pc>:}
 {\lone}*=<14pt>{}:@{=}^<>(.5){} _<>(.5){\Pi_7} [d] {
       \pars{R \lpar P' \lpar U_1} }}
    \endxy}
\qquad
\textrm{and}
\qquad
 \vcenter{\xy \xygraph{[]!{0;<2.2pc,0pc>:}
 {\lone}*=<14pt>{}:@{=}^<>(.5){} _<>(.5){\Pi_8} [d] {
       T } }
    \endxy}
\quad 
\\[8pt]
$$
together with the derivation
%%%%%%%%%%%%%%%%%%%%%%%%%%%%%%%%%%%%%%%
%%%%%%%%%%%%%%%%%%%%%%%%%%%%%%%%%%%%%%%

$$
\qquad \qquad
\vcenter{\xy\xygraph{[]!{0;<5pc,0pc>:}
{
\dernote{\unittwodown}{\quad}{\pars{\aprs{P' \ltens \lone} \lpar  U_1}}
{\leaf{
\pars{P' \lpar  U_1}
}}
  }-@{=}^<>(.5){} _<>(.5){\Delta'}[d] {
\vcenter{\xy\xygraph{[]!{0;<2.6pc,0pc>:}
{
\pars{\aprs{P' \ltens P''} \lpar  U_1 \lpar  U_2}
  }-@{=}^<>(.5){} _<>(.5){\Delta_1}[d] {
\pars{\aprs{P' \ltens P''} \lpar  U}
}
    }\endxy}
}
    }\endxy}
\quad,
\\[8pt]
$$
where $\Delta'$ is  delivered by 
Lemma \ref{lemma:reslazyswitch:independence}
with proof $\Pi_6$;\\
%%%%%%%%%%%%%%%%%%%%%%%%%%%%%%%%%%%%%%%
%%%%%%%%%%%%%%%%%%%%%%%%%%%%%%%%%%%%%%%
%%%%%%%%%%%%%%%%%%%%%%%%%%%%%%%%%%%%%%%
\item%[($iii.ii.$)]
%%%%%%%%%%%%%%%%%%%%%%%%%%%%%%%%%%%%%%%
%%%%%%%%%%%%%%%%%%%%%%%%%%%%%%%%%%%%%%%
%%%%%%%%%%%%%%%%%%%%%%%%%%%%%%%%%%%%%%%
or

$$
\qquad \qquad
 \vcenter{\xy \xygraph{[]!{0;<2.2pc,0pc>:}
 {\lone}*=<14pt>{}:@{=}^<>(.5){} _<>(.5){\Pi_9} [d] {
       \pars{T \lpar P' \lpar U_1} }}
    \endxy}
\qquad
\textrm{and}
\qquad
 \vcenter{\xy \xygraph{[]!{0;<2.2pc,0pc>:}
 {\lone}*=<14pt>{}:@{=}^<>(.5){} _<>(.5){\Pi_{10}} [d] {
       R } }
    \endxy}
\quad 
\\[8pt]
$$
together with the derivation in case ($iii.i.$) above;\\
%%%%%%%%%%%%%%%%%%%%%%%%%%%%%%%%%%%%%%%
%%%%%%%%%%%%%%%%%%%%%%%%%%%%%%%%%%%%%%%
%%%%%%%%%%%%%%%%%%%%%%%%%%%%%%%%%%%%%%%
%%%%%%%%%%%%%%%%%%%%%%%%%%%%%%%%%%%%%%%
%%%%%%%%%%%%%%%%%%%%%%%%%%%%%%%%%%%%%%%
\item%[($iii.iii.$)]
%%%%%%%%%%%%%%%%%%%%%%%%%%%%%%%%%%%%%%%
%%%%%%%%%%%%%%%%%%%%%%%%%%%%%%%%%%%%%%%
%%%%%%%%%%%%%%%%%%%%%%%%%%%%%%%%%%%%%%%
or

$$
\qquad \qquad
\vcenter{\xy\xygraph{[]!{0;<2.2pc,0pc>:}
{\pars{P_1 \lpar  P_2}
  }-@{=}^<>(.5){} _<>(.5){\Delta_2}[d] {
\pars{P' \lpar  U_1}
}
    }\endxy}
\qquad 
,
\qquad
 \vcenter{\xy \xygraph{[]!{0;<2.2pc,0pc>:}
 {\lone}*=<14pt>{}:@{=}^<>(.5){} _<>(.5){\Pi_{11}} [d] {
       \pars{R \lpar P_1} }}
    \endxy}
\qquad
\textrm{and}
\qquad
 \vcenter{\xy \xygraph{[]!{0;<2.2pc,0pc>:}
 {\lone}*=<14pt>{}:@{=}^<>(.5){} _<>(.5){\Pi_{12}} [d] {
       \pars{ T \lpar  P_2 }} }
    \endxy}
\quad .
\\[8pt]
$$
In this case, by using $\Delta_2$, we can construct the derivation

$$
\vcenter{\xy\xygraph{[]!{0;<6.6pc,0pc>:}
{\pars{P_1 \lpar  P_2}
  }-@{=}^<>(.5){} _<>(.5){\Delta_2}[d] {
\vcenter{\xy\xygraph{[]!{0;<5pc,0pc>:}
{
\dernote{\unittwodown}{\quad\,}{\pars{\aprs{P' \ltens \lone} \lpar  U_1}}{\leaf{\pars{P' \lpar  U_1}}}
  }-@{=}^<>(.5){} _<>(.5){\Delta'}[d] {
\vcenter{\xy\xygraph{[]!{0;<2.5pc,0pc>:}
{
\pars{\aprs{P' \ltens P''} \lpar  U_1 \lpar  U_2}
  }-@{=}^<>(.5){} _<>(.5){\Delta_1}[d] {
\pars{\aprs{P' \ltens P''} \lpar  U}
}
    }\endxy}
}
    }\endxy}
}
    }\endxy},
\\[8pt]
$$
where $\Delta'$ is  delivered by 
Lemma \ref{lemma:reslazyswitch:independence}
with proof $\Pi_6$.\\
%%%%%%%%%%%%%%%%%%%%%%%%%%%%%%%%%%%%%%%
%%%%%%%%%%%%%%%%%%%%%%%%%%%%%%%%%%%%%%%
%%%%%%%%%%%%%%%%%%%%%%%%%%%%%%%%%%%%%%%
\end{enumerate}

%%%%%%%%%%%%%%%%%%%%%%%%%
%%%%%%%%%%%%%%%%%%%%%%%%%%%%
%%%%%%%%%%%%%%%%%%%%%%%%%%%%
%%%%%%%%%%%%%%%%%%%%%%%%%%%%

\item
   $P \approx \aprs{P' \ltens P''}$  
   such that the bottom most rule instance is given with: 

$$
\dernote{\reslazyswir}{\quad .}
    {\pars{\aprs{R \ltens T} \lpar  \aprs{P' \ltens P''}}}
    {\leaf{ \aprs{\pars{\aprs{R \ltens T} \lpar P'} \ltens P''}
}}
\\[8pt]
$$
By Proposition \ref{proposition:multiplicative:conjunction}, 
we have the proofs

$$
\qquad \qquad
 \vcenter{\xy \xygraph{[]!{0;<2.2pc,0pc>:}
 {\lone}*=<14pt>{}:@{=}^<>(.5){} _<>(.5){\Pi_{1}} [d] {
       P''  }}
    \endxy}
\qquad
 \textrm{ and  }
\qquad
 \vcenter{\xy \xygraph{[]!{0;<2.2pc,0pc>:}
 {\lone}*=<14pt>{}:@{=}^<>(.5){} _<>(.5){\Pi_{2}} [d] {
       \pars{\aprs{R \ltens T} \lpar P'}
 }}
    \endxy}
\quad .
\\[8pt]
$$
%%%%%%%%%%%%%%%%%%%%%%
%%%%%%%%%%%%%%%%%%%%%%
We apply the induction hypothesis to $\Pi_2$ and obtain \\

%%%%%%%%%%%%%%%%%%%%%%
%%%%%%%%%%%%%%%%%%%%%%
%%%%%%%%%%%%%%%%%%%%%%
\begin{enumerate}
%%%%%%%%%%%%%%%%%%%%%%%
%%%%%%%%%%%%%%%%%%%%%%%
\item[(i)]
either

$$
\qquad \qquad
 \vcenter{\xy \xygraph{[]!{0;<2.2pc,0pc>:}
 {\lone}*=<14pt>{}:@{=}^<>(.5){} _<>(.5){\Pi_{3}} [d] {
       \pars{ R \lpar P'} 
 }}
    \endxy}
\qquad
 \textrm{ and  }
\qquad
 \vcenter{\xy \xygraph{[]!{0;<2.2pc,0pc>:}
 {\lone}*=<14pt>{}:@{=}^<>(.5){} _<>(.5){\Pi_{4}} [d] {
       T
 }}
    \endxy}
\quad;
\\[8pt]
$$
%%%%%%%%%%%%%%%%%%%%%%%%%%
%%%%%%%%%%%%%%%%%%%%%%%%%%
\item[(ii)]
%%%%%%%%%%%%%%%%%%%%%%%%%%
%%%%%%%%%%%%%%%%%%%%%%%%%%
or

$$
\qquad \qquad
 \vcenter{\xy \xygraph{[]!{0;<2.2pc,0pc>:}
 {\lone}*=<14pt>{}:@{=}^<>(.5){} _<>(.5){\Pi_{5}} [d] {
       \pars{ T \lpar P'} 
 }}
    \endxy}
\qquad
 \textrm{ and  }
\qquad
 \vcenter{\xy \xygraph{[]!{0;<2.2pc,0pc>:}
 {\lone}*=<14pt>{}:@{=}^<>(.5){} _<>(.5){\Pi_{6}} [d] {
       R
 }}
    \endxy}
\quad ;
\\[8pt]
$$
%%%%%%%%%%%%%%%%%%%%%%%%%%
%%%%%%%%%%%%%%%%%%%%%%%%%%
\item[(iii)]
%%%%%%%%%%%%%%%%%%%%%%%%%%
%%%%%%%%%%%%%%%%%%%%%%%%%%
or

$$
\qquad \qquad
 \vcenter{\xy \xygraph{[]!{0;<2.2pc,0pc>:}
 {\pars{ P'_1 \lpar P'_2}}*=<14pt>{}:@{=}^<>(.5){} _<>(.5){\Delta_1} [d] {
       P' 
 }}
    \endxy}
\qquad
 \textrm{, }
\qquad
 \vcenter{\xy \xygraph{[]!{0;<2.2pc,0pc>:}
 {\lone}*=<14pt>{}:@{=}^<>(.5){} _<>(.5){\Pi_{7}} [d] {
       \pars{ R \lpar P'_1} 
 }}
    \endxy}
\qquad
 \textrm{ and  }
\qquad
 \vcenter{\xy \xygraph{[]!{0;<2.2pc,0pc>:}
 {\lone}*=<14pt>{}:@{=}^<>(.5){} _<>(.5){\Pi_{8}} [d] {
       \pars{T \lpar P'_2}
 }}
    \endxy}
\quad ,
\\[8pt]
$$
%%%%%%%%%%%%%%%%%%%%%%%%%%
%%%%%%%%%%%%%%%%%%%%%%%%%%
%%%%%%%%%%%%%%%%%%%%%%%%%%
%%%%%%%%%%%%%%%%%%%%%%%%%%
\end{enumerate}
which, together with $\Pi_1$, suffice to construct the required proofs and derivations
 as in the cases above.
\qed
%%%%%%%%%%%%%%%%%%%%%%%%%
\end{enumerate}
%%%%%%%%%%%%%%%%%%%%
\end{enumerate}

%%%%%%%%%%%%%%%%%%%%%%%%%%%%%%%%%
%%%%%%%%%%%%%%%%%%%%%%%%%%%%%%%%%
%%%%%%%%%%%%%%%%%%%%%%%%%%%%%%%%%
%%%%%%%%%%%%%%%%%%%%%%%%%%%%%%%%%
%%%%%%%%%%%%%%%%%%%%%%%%%%%%%%%%%
%%%%%%%%%%%%%%%%%%%%%%%%%%%%%%%%%
%%%%%%%%%%%%%%%%%%%%%%%%%%%%%%%%%
%%%%%%%%%%%%%%%%%%%%%%%%%%%%%%%%%
%%%%%%%%%%%%%%%%%%%%%%%%%%%%%%%%%
%%%%%%%%%%%%%%%%%%%%%%%%%%%%%%%%%
%%%%%%%%%%%%%%%%%%%%%%%%%%%%%%%%%
%%%%%%%%%%%%%%%%%%%%%%%%%%%%%%%%%
%%%%%%%%%%%%%%%%%%%%%%%%%%%%%%%%%
%%%%%%%%%%%%%%%%%%%%%%%%%%%%%%%%%
%%%%%%%%%%%%%%%%%%%%%%%%%%%%%%%%%
%%%%%%%%%%%%%%%%%%%%%%%%%%%%%%%%%
%%%%%%%%%%%%%%%%%%%%%%%%%%%%%%%%%
%%%%%%%%%%%%%%%%%%%%%%%%%%%%%%%%%
%%%%%%%%%%%%%%%%%%%%%%%%%%%%%%%%%
%%%%%%%%%%%%%%%%%%%%%%%%%%%%%%%%%

\begin{thm}[Context Reduction for $\MLSdli$]
\label{theorem:cont:reduc:deep:res:swir}
For all structures $R$ and for all contexts $S\cons{\;\;}$ 
if $S\cons{R}$ is provable in $\MLSdli$ then 
\begin{enumerate}[label=(\roman*)]
\item%[(i)]
either for all structures $X$ there exist derivations

$$
\vcenter{\xy\xygraph{[]!{0;<2.2pc,0pc>:}
{ X
  }-@{=}^<>(.5){\MLSdli} _<>(.5){}[d] {
S\cons{X}
}
    }\endxy}
\qquad \textrm{ and }
\qquad 
 \vcenter{\xy\xygraph{[]!{0;<2.2pc,0pc>:}
{\lone
  }-@{=}^<>(.5){\MLSdli} _<>(.5){}[d] {
  R
}
    }\endxy}  ;
\\[8pt]
$$
\item%[(ii)]
or there exists 
a $U$ such that for all structures $X$ there exist derivations
$$
\vcenter{\xy\xygraph{[]!{0;<2.2pc,0pc>:}
{\pars{X \lpar U}
  }-@{=}^<>(.5){\MLSdli} _<>(.5){}[d] {
S\cons{X}
}
    }\endxy}
\qquad \textrm{ and }
\qquad 
 \vcenter{\xy\xygraph{[]!{0;<2.2pc,0pc>:}
{\lone
  }-@{=}^<>(.5){\MLSdli} _<>(.5){}[d] {
 \pars{R \lpar  U}
}
    }\endxy}  .
\\[8pt]
$$
\end{enumerate}
\end{thm}

\proof
By induction on the size of $S\cons{\;\;}$.
The base case is given with empty context, which is covered by (i).
There are two inductive cases as these give the only possibilities for $S\cons{\;\;}$ 
with respect to its structure:\\
\begin{enumerate}
\item
$S\cons{\;\;} \approx \aprs{S'\cons{\;\;} \ltens  P}$, for some 
$P$. 
From Proposition \ref{proposition:multiplicative:conjunction}, it follows that 
there must be  
proofs  of $S'\cons{R}$ and $P$, thus we have 
%%%%%%%%%%%%

$$
\vcenter{\xy\xygraph{[]!{0;<2.4pc,0pc>:}
{S\cons{X} 
  }-@{=}^<>(.5){} _<>(.5){\Pi}[d] {
\aprs{S'\cons{X} \ltens P}
}  }\endxy}
\quad ,
\\[8pt]
$$
%%%%%%%%%%%%
where $\Pi$ is the proof of $P$. By applying the induction hypothesis on $S'\cons{R}$, 
we can construct the desired derivations. \\
 
\item
$S\cons{\;\;} \approx \pars{S'\cons{\;\;} \lpar P}$, 
for some $P$ such that $S'\cons{\;\;}$ is not a par: 
if $S'\cons{\;\;}$ is the empty context, 
that is, $S'\cons{\lbot} = \lbot$, 
then the theorem is proved. Otherwise,  it must be that,
for some $T$, 
%%%%%%%%%%%%%%%%%%%%%%%%%%%%%%
$S'\cons{\;\;} \approx \aprs{S''\cons{\;\;} \ltens T}$. 
 By Theorem \ref{theorem:split:MLSdi}  there exist\\
\begin{enumerate}
\item
 either
$$
\vcenter{\xy\xygraph{[]!{0;<2.2pc,0pc>:}
{\lone
  }-@{=}^<>(.5){} _<>(.5){\Pi_1}[d] {
\pars{S''\cons{R}, P}
}
    }\endxy}
\qquad
\textrm{ and } 
\qquad
\vcenter{\xy\xygraph{[]!{0;<2.2pc,0pc>:}
{\lone
  }-@{=}^<>(.5){} _<>(.5){\Pi_2}[d] {
T
}
    }\endxy} \quad ;
%\\[8pt]
$$
%%%%%%%%%%%%%%%%%%%%%%%%%%%
%%%%%%%%%%%%%%%%%%%%%%%%%%%
%%%%%%%%%%%%%%%%%%%%%%%%%%%
%%%%%%%%%%%%%%%%%%%%%%%%%%%
\item
 or
$$
\vcenter{\xy\xygraph{[]!{0;<2.2pc,0pc>:}
{\lone
  }-@{=}^<>(.5){} _<>(.5){\Pi_3}[d] {
\pars{T, P}
}
    }\endxy}
\qquad
\textrm{ and } 
\qquad
\vcenter{\xy\xygraph{[]!{0;<2.2pc,0pc>:}
{\lone
  }-@{=}^<>(.5){} _<>(.5){\Pi_4}[d] {
S''\cons{R}
}
    }\endxy} \quad ;
\\[8pt]
$$
%%%%%%%%%%%%%%%%%%%%%%%%%%%
%%%%%%%%%%%%%%%%%%%%%%%%%%%
%%%%%%%%%%%%%%%%%%%%%%%%%%%
%%%%%%%%%%%%%%%%%%%%%%%%%%%
\item 
or
$$
\vcenter{\xy\xygraph{[]!{0;<2.2pc,0pc>:}
{\pars{P_1 \lpar  P_2}
  }-@{=}^<>(.5){} _<>(.5){\Delta}[d] {
P
}
    }\endxy}
\quad 
,
\quad
 \vcenter{\xy \xygraph{[]!{0;<2.2pc,0pc>:}
 {\lone}*=<14pt>{}:@{=}^<>(.5){} _<>(.5){\Pi_5} [d] {
       \pars{S''\cons{R} \lpar P_1} }}
    \endxy}
\quad
\textrm{ and }
\quad
 \vcenter{\xy \xygraph{[]!{0;<2.2pc,0pc>:}
 {\lone}*=<14pt>{}:@{=}^<>(.5){} _<>(.5){\Pi_6} [d] {
       \pars{ T \lpar  P_2 }} }
    \endxy}
\quad .
\\[8pt]
$$
\end{enumerate}
If (a) is the case, then we have 

$$
\vcenter{\xy\xygraph{[]!{0;<3.4pc,0pc>:}
{
\dernote{\unittwodown}{\quad \;}{\pars{\aprs{S''\cons{X} \ltens  \lone} \lpar  P}}
{\leaf{\pars{S''\cons{X} \lpar  P}
}}
  }-@{=}^<>(.5){} _<>(.5){\Pi_2}[d] {
\pars{\aprs{S''\cons{X} \ltens T} \lpar  P}
}
    }\endxy}
\quad ,
\\[8pt]
$$ 
which is proved by applying the induction hypothesis on proof $\Pi_1$.\\

If $(b)$ is the case, then we have 

$$
\vcenter{\xy\xygraph{[]!{0;<2.4pc,0pc>:}
{
S''\cons{X} 
  }-@{=}^<>(.5){} _<>(.5){\Delta'}[d] {
\pars{\aprs{S''\cons{X} \ltens T} \lpar  P}
}
    }\endxy}
\quad , 
\\[8pt]
$$ 
where $\Delta'$ is  the derivation delivered by 
Lemma \ref{lemma:reslazyswitch:independence}
with proof $\Pi_3$.
We prove by applying the induction hypothesis to the proof $\Pi_4$ of $S''\cons{R}$.\\

If $(c)$ is the case, then by applying the induction hypothesis to $\Pi_5$, 
we obtain a derivation $\Delta_1$,  and we can then construct the derivation

$$
\qquad \quad
\vcenter{\xy\xygraph{[]!{0;<6.6pc,0pc>:}
{\pars{X \lpar  U}
  }-@{=}^<>(.5){} _<>(.5){\Delta_1}[d] {
\vcenter{\xy\xygraph{[]!{0;<5pc,0pc>:}
{\dernote{\unittwodown}{\quad}{\pars{\aprs{S''\cons{X} \ltens \lone}\lpar  P_1}}
{\leaf{\pars{S''\cons{X} \lpar  P_1}}}
  }-@{=}^<>(.5){} _<>(.5){\Delta''}[d] {
\vcenter{\xy\xygraph{[]!{0;<2.5pc,0pc>:}
{
\pars{\aprs{S'\cons{X} \ltens  T} \lpar  P_1 \lpar  P_2}
  }-@{=}^<>(.5){} _<>(.5){\Delta}[d] {
\pars{\aprs{S''\cons{X} \ltens T} \lpar  P}
}
    }\endxy}
}
    }\endxy}
}
    }\endxy}
\qquad
\qquad \textrm{ and }
\qquad
 \vcenter{\xy \xygraph{[]!{0;<2.2pc,0pc>:}
 {\lone}*=<14pt>{}:@{=}^<>(.5){} _<>(.5){} [d] {
       \pars{R \lpar U} }}
    \endxy}
\quad ,
\\[8pt]
$$
where $\Delta''$ is the derivation delivered by 
Lemma \ref{lemma:reslazyswitch:independence}
with proof $\Pi_6$. 
\qed
%%%%%%%%%%%%%%%%%%%%%%%%%%%
\end{enumerate}

%%%%%%%%%%%%%%%%%%%%%%%%%%%%%%%%%
%%%%%%%%%%%%%%%%%%%%%%%%%%%%%%%%%
%%%%%%%%%%%%%%%%%%%%%%%%%%%%%%%%%
%%%%%%%%%%%%%%%%%%%%%%%%%%%%%%%%%
%%%%%%%%%%%%%%%%%%%%%%%%%%%%%%%%%
%%%%%%%%%%%%%%%%%%%%%%%%%%%%%%%%%
%%%%%%%%%%%%%%%%%%%%%%%%%%%%%%%%%
%%%%%%%%%%%%%%%%%%%%%%%%%%%%%%%%%
%%%%%%%%%%%%%%%%%%%%%%%%%%%%%%%%%
%%%%%%%%%%%%%%%%%%%%%%%%%%%%%%%%%
%%%%%%%%%%%%%%%%%%%%%%%%%%%%%%%%%
%%%%%%%%%%%%%%%%%%%%%%%%%%%%%%%%%
%%%%%%%%%%%%%%%%%%%%%%%%%%%%%%%%%
%%%%%%%%%%%%%%%%%%%%%%%%%%%%%%%%%

We are now ready to show that the systems $\MLS$ and $\MLSdli$ 
prove the same structures.

\begin{thm} \label{MLSdli:equivalent}
The systems $\MLSdl$ and  $\MLSdli$  are equivalent.
\end{thm}

\proof
Observe that every proof in $\MLSdli$ is also a proof in $\MLSdl$.
For the other direction, single out the upper-most instance of 
the switch rule in the $\MLSdl$ proof, which is not an instance of 
the rule $\deeplazyintswir$:

$$
\vcenter{\xy \xygraph{[]!{0;<3pc,0pc>:}
 {\lone}*=<14pt>{}:@{=}^<>(.5){\, \MLSdli} _<>(.5){} [d] {
\dernote{\mathsf{dls}}{\quad }
        {S\pars{\aprs{R \ltens T} \lpar U}}
   {\leaf{ S\aprs{ \pars{R \lpar U} \ltens  T} } }
}}
    \endxy}
\\[8pt]
$$
From Theorem \ref{theorem:cont:reduc:deep:res:swir}, for any structure $ X$,  we have\\
\begin{enumerate}
%%%%%%%%%%%%%%%%%%%%%
%%%%%%%%%%%%%%%%%%%%%
%%%%%%%%%%%%%%%%%%%%%
\item[(i)] 
%%%%%%%%%%%%%%%%%%%%%
%%%%%%%%%%%%%%%%%%%%%
either

$$
\vcenter{\xy\xygraph{[]!{0;<2.2pc,0pc>:}
{X
  }-@{=}^<>(.5){\, \MLSdli} _<>(.5){\Delta}[d] {
S\cons{X}
}}\endxy}
\qquad 
\textrm{ and  }
\qquad 
\vcenter{\xy \xygraph{[]!{0;<2.2pc,0pc>:}
 {\lone}*=<14pt>{}:@{=}^<>(.5){\, \MLSdli} _<>(.5){} [d] {
       \aprs{\pars{R \lpar U} \ltens  T}  }}
    \endxy}
\quad ;
\\[8pt]
$$

\item[(ii)] 
%%%%%%%%%%%%%%%%%%%%%
%%%%%%%%%%%%%%%%%%%%%
or

$$
\vcenter{\xy\xygraph{[]!{0;<2.2pc,0pc>:}
{\pars{X \lpar  V}
  }-@{=}^<>(.5){\, \MLSdli} _<>(.5){\Delta}[d] {
S\cons{X}
}}\endxy}
\qquad 
\textrm{ such that  }
\qquad 
\vcenter{\xy \xygraph{[]!{0;<2.2pc,0pc>:}
 {\lone}*=<14pt>{}:@{=}^<>(.5){\, \MLSdli} _<>(.5){} [d] {
       \pars{\aprs{\pars{R \lpar U} \ltens  T} \lpar V} }}
    \endxy}
\quad .
\\[8pt]
$$
%%%%%%%%%%%%%%%%%%%%%
%%%%%%%%%%%%%%%%%%%%%
%%%%%%%%%%%%%%%%%%%%%
\end{enumerate}
%%%%%%%%%%%%%%%%%%%%%
%%%%%%%%%%%%%%%%%%%%%
%%%%%%%%%%%%%%%%%%%%%
We prove case (ii) as the proof of case (i) is similar.
We apply Theorem \ref{theorem:split:MLSdi}. \\
%%%%%%%%%%%%%%%%%%%%%%%%%%%%%%%
%%%%%%%%%%%%%%%%%%%%%%%%%%%%%%%
%%%%%%%%%%%%%%%%%%%%%%%%%%%%%%%
%%%%%%%%%%%%%%%%%%%%%%%%%%%%%%% 
%% 1
\begin{enumerate}[label=(ii.\roman*)]
\item%[($ii.i.$)]
We get either

$$
\vcenter{\xy\xygraph{[]!{0;<2.2pc,0pc>:}
{\lone
  }-@{=}^<>(.5){\, \MLSdli} _<>(.5){\Pi_1}[d] {
\pars{R \lpar U \lpar V} 
}}\endxy}
\qquad 
\textrm{ and } 
\qquad
\vcenter{\xy \xygraph{[]!{0;<2.2pc,0pc>:}
 {\lone}*=<14pt>{}:@{=}^<>(.5){\, \MLSdli} _<>(.5){\Pi_2} [d] {
       T}}
    \endxy}
\quad ;
\\[8pt]
$$
%%%%%%%%%%%%%%%%%%%%%%%%%%%%%%%
%%%%%%%%%%%%%%%%%%%%%%%%%%%%%%% 
%% 2
\item%[($ii.ii.$)]
or 

$$
\vcenter{\xy\xygraph{[]!{0;<2pc,0pc>:}
{\lone
  }-@{=}^<>(.5){\, \MLSdli} _<>(.5){\Pi_3}[d] {
\pars{R \lpar U} 
}}\endxy}
\qquad 
\textrm{ and } 
\qquad
\vcenter{\xy \xygraph{[]!{0;<2pc,0pc>:}
 {\lone}*=<14pt>{}:@{=}^<>(.5){\, \MLSdli} _<>(.5){\Pi_4} [d] {
       \pars{T \lpar V} }}
    \endxy}
\quad ;
\\[8pt]
$$
%%%%%%%%%%%%%%%%%%%%%%%%%%%%%%%
%%%%%%%%%%%%%%%%%%%%%%%%%%%%%%% 
%% 3
\item%[($ii.iii.$)]
or
$$
\vcenter{\xy\xygraph{[]!{0;<2pc,0pc>:}
{\pars{K_1 \lpar  K_2}
  }-@{=}^<>(.5){\, \MLSdli} _<>(.5){\Delta_1}[d] {
V
}}\endxy}
\qquad , 
\qquad
\vcenter{\xy \xygraph{[]!{0;<2pc,0pc>:}
 {\lone}*=<14pt>{}:@{=}^<>(.5){\, \MLSdli} _<>(.5){\Pi_5} [d] {
       \pars{R \lpar U \lpar K_1} }}
    \endxy}
\qquad 
\textrm{ and } 
\qquad
\vcenter{\xy \xygraph{[]!{0;<2pc,0pc>:}
 {\lone}*=<14pt>{}:@{=}^<>(.5){\, \MLSdli} _<>(.5){\Pi_6} [d] {
       \pars{K_2 \lpar  T} }}
    \endxy}
\quad
.
\\[8pt]
$$
\end{enumerate}
We can  construct the following proofs for the cases (ii.i), (ii.ii) and (ii.iii), respectively,
where $\Delta_2$ is the derivation delivered by 
Lemma \ref{lemma:reslazyswitch:independence}
with the proof $\Pi_4$, and 
$\Delta_3$ is the derivation delivered by 
Lemma \ref{lemma:reslazyswitch:independence}
with the proof $\Pi_6$. 

$$
\vcenter{\xy\xygraph{[]!{0;<6.6pc,0pc>:}
{
\lone
  }-@{=}^<>(.5){\, \MLSdli} _<>(.5){\Pi_1}[d] {
\vcenter{\xy\xygraph{[]!{0;<5pc,0pc>:}
{
\dernote{\unittwodown}{\quad\;}{\pars{\aprs{R \ltens \lone} \lpar U \lpar V}}
{ \leaf{\pars{R  \lpar U \lpar V}}}
  }-@{=}^<>(.5){\, \MLSdli} _<>(.5){\Pi_2}[d] {
\vcenter{\xy\xygraph{[]!{0;<2.4pc,0pc>:}
{
\pars{\aprs{R \ltens T} \lpar  U \lpar  V}
  }-@{=}^<>(.5){\, \MLSdli} _<>(.5){\Delta}[d] {
 S\pars{\aprs{R \ltens T} \lpar U}
}}\endxy}
}}\endxy}
}}\endxy}
%%%%%%%%%%%%%%%%%%%%%
%%%%%%%%%%%%%%%%%%%%%
\qquad \;
%%%%%%%%%%%%%%%%%%%%%
%%%%%%%%%%%%%%%%%%%%%
\vcenter{\xy\xygraph{[]!{0;<6.6pc,0pc>:}
{
\lone
  }-@{=}^<>(.5){\, \MLSdli} _<>(.5){\Pi_3}[d] {
\vcenter{\xy\xygraph{[]!{0;<5pc,0pc>:}
{
\dernote{\unittwodown}{\quad\,}{\pars{\aprs{R \ltens \lone}  \lpar U}}
{\leaf{
 \pars{R  \lpar U}
}}
  }-@{=}^<>(.5){\, \MLSdli} _<>(.5){\Delta_2}[d] {
\vcenter{\xy\xygraph{[]!{0;<2.4pc,0pc>:}
{
\pars{\aprs{R \ltens T} \lpar  U \lpar  V}
  }-@{=}^<>(.5){\, \MLSdli} _<>(.5){\Delta}[d] {
 S\pars{\aprs{R \ltens T} \lpar U}
}}\endxy}
}}\endxy}
}}\endxy}
%%%%%%%%%%%%%%%%%%%%%
%%%%%%%%%%%%%%%%%%%%%
\qquad \;
%%%%%%%%%%%%%%%%%%%%%
%%%%%%%%%%%%%%%%%%%%%
\vcenter{\xy \xygraph{[]!{0;<7.6pc,0pc>:}
 {\lone}*=<14pt>{}:@{=}^<>(.5){\, \MLSdli} _<>(.5){\Pi_5} [d] {
\vcenter{\xy\xygraph{[]!{0;<6.2pc,0pc>:}
{\dernote{\unittwodown}{\quad\,}{\pars{\aprs{R \ltens \lone} \lpar  U \lpar  K_1}}
{\leaf{
\pars{R \lpar  U \lpar  K_1}}}
  }-@{=}^<>(.5){\, \MLSdli} _<>(.5){\Delta_3}[d] {
\vcenter{\xy\xygraph{[]!{0;<4pc,0pc>:}
{
 \pars{\aprs{R \ltens T} \lpar U \lpar K_1 \lpar K_2}
  }-@{=}^<>(.5){\, \MLSdli} _<>(.5){\Delta_1}[d] {
\vcenter{\xy\xygraph{[]!{0;<2.4pc,0pc>:}
{
\pars{\aprs{R \ltens T} \lpar  U \lpar  V}
  }-@{=}^<>(.5){\, \MLSdli} _<>(.5){\Delta}[d] {
 S\pars{\aprs{R \ltens T} \lpar U}
}}\endxy}
}}\endxy}
}}\endxy}
}}
    \endxy}
\\[8pt]
$$
We repeat the  procedure above inductively until all the instances 
of the rule $\mathsf{dls}$, which are not instances of the rule 
$\deeplazyintswir$, are removed from the proof.
\qed

\begin{cor}
\label{corollary:equivalent} %%%
The systems $\MLS$, $\MLSu$, $\MLSl$, $\MLSd$, $\MLSdl$  and $\MLSdli$ are equivalent.
\end{cor}

\proof
First apply 
Proposition \ref{proposition:deep:switch},
and then Theorem \ref{MLSdli:equivalent}.
\qed

%%%%%%%%%%%%%%%%%%%%%%%%%%%%%%%%%%%%%%
%%%%%%%%%%%%%%%%%%%%%%%%%%%%%%%%%%%%%%

\begin{cor}
\label{corollary:equivalent} %%%%
The systems 
$\MLSu$, $\MLSl$, $\MLSd$,  $\MLSi$, 
$\MLSdl$, $\MLSli$, $\MLSdi$  and $\MLSdli$ are equivalent.
\end{cor}

\proof
Given that every instance of the rule $\deeplazyintswir$
is an instance of the rules
$\mathsf{dis}$,
$\mathsf{lis}$ and
$\mathsf{is}$, the conditions (2) and (3)
in Definition \ref{definition:deep:lazy:int:swir} 
can be removed from $\MLSdli$ 
without losing completeness. 
\qed

\begin{rem}
Lemma \ref{lemma:reslazyswitch:independence},
Theorem \ref{theorem:split:MLSdi},
Theorem \ref{theorem:cont:reduc:deep:res:swir}
and Theorem \ref{MLSdli:equivalent} 
can be proved by using the same argument also for 
the systems $\MLSi$, $\MLSdi$ and $\MLSli$. 
\end{rem}

%%%%%%%%%%%%%%%%%%%%%%%%%%%%%%%%%%%%%%
%%%%%%%%%%%%%%%%%%%%%%%%%%%%%%%%%%%%%%
%%%%%%%%%%%%%%%%%%%%%%%%%%%%%%%%%%%%%%
%%%%%%%%%%%%%%%%%%%%%%%%%%%%%%%%%%%%%%
%%%%%%%%%%%%%%%%%%%%%%%%%%%%%%%%%%%%%%
%%%%%%%%%%%%%%%%%%%%%%%%%%%%%%%%%%%%%%
%%%%%%%%%%%%%%%%%%%%%%%%%%%%%%%%%%%%%%
%%%%%%%%%%%%%%%%%%%%%%%%%%%%%%%%%%%%%%
%%%%%%%%%%%%%%%%%%%%%%%%%%%%%%%%%%%%%%
%%%%%%%%%%%%%%%%%%%%%%%%%%%%%%%%%%%%%%
%%%%%%%%%%%%%%%%%%%%%%%%%%%%%%%%%%%%%%
%%%%%%%%%%%%%%%%%%%%%%%%%%%%%%%%%%%%%%
%%%%%%%%%%%%%%%%%%%%%%%%%%%%%%%%%%%%%%
%%%%%%%%%%%%%%%%%%%%%%%%%%%%%%%%%%%%%%
%%%%%%%%%%%%%%%%%%%%%%%%%%%%%%%%%%%%%%
%%%%%%%%%%%%%%%%%%%%%%%%%%%%%%%%%%%%%%
%%%%%%%%%%%%%%%%%%%%%%%%%%%%%%%%%%%%%%
%%%%%%%%%%%%%%%%%%%%%%%%%%%%%%%%%%%%%%
%%%%%%%%%%%%%%%%%%%%%%%%%%%%%%%%%%%%%%
%%%%%%%%%%%%%%%%%%%%%%%%%%%%%%%%%%%%%%
%%%%%%%%%%%%%%%%%%%%%%%%%%%%%%%%%%%%%%
%%%%%%%%%%%%%%%%%%%%%%%%%%%%%%%%%%%%%%
%%%%%%%%%%%%%%%%%%%%%%%%%%%%%%%%%%%%%%
%%%%%%%%%%%%%%%%%%%%%%%%%%%%%%%%%%%%%%
%%%%%%%%%%%%%%%%%%%%%%%%%%%%%%%%%%%%%%
%%%%%%%%%%%%%%%%%%%%%%%%%%%%%%%%%%%%%%
%%%%%%%%%%%%%%%%%%%%%%%%%%%%%%%%%%%%%%
%%%%%%%%%%%%%%%%%%%%%%%%%%%%%%%%%%%%%%
%%%%%%%%%%%%%%%%%%%%%%%%%%%%%%%%%%%%%%
%%%%%%%%%%%%%%%%%%%%%%%%%%%%%%%%%%%%%%
%%%%%%%%%%%%%%%%%%%%%%%%%%%%%%%%%%%%%%
%%%%%%%%%%%%%%%%%%%%%%%%%%%%%%%%%%%%%%
%%%%%%%%%%%%%%%%%%%%%%%%%%%%%%%%%%%%%%

\subsection{Interaction and Cut Elimination}
\label{subsection:cut:elimination}

The restrictions that we have imposed on the switch rule, which is the rule responsible for 
context management, succeed in controlling the nondeterminism  in proof search. In the following, 
we show that these restrictions do not sacrifice a clean proof theory, as they permit 
an independent cut elimination theorem, which does not rely on the unrestricted system $\MLS$.
Splitting theorem (Theorem \ref{theorem:split:MLSdi}) 
provides a general procedure for 
proving cut-elimination in many deep inference 
systems \cite{Gug02}. 
However, the restrictions that we impose on the 
switch rule require special care.  In order to obtain a constructive proof of 
cut elimination for $\MLSdli$, we need the following lemma.

\begin{lem}  \label{lemma:cut:elimination}
For any contexts $S_1\cons{\;\;}$ and $S_2\cons{\;\;}$, 
if $S_1\cons{\lone}$ and $S_2\cons{\lone}$  are provable in $\MLSdli$ then 
$\pars{S_1\cons{a} \lpar  S_2\cons{\neg{a}}}$ is provable in $\MLSdli$.
\end{lem}

\proof
We prove with induction on the size of $\S_1\cons{\lone}$.
There are the following possibilities for the structure of  
$S_1\cons{\;\;}$.
%%%%%%%%%%%%%%%
%%%%%%%%%%%%%%%
%%%%%%%%%%%%%%%
%%%%%%%%%%%%%%%
%%%%%%%%%%%%%%%
\begin{itemize}
%%%%%%%%%%%%%%%%%%%%%%%%%%%%%%%%%%%%%%
%%%%%%%%%%%%%%%%%%%%%%%%%%%%%%%%%%%%%%
%%%%%%%%%%%%%%%%%%%%%%%%%%%%%%%%%%%%%%
%%%%%%%%%%%%%%%%%%%%%%%%%%%%%%%%%%%%%%
\item
If, for some structure $P$, $S_1\cons{\;\;} \approx \aprs{\cons{\;\;} \ltens P}$, 
by Proposition \ref{proposition:multiplicative:conjunction},
we have that $P$ must have a proof $\Pi$ in $\MLSdli$. We construct 
the following proof:

$$
\vcenter{\xy \xygraph{[]!{0;<6.6pc,0pc>:}
 {\lone}*=<14pt>{}:@{=}^<>(.5){} _<>(.5){} [d] {
%%%%%%%%%%%%%%%%%%%
%%%%%%%%%%%%%%%%%%%
%%%%%%%%%%%%%%%%%%%
\vcenter{\xy\xygraph{[]!{0;<5.4pc,0pc>:}
{  
\dernote{\ruleaidown}{\qquad}{
S_2\pars{a \lpar \neg{a}}}
{\leaf{S_2\cons{\lone}}}
  }-@{=}^<>(.5){\; \{ \deeplazyintswir\}} _<>(.5){}[d] {
%%%%%%%%%%%%%%%%%%%%%%%%%%
%%%%%%%%%%%%%%%%%%%%%%%%%%
\vcenter{\xy\xygraph{[]!{0;<3.2pc,0pc>:}
{
\dernote{\unittwodown}{\quad \;}{
\pars{ \aprs{a \ltens \lone}  \lpar  S_2\cons{\neg{a}}}}
{\leaf{\pars{ a  \lpar  S_2\cons{\neg{a}}}}}
  }-@{=}^<>(.5){} _<>(.5){\Pi}[d] {
\pars{ \aprs{a \ltens P}  \lpar S_2\cons{\neg{a}}}
}
    }\endxy}
%%%%%%%%%%%%%%%%%%%%%%%%%%%
%%%%%%%%%%%%%%%%%%%%%%%%%%%
}
    }\endxy}
%%%%%%%%%%%
%%%%%%%%%%%
%%%%%%%%%%%
}
    }\endxy}
\\[8pt]
$$

\item
%%%%%%%%%%%%%%%%%%%%%%%%%%%%%%%%%%%%%%
%%%%%%%%%%%%%%%%%%%%%%%%%%%%%%%%%%%%%%
%%%%%%%%%%%%%%%%%%%%%%%%%%%%%%%%%%%%%%
%%%%%%%%%%%%%%%%%%%%%%%%%%%%%%%%%%%%%%
If, for some structure $P$  and context $S_1'$, $S_1\cons{\;\;} \approx \aprs{P \ltens  S'_1\cons{\;\;}}$,
by Proposition \ref{proposition:multiplicative:conjunction},
we have that $P$ has a proof $\Pi$ in $\MLSdli$. 
We construct

$$
\vcenter{\xy \xygraph{[]!{0;<4.4pc,0pc>:}
 {\lone}*=<14pt>{}:@{=}^<>(.5){} _<>(.5){\Pi'} [d] {
%%%%%%%%%%%%%%%%%%%
%%%%%%%%%%%%%%%%%%%
%%%%%%%%%%%%%%%%%%%
\vcenter{\xy\xygraph{[]!{0;<3.2pc,0pc>:}
{
\dernote{\unittwodown}{\quad\;}{\pars{\aprs{\lone \ltens  S'_1\cons{a} } \lpar  S_2\cons{\neg{a}}}}{
\leaf{\pars{ S'_1\cons{a}  \lpar  S_2\cons{\neg{a}}}
}} 
 }-@{=}^<>(.5){} _<>(.5){\Pi}[d] {
\pars{ \aprs{P \ltens  S'_1\cons{a} }  \lpar  S_2\cons{\neg{a}}}
}
    }\endxy}
%%%%%%%%%%%
%%%%%%%%%%%
%%%%%%%%%%%
}
    }\endxy}
\\[8pt]
$$
where proof $\Pi'$ is delivered by the induction hypothesis.\\
%%%%%%%%%%%%%%%%%%%%%%%%%%%%%%
%%%%%%%%%%%%%%%%%%%%%%%%%%%%%%
%%%%%%%%%%%%%%%%%%%%%%%%%%%%%%

\item
If, for some structure $P$, $S_1\cons{\;\;} \approx \pars{\cons{\;\;} \lpar  P}$ 
then we can then construct the proof

$$
\vcenter{\xy \xygraph{[]!{0;<5.8pc,0pc>:}
 {\lone}*=<14pt>{}:@{=}^<>(.5){} _<>(.5){\Pi_1} [d] {
%%%%%%%%%%%%%%%%%%%
%%%%%%%%%%%%%%%%%%%
\vcenter{\xy \xygraph{[]!{0;<4.4pc,0pc>:}
 {\pars{P \lpar \lone}}*=<14pt>{}:@{=}^<>(.5){} _<>(.5){\Pi_2} [d] {
%%%%%%%%%%%%%%%%%%%
%%%%%%%%%%%%%%%%%%%
%%%%%%%%%%%%%%%%%%%
\vcenter{\xy\xygraph{[]!{0;<3pc,0pc>:}
{
\dernote{\ruleaidown}{\;\,}{
\pars{ P \lpar  S_2\pars{a \lpar \neg{a}}}}
{\leaf{\pars{ P \lpar  S_2\cons{\lone}}}}
  }-@{=}^<>(.5){\{ \deeplazyintswir \}} _<>(.5){}[d] {
\quad \, \pars{a \lpar P \lpar  S_2\cons{\neg{a}}}
}
    }\endxy}
%%%%%%%%%%%
%%%%%%%%%%%
%%%%%%%%%%%
}
    }\endxy}
%%%%%%%%%%%
%%%%%%%%%%%
}
    }\endxy}
\\[8pt]
$$
where proofs $\Pi_1$ and $\Pi_2$
are the proofs of $S_1\cons{\lone}$  and $S_2\cons{\lone}$. \\

\item
If  $S_1\cons{\;\;} \approx \pars{P \lpar  S'_1\cons{\;\;}}$, for some structure $P$, 
we can assume that  $S'_1\cons{\;\;} \approx \aprs{Q \ltens S''_1\cons{\;\;}}$, 
because otherwise we have one of the previous cases. 
We apply Theorem \ref{theorem:split:MLSdi}
to the proof of $\pars{P \lpar \aprs{Q \ltens S''_1\cons{\lone}}}$ and we obtain\\[20pt]
%%%%%%%%%
%%%%%%%%%

\begin{enumerate}
%%%%%%%%%%
\item
either

$$
\vcenter{\xy \xygraph{[]!{0;<2.2pc,0pc>:}
 {\lone}*=<14pt>{}:@{=}^<>(.5){} _<>(.5){\Pi_1} [d] {
\pars{Q \lpar P}
}}
    \endxy}
\qquad
 \textrm{ and }
\qquad
\vcenter{\xy \xygraph{[]!{0;<2.2pc,0pc>:}
 {\lone}*=<14pt>{}:@{=}^<>(.5){} _<>(.5){\Pi_2} [d] {
S''_1\cons{\lone}
}}
    \endxy}
\quad ;
\\[8pt]
$$
%%%%%%
%%%%%%
\item
or

$$
\vcenter{\xy \xygraph{[]!{0;<2.2pc,0pc>:}
 {\lone}*=<14pt>{}:@{=}^<>(.5){} _<>(.5){\Pi_3} [d] {
Q
}}
    \endxy}
\qquad \textrm{ and }
\qquad
\vcenter{\xy \xygraph{[]!{0;<2.2pc,0pc>:}
 {\lone}*=<14pt>{}:@{=}^<>(.5){} _<>(.5){\Pi_4} [d] {
\pars{S''_1\cons{\lone} \lpar P}
}}
    \endxy}
\quad ;
\\[8pt]
$$
%%%%%%%
%%%%%%%
\item
or

$$
\vcenter{\xy\xygraph{[]!{0;<2.2pc,0pc>:}
{\pars{P_1 \lpar  P_2}
  }-@{=}^<>(.5){} _<>(.5){\Delta}[d] {
P
}
    }\endxy}
\qquad ,
\qquad
\vcenter{\xy \xygraph{[]!{0;<2.2pc,0pc>:}
 {\lone}*=<14pt>{}:@{=}^<>(.5){} _<>(.5){\Pi_5} [d] {
\pars{Q \lpar P_1}
}}
    \endxy}\; \ \textrm{ and }
\qquad
\vcenter{\xy \xygraph{[]!{0;<2.2pc,0pc>:}
 {\lone}*=<14pt>{}:@{=}^<>(.5){} _<>(.5){\Pi_6} [d] {
\pars{S''_1\cons{\lone} \lpar P_2}
}}
    \endxy}
\quad.
\\[8pt]
$$
\end{enumerate}
%%%%%%%%%%%
%%%%%%%%%%%
We can then construct the following derivations for the cases $(1)$ and $(2)$

$$
\vcenter{\xy\xygraph{[]!{0;<4.6pc,0pc>:}
{
\lone
  }-@{=}^<>(.5){} _<>(.5){\Pi_7}[d] {
%%%%%%%%%%%%%%%%%%%%%%%%%%%%
%%%%%%%%%%%%%%%%%%%%%%%%%%%%
\vcenter{\xy\xygraph{[]!{0;<3.4pc,0pc>:}
{
\dernote{\unittwodown}{\quad}{\pars{\aprs{S''_1\cons{a} \ltens \lone } \lpar  S_2\cons{\neg{a}}}}{
\leaf{
\pars{S''_1\cons{a} \lpar  S_2\cons{\neg{a}}}
}} 
 }-@{=}^<>(.5){} _<>(.5){\Delta'}[d] {
%%%%%%%%%%%%%%%%%%%%%%%%
\pars{P \lpar  \aprs{Q \ltens S''_1\cons{a}} \lpar  S_2\cons{\neg{a}}}
%%%%%%%%%%%%%%%%%%%%%%%%
}
    }\endxy}
%%%%%%%%%%%%%%%%%%%%%%%%%%
%%%%%%%%%%%%%%%%%%%%%%%%%%
}
    }\endxy}
%%%%%%%%%%%%%%%%%%%%%%%%%%%
%%%%%%%%%%%%%%%%%%%%%%%%%%%
%%%%%%%%%%%%%%%%%%%%%%%%%%%
%%%%%%%%%%%%%%%%%%%%%%%%%%%
%%%%%%%%%%%%%%%%%%%%%%%%%%%
%%%%%%%%%%%%%%%%%%%%%%%%%%%
\qquad
%%%%%%%%%%%%%%%%%%%%%%%%%%%
%%%%%%%%%%%%%%%%%%%%%%%%%%%
%%%%%%%%%%%%%%%%%%%%%%%%%%%
%%%%%%%%%%%%%%%%%%%%%%%%%%%
%%%%%%%%%%%%%%%%%%%%%%%%%%%
%%%%%%%%%%%%%%%%%%%%%%%%%%%
\vcenter{\xy\xygraph{[]!{0;<4.8pc,0pc>:}
{
\lone
  }-@{=}^<>(.5){} _<>(.5){\Pi_8}[d] {
%%%%%%%%%%%%%%%%%%%%%%%%%%%%
%%%%%%%%%%%%%%%%%%%%%%%%%%%%
\vcenter{\xy\xygraph{[]!{0;<3.4pc,0pc>:}
{
\dernote{\unittwodown}{\quad \;}{\pars{P \lpar  \aprs{\lone \ltens S''_1\cons{a} }\lpar  S_2\cons{\neg{a}}}}
{\leaf{
\pars{P \lpar  S''_1\cons{a} \lpar  S_2\cons{\neg{a}}}
}}
  }-@{=}^<>(.5){} _<>(.5){\Pi_3}[d] {
%%%%%%%%%%%%%%%%%%%%%%%%
\pars{P \lpar  \aprs{Q \ltens S''_1\cons{a}} \lpar  S_2\cons{\neg{a}}}
%%%%%%%%%%%%%%%%%%%%%%%%
}
    }\endxy}
%%%%%%%%%%%%%%%%%%%%%%%%%%
%%%%%%%%%%%%%%%%%%%%%%%%%%
}
    }\endxy}
\\[8pt]
$$
where $\Delta'$ is the derivation delivered by 
Lemma \ref{lemma:reslazyswitch:independence} 
with proof $\Pi_1$ and 
the proofs $\Pi_7$ and $\Pi_8$ are delivered by the induction hypothesis.
For the case (3), we can construct the following proof 

$$
\vcenter{\xy \xygraph{[]!{0;<6pc,0pc>:}
 {\lone}*=<14pt>{}:@{=}^<>(.5){} _<>(.5){\Pi_9} [d] {
%%%%%%%%%%%%%%%%%%%%%
%%%%%%%%%%%%%%%%%%%%%
%%%%%%%%%%%%%%%%%%%%%
\vcenter{\xy\xygraph{[]!{0;<4.6pc,0pc>:}
{
\dernote{\unittwodown}{\quad}{\pars{P_2 \lpar \aprs{\lone \ltens S''_1\cons{a}} \lpar  S_2\cons{\neg{a}}}}
{\leaf{
\pars{P_2 \lpar S''_1\cons{a} \lpar  S_2\cons{\neg{a}}}
}} 
 }-@{=}^<>(.5){} _<>(.5){\Delta''}[d] {
%%%%%%%%%%%%%%%%%%%%%%%%%%%%
%%%%%%%%%%%%%%%%%%%%%%%%%%%%
\vcenter{\xy\xygraph{[]!{0;<2.6pc,0pc>:}
{\pars{P_1 \lpar P_2 \lpar  \aprs{Q  \ltens S''_1\cons{a}} \lpar  S_2\cons{\neg{a}}}
  }-@{=}^<>(.5){} _<>(.5){\Delta}[d] {
%%%%%%%%%%%%%%%%%%%%%%%%
\pars{P \lpar  \aprs{Q \ltens S''_1\cons{a}} \lpar  S_2\cons{\neg{a}}}
%%%%%%%%%%%%%%%%%%%%%%%%
}
    }\endxy}
%%%%%%%%%%%%%%%%%%%%%%%%%%
%%%%%%%%%%%%%%%%%%%%%%%%%%
}
    }\endxy}
%%%%%%%%%%%%%%%%%%%%%%%%%%%
%%%%%%%%%%%%%%%%%%%%%%%%%%%
%%%%%%%%%%%%%%%%%%%%%%%%%%%
}
    }\endxy}
\\[8pt]
$$
where $\Delta''$ is the derivation delivered by 
Lemma \ref{lemma:reslazyswitch:independence} 
with proof $\Pi_5$ and 
the proof $\Pi_9$ is delivered by the induction hypothesis.
\qed
\end{itemize}

%%%%%%%%%%%%%%%%%%%%%%%%%%%%%%%%%%%%%%
%%%%%%%%%%%%%%%%%%%%%%%%%%%%%%%%%%%%%%
%%%%%%%%%%%%%%%%%%%%%%%%%%%%%%%%%%%%%%
%%%%%%%%%%%%%%%%%%%%%%%%%%%%%%%%%%%%%%
%%%%%%%%%%%%%%%%%%%%%%%%%%%%%%%%%%%%%%
%%%%%%%%%%%%%%%%%%%%%%%%%%%%%%%%%%%%%%
%%%%%%%%%%%%%%%%%%%%%%%%%%%%%%%%%%%%%%
%%%%%%%%%%%%%%%%%%%%%%%%%%%%%%%%%%%%%%
%%%%%%%%%%%%%%%%%%%%%%%%%%%%%%%%%%%%%%
%%%%%%%%%%%%%%%%%%%%%%%%%%%%%%%%%%%%%%
%%%%%%%%%%%%%%%%%%%%%%%%%%%%%%%%%%%%%%
%%%%%%%%%%%%%%%%%%%%%%%%%%%%%%%%%%%%%%
%%%%%%%%%%%%%%%%%%%%%%%%%%%%%%%%%%%%%%
%%%%%%%%%%%%%%%%%%%%%%%%%%%%%%%%%%%%%%
%%%%%%%%%%%%%%%%%%%%%%%%%%%%%%%%%%%%%%
%%%%%%%%%%%%%%%%%%%%%%%%%%%%%%%%%%%%%% 
%%%%%%%%%%%%%%%%%%%%%%%%%%%%%%%%%%%%%%

\begin{lem}
\label{lemma:uniqueness}
Let  $S\pars{a \lpar a}$ be a structure such that atoms $a$ and $\neg a$  
do not occur in $S\cons{\;\;}$. 
$S\pars{a \lpar \neg a}$ has a proof  in $\MLSdli$  if and only if 
$\pars{a \lpar S\cons{ \neg a}}$ has a proof.
 \end{lem}

\proof
For the proof of the if part, 
we have the derivation 

$$
\vcenter{\xy\xygraph{[]!{0;<2.4pc,0pc>:}
{S\pars{a \lpar \neg a}
  }-@{=}^<>(.5){\{\deeplazyintswir\}} _<>(.5){}[d] {
%%%%%%%%%%%%%%%%%%%%%%%%
\pars{a \lpar S\cons{ \neg a}}
%%%%%%%%%%%%%%%%%%%%%%%%
}
    }\endxy}\;.
\\[8pt]
%%%%%%%%%%%%%%%%%%%%%%%%%%
%%%%%%%%%%%%%%%%%%%%%%%%%%
$$
For the only if part,  we construct a proof of $S\pars{a \lpar \neg a}$ from 
the proof $\Pi$ of $\pars{a \lpar S\cons{\neg a}}$ by inductive case analysis, 
which results in removing the instances of the rule $\deeplazyintswir$. 
\qed

%%%%%%%%%%%%%%%%%%%%%%%%%%%%%%%%%%%%%%
%%%%%%%%%%%%%%%%%%%%%%%%%%%%%%%%%%%%%%
%%%%%%%%%%%%%%%%%%%%%%%%%%%%%%%%%%%%%%
%%%%%%%%%%%%%%%%%%%%%%%%%%%%%%%%%%%%%%
%%%%%%%%%%%%%%%%%%%%%%%%%%%%%%%%%%%%%%
%%%%%%%%%%%%%%%%%%%%%%%%%%%%%%%%%%%%%%
%%%%%%%%%%%%%%%%%%%%%%%%%%%%%%%%%%%%%%
%%%%%%%%%%%%%%%%%%%%%%%%%%%%%%%%%%%%%%
%%%%%%%%%%%%%%%%%%%%%%%%%%%%%%%%%%%%%%
%%%%%%%%%%%%%%%%%%%%%%%%%%%%%%%%%%%%%%

\begin{thm}[Cut Elimination]
\label{theorem:cut:elimination}
The rule $\ruleaiup$ is admissible for $\MLSdli$.
\end{thm}

\proof
We single out the top-most instance of the rule $\ruleaiup$.

$$
\vcenter{\xy \xygraph{[]!{0;<3.2pc,0pc>:}
 {\lone}*=<14pt>{}:@{=}^<>(.5){\MLSdli} _<>(.5){\Pi_1} [d] {
\dernote{\ruleaiup}{}{S\cons{R}}{
\leaf{S\pars{R \lpar \aprs{a \ltens \neg{a}}}}}
}}
    \endxy}
\\[8pt]
$$
We first apply  
Theorem \ref{theorem:cont:reduc:deep:res:swir}
and then Theorem \ref{theorem:split:MLSdi} and obtain 

$$
\vcenter{\xy\xygraph{[]!{0;<2.4pc,0pc>:}
{\pars{S_1 \lpar  S_2}
  }-@{=}^<>(.5){} _<>(.5){\Delta}[d] {
S\cons{R}
}
    }\endxy}
\qquad  ,
\qquad
\vcenter{\xy \xygraph{[]!{0;<2.4pc,0pc>:}
 {\lone}*=<14pt>{}:@{=}^<>(.5){} _<>(.5){\Pi_1} [d] {
\pars{a \lpar S_1}
}}
    \endxy}
\qquad 
\textrm{ and }
\qquad
\vcenter{\xy \xygraph{[]!{0;<2.4pc,0pc>:}
 {\lone}*=<14pt>{}:@{=}^<>(.5){} _<>(.5){\Pi_2} [d] {
\pars{\neg{a} \lpar S_2}
}}
    \endxy}
\quad  .
\\[8pt]
$$
In order for a structure to have a proof in $\MLSdli$, that structure must contain pairs of dual atoms.
Because there are the proofs $\Pi_1$ and $\Pi_2$, 
it follows that we find  $S'_1\cons{\;\;}$ and $S'_2\cons{\;\;}$,
such that $S_1 = S'_1\cons{\neg{a}}$ and $S_2 = S'_2\cons{a}$,
where we can uniquely mark $a$ and $\neg a$. From Lemma \ref{lemma:uniqueness}, 
it follows that there are proofs 
$$
\vcenter{\xy \xygraph{[]!{0;<2.2pc,0pc>:}
 {\lone}*=<14pt>{}:@{=}^<>(.5){} _<>(.5){\Pi_3} [d] {
S'_1\cons{\lone}
}}
    \endxy} 
\qquad  
\textrm{ and }
\qquad
\vcenter{\xy \xygraph{[]!{0;<2.2pc,0pc>:}
 {\lone}*=<14pt>{}:@{=}^<>(.5){} _<>(.5){\Pi_4} [d] {
S'_2\cons{\lone}
}}
    \endxy}
\quad .
$$
We can then construct 

$$
\vcenter{\xy \xygraph{[]!{0;<3.2pc,0pc>:}
 {\lone}*=<14pt>{}:@{=}^<>(.5){} _<>(.5){\Pi_3} [d] {
%%%%%%%%%%%%%%%%%%%%%%
%%%%%%%%%%%%%%%%%%%%%%
\vcenter{\xy\xygraph{[]!{0;<2.2pc,0pc>:}
{\pars{S'_1\cons{\neg{a}} \lpar  S'_2\cons{a}}
  }-@{=}^<>(.5){} _<>(.5){\Delta}[d] {
S\cons{R}
}
    }\endxy}
%%%%%%%%%%%%%%%%%%%%%%%
%%%%%%%%%%%%%%%%%%%%%%%
}}
    \endxy}
\\[8pt]
$$
where $\Pi_3$ is delivered by 
Lemma \ref{lemma:cut:elimination}
from the proofs $\Pi_1$ and $\Pi_2$.
By repeating this procedure inductively, starting from the top, 
we  remove one by one all the instances of the rule $\ruleaiup$.
\qed

\begin{rem}
For the case of the systems $\MLSi$, $\MLSdi$, $\MLSli$ and $\MLSdli$, 
it is not possible to give a decomposition of proofs \cite{StraTh,SG11},
 where the instances of the rule $\ruleaidown$ are permuted above the instances
of the switch rule 
as the structure in Example \ref{example:there:are:cases} demonstrates.
In order to prove this
structure, in bottom-up proof construction,
we need an instance of rule $\ruleaidown$ prior to any instance of the rule 
$\mathsf{is}$.  This is because the unit $\bot$ requires its dual structure 
$\lone$, with which it can interact. This can be delivered only by an instance 
of the rule $\ruleaidown$, and only after this, an instance of $\mathsf{is}$ 
can be applied.
However, in $\BV$, which extends multiplicative linear logic
with the rules mix and nullary mix \cite{Gug02,Kah08}, such a decomposition 
result can be given, because the units can be omitted in $\BV$, and 
this kind of behavior cannot emerge.
\end{rem}

\subsection{Interaction and Proof Complexity}

We have shown that the  conditions we impose on the switch rule provide 
a more immediate access to the shorter proofs that are made available by 
deep inference. 
We now show that these conditions
 do not introduce any additional complexity as they succeed in 
polynomially simulating proofs in other formalisms such as the sequent calculus or proof nets. 
For the case of multiplicative linear 
logic this is not surprising as this logic is known to be NP-complete \cite{Ka91,LiWi94},
 thus in any reasonable formalism, providing polynomial time 
proofs of multiplicative linear logic formulae 
boils down to correctly guessing the right pairing of the atoms.

\begin{defi} 
A system $\sysS$  polynomially simulates proof system $\sysS'$
if there is a polynomial $p$ such that for every proof $\Pi$ 
 in $\sysS$ there is a proof $\Pi'$ in 
$\sysS'$ of the same formula such that $|\Pi'| \leq p(|\Pi|)$.
\end{defi}

\begin{cor}
\label{corollary:p:simulates}
The system $\MLSdli$ polynomially simulates any proof system for multiplicative linear logic.
\end{cor}

\proof
From Corollary \ref{corollary:equivalent} we have that system $\MLSdli$ is 
complete for multiplicative linear logic.  
From Proposition \ref{proposition:MLL:inNP} it follows that the size of 
any proof in system $\MLSdli$ is bounded by $\mathcal{O} ({| \occ \, R |}^2 )$.
\qed

\begin{rem}
Corollary \ref{corollary:p:simulates} sets a polynomial bound on the size of the $\MLSdli$ proofs, 
however it does not provide a constructive procedure for translating proofs in other formalisms. 
This can be achieved by using the existing proof as an oracle for 
the construction of the deep inference proof. For 
this, it suffices to label the formula such that this labeling results in 
pairwise distinctness of the atoms with respect to 
the pairing of the dual atoms at the axioms.
On these structures, we can use the restriction of the rule $\deeplazyintswir$ 
that we call 
%\emph{confluent} $\deeplazyintswir$ (denoted with 
$\mathsf{cdlis}$
%)
such that

$$
        \vcenter{\dernote{\mathsf{dlis}}{\quad }
        {S\pars{U \lpar \aprs{R \ltens T}   }}
        {\leaf{S\aprs{\pars{U \lpar  R} \ltens T}}}}
\\[8pt]
$$
is an instance of $\mathsf{cdlis}$ if  $\at\, \overline{T} \, \cap \,   \at\, U = \emptyset$. 
By applying the rules $\unitonedown$, $\unittwodown$, $\ruleaidown$ and $\mathsf{cdlis}$
exhaustively on the labeled structure, we can construct the desired proof as all these rules preserve the 
relation $\downarrow$ between dual atoms.
However, it is important to note that the rule $\mathsf{cdlis}$  is not complete for generic structures
as illustrated by the structure $\pars{a \lpar  a \lpar \aprs{\neg{a} \ltens \neg{a}}}$.
\end{rem}

\subsection{Relation with Focusing}

Focusing as a technique was introduced by Andreoli \cite{And92,And01} 
to structure the sequent calculus proofs for 
controlling the nondeterminism in proof search. 
Focusing is based on the intuition that the sequent rules 
have certain permutative affinities, which can be exploited in proof search
by separating the inference rules into 
groups of synchronous and asynchronous rules.
Asynchronous rules, which are invertible,
are then applied first whenever they are applicable, as 
they do not require back-tracking in proof search.
By assigning polarities to the inference rules that manage  
different logical operators, inference rules of the same polarity are 
applied always in groups.  In the context of multiplicative linear logic,
this machinery has the implication that $\lpar$ connective is 
mapped to meta-level sequents prior to handling of any $\ltens$
connective, and any atom with a positive polarity is paired immediately 
with a negative atom at an instance of an identity rule that 
closes the sequent calculus proof branch. 
As a result of this,  proof 
construction is separated into alternating deterministic 
and non-deterministic phases.

Synthetic connectives \cite{Cha08} as a method, which operates within 
focusing technique, uses an idea similar to  deep and lazy conditions in 
Definition \ref{definition:deep:lazy:int:swir}, however within the 
sequent calculus setting. Given that focusing technique aims at 
processing connectives of the same polarity until only 
subformulae of the opposite polarities are left, this idea is based on  
considering $n$-ary connectives, whose immediate subformulae are 
necessarily of the opposite polarity.  
For example, in a synthetic tensor $A_1 \ltens...\ltens A_n$, each $A_i$ 
cannot itself be a tensor; it must be either a par, or an atom.  
The deep and lazy conditions of the rule $\deeplazyintswir$ 
impose similar  alternations, however
within the deep inference setting.  While focusing in the sequent calculus must 
process connectives at the top level, deep inference may operate at any depth. 
In particular, deep lazy switch can be applied at the highest possible depth, 
directly where the subformulae of opposite polarity are to be found.

Focusing technique has been adapted to the deep inference proofs in 
 linear logic \cite{CGS12}. In the deep inference setting, because 
there is no distinction between meta-level 
and object level, the invertible inference rules that map the logical operators 
to sequents become redundant. Apart from this, in the focused 
deep inference system, 
a mechanism that treats the units explicitly by means of inference rules 
as in this paper is used.  This makes it possible to remove the units 
$\lone$ and $\bot$
whenever it is possible to apply the corresponding  inference 
rules that are invertible. The rest of 
the nondeterminism is annotated with syntactic expressions that
makes the considered synchronous connectives explicit, so that backtracking 
can be managed to enter another focused choice in case of a bad decision.

\begin{rem}
\label{similar:to:focusing}
In Definition \ref{definition:deep:lazy:int:swir}, the restrictions  
$(2)$ and $(3)$ implement a mechanism that is similar to 
the focusing technique. 
This is because the restriction $(2)$
does not allow the structure $U$ to be $\lbot$ in an instance of $\deepreslazyswir$, 
and this leaves the rule $\unitonedown$ as the only applicable rule to the 
$\lbot$ structures that are connected  with par.  
Similarly, the restriction $(3)$
does not allow the structure $R$ to be $\lone$ in an instance of $\deepreslazyswir$, 
which leaves the rule $\unittwodown$ as the only applicable rule to the 
$\lone$ structures that are connected  with copar.  This introduces a mechanism 
that results in the removal of the units $\lone$ and $\lbot$ during proof construction
whenever it is possible, since the rules $\unitonedown$ and $\unittwodown$,
are the only applicable rules on these structures.
\end{rem}

Besides the similarity described above and the similarity addressed 
in Remark \ref{similar:to:focusing},
the focusing approach presented in \cite{CGS12} is orthogonal to the 
approach of this paper, thus 
focusing technique and our technique
should be applicable in a complementary manner. 
It should therefore be possible to benefit from both approaches 
for obtaining the shorter proofs that are made available by deep inference.

%%%%%%%%%%%%%%%%%%%%%%%%%%%%%%%%%%%%%%%%%%%%%%%%%%%%
%%%%%%%%%%%%%%%%%%%%%%%%%%%%%%%%%%%%%%%%%%%%%%%%%%%%
%%%%%%%%%%%%%%%%%%%%%%%%%%%%%%%%%%%%%%%%%%%%%%%%%%%%
 
\section{Interaction and Depth in Classical Logic}
\label{section:int:depth:KSg}

Deep inference systems  for different logics follow a common scheme where  context management 
in proof construction is performed by  the switch rule. In this section,  we show that
the ideas above can be carried over to classical logic, that is,
the switch rule  can be replaced with the 
deep lazy interaction switch  rule 
without loosing completeness in a 
deep inference system \cite{BruTh} for classical logic.

\begin{defi} \label{definition:KSg:structures}   
There are countably many 
\emph{atomic propositions} which are 
denoted by $a$, $b$, $c$,\ldots 
Classical logic \emph{structures} are generated by
$$
  R \grammareq 
  \false \, \mid \, \true \, \mid \, 
   a \mid \;
   \pars{\, R\, \lor \, R \, } \; \mid \; 
   \aprs{\, R\, \land \, R \, } \; \mid \; 
  \neg{R} 
$$
where $\false$ and $\true$ are the units false and true, 
respectively.
\index{disjunction}
$ \pars{R \lor R}$ is a \emph{disjunction} and  
\index{conjunction}
$ \aprs{R \land R}$ is a \emph{conjunction}. 
$\neg{R}$ is the \emph{negation} of the structure $R$.
Classical logic structures are considered equivalent modulo
relation $\approx$, which is the smallest congruence relation 
induced by the equational system consisting of the equations
for associativity and commutativity for disjunction and conjunction 
and De Morgan equations for negation.
While writing the classical logic structures we apply the same 
conventions as those for multiplicative linear logic structures:
we denote the structures in the same equivalence class   
by picking a  structure from the equivalence class. 
If there is no ambiguity, when writing the structures, 
we drop the superfluous brackets by resorting 
to the equations for associativity.
\end{defi}

\begin{rem} 
As for multiplicative linear logic structures, 
we consider classical logic structures in 
negation normal form by applying the De Morgan equations 
for negation to push the negation symbol to the 
atoms. Because the bottom-up instances of the inference rules, defined below,
 do not introduce new negation symbols, 
we  consider the classical logic structures to be in negation 
normal form, and we remove the De Morgan equations 
from the relation $\approx$. 
\end{rem}

\renewcommand{\rulebox}[1]{\mbox{$#1$}}
\begin{figure}[t]
%    \fbox{
%      \parbox{0.99\textwidth}{
%%%%%%%%%%%%%%%%%%%%%%%%%%%%%%%%%%%%%%%%%%%%%%
\begin{center}
{
\small
\framebox[5.2in]
{
\begin{minipage}[t]{5.2in}

\begin{center}
\vspace{2mm}
%%%%%%%%%%%%%%%%%%%%%%%%%%%%%%%%%%%%%%%%%%%%%%%
$$
\begin{array}{cccc}
\dernote{\ruleaidown}{\;  }
        {S\pars{a \lor \bar a}}
        {\leaf{S\cons{\true}}}
         \quad & \quad 
\dernote{\swir}{\; }
        {S\pars{U \lor \aprs{R \land T} }}
        {\leaf{S\aprs{\pars{U \lor R} \land T}}}
        \quad & \quad
\dernote{\rulewdown}{\;  \textrm{} }
        {S\cons{R}}
        {\leaf{S\cons{\false}}}
         \quad & \quad
\dernote{\rulecdown}{\; }
        {S\cons{R}}
        {\leaf{S\pars{R \lor R}}}
\\[12pt]
\dernote{\unitonedown\!}{}{S\pars{R \lor \false}}{\leaf{S\cons{R}}}
&
\dernote{\unittwodown\!}{}{S\aprs{R \land \true}}{\leaf{S\cons{R}}}
&
\dernote{\unitthreedown\!}{}{S\pars{ \true \lor \true}}{\leaf{S\cons{\true}}}
&
\dernote{\unitfourdown\!}{}{S\aprs{\false \land \false}}{\leaf{S\cons{\false}}}
\end{array}
$$
%%%%%%%%%%%%%%%%%%%%%%%%%%%%%%%%%%%%%%%%%%%%%%%%%%
\vspace{2mm}

\end{center}
\end{minipage}
}  }

\end{center}
%%%%%%%%%%%%%%%%%%%%%%%%%%%%%%%%%%%%%%%%%%%%%%%%%%%
%        }   }
 %   \vspace{-2mm}        
    \caption{Deep Inference System $\KSu$}
    \label{figure:KSu}
%\vspace{-4mm}
\end{figure}

\begin{defi}    \label{System:KSg}
The system $\KSu$ for classical logic is the system depicted in Figure \ref{figure:KSu}. 
The rules in the upper row, which we borrow from \cite{BruTh}, 
are called \emph{atomic interaction},
\emph{switch},
\emph{weakening} and
\emph{contraction}.
The rules in the bottom row are the invertible rules corresponding to 
the equations for unit given in \cite{BruTh}. We call these rules 
unit one, unit two, unit three and unit four. 
\end{defi}

\begin{defi}
A \emph{proof} $\Pi$ in  $\KSu$
is a finite  derivation whose premise is the unit $\true$.
The size $|\Pi|$ of a proof
$\Pi$ is the number of unit and atom occurrences appearing in it.
\end{defi}

\begin{defi} \label{definition:switch:KSg:systems}
\emph{System} $\KSu$ \emph{with deep lazy interaction switch}, or 
$\KSdli$ 
is the  system  obtained by replacing the switch rule with the rule 
$\deeplazyintswir$ as defined in Definition \ref{definition:deep:lazy:int:swir}, 
however here copar is replaced with conjunction and par is replaced with disjunction. 
We define the other systems obtained by replacing in  $\KSu$ 
the switch rule with one of the rules given in 
Figure \ref{figure:switch} similarly by adding the prefix of the switch 
rule as a suffix to the name of the system. For example,
the \emph{system} $\KSu$ \emph{with deep interaction switch}, or 
$\KSdi$ 
is the  system that replaces the rule $\swir$ in $\KSu$ with the rule $\deepintswir$. 
\end{defi}

In the following,  we show that proofs in $\KSu$ 
can be constructed as   proofs  consisting of separate  
phases such that in each phase only certain inference rules are used. 
For this purpose, we use a semantic argument, 
similar to the one given in \cite{BruTh}. Due to the conjunctive normal forms 
that we obtain, this also provides a completeness  argument for $\KSu$.
The proofs are given by first simulating 
the construction of the sequent calculus proofs, as 
in the left-most derivation below, and then obtaining their 
permutations that are  available due to deep inference.
The decomposed structure of the proofs obtained this way 
becomes  instrumental  in showing that $\KSdli$ is complete.

\begin{thm}  \label{theorem:semantic:cut:elim}
If a structure $R$ has a proof in $\KSu$, then there exist structures
$R_1$, $R_2$, $R_3$, $R_4$, $R_1'$, $R_2'$, and $R_3'$ 
and proofs of the following forms. In other words, every structure 
that has a proof in $\KSu$ has also proofs consisting of different phases 
with distinct inference rules as depicted below, where a proof 
phase can be empty, that is, there can be proofs with  
no instance of a particular rule.

$$
\begin{array}{c}
%%%%%%%%%%%%%
%%%%%%%%%%%%%
%%%%%%%%%%%%%
%%%%%%%%%%%%%
\vcenter{\xy\xygraph{[]!{0;<1.1pc,0pc>:}
 {                             \true
 }-@{=}^<>(.5){  
                \, \{ \, \unitonedown \, , \, \unittwodown \}  
                                  } _<>(.5){ \Delta_4}
                                   [dd] {
                                   R_4
                               }
         -@{=}^<>(.5){ 
             \,  \{ \,  \rulewdown\,  \} 
                       } _<>(.5){  \Delta_3}
                                 [dd] {
                                 R_3
                                    }
         -@{=}^<>(.5){ 
             \,  \{ \, \ruleaidown\,  \} 
                       } _<>(.5){  \Delta_2}
                                 [dd] {
                                 R_2
                                 }
         -@{=}^<>(.5){\, 
         \{ \, \swir  , \,  \rulecdown \,\}
             }_<>(.5){\Delta_1} [dd] {R}
       }\endxy} 
\end{array}
\;
\stackrel{i.}{\leadsto}
\quad \; 
\begin{array}{c}
\vcenter{\xy\xygraph{[]!{0;<1.1pc,0pc>:}
 {                             \true
 }-@{=}^<>(.5){  
                \,  \{ \, \unitonedown \, , \, \unittwodown \}  
                                  } _<>(.5){ \Delta_4}
                                   [dd] {
                                   R_4
                                 }
         -@{=}^<>(.5){ 
             \,  \{ \, \rulewdown\,  \} 
                       } _<>(.5){  \Delta_3}
                                 [dd] {
                                 R_3
                                    }
         -@{=}^<>(.5){ 
             \,  \{ \, \ruleaidown\,  \} 
                       } _<>(.5){  \Delta_2}
                                 [dd] {
                                 R_2
                                 }
         -@{=}^<>(.5){ 
             \,  \{ \, \swir \,  \} 
                       } _<>(.5){\Delta_{1,b}}
                                 [dd] {
                                 R_1
                                 }
         -@{=}^<>(.5){\, 
         \{ \, \rulecdown \,\}
             }_<>(.5){\Delta_{1,a}} [dd] {R}
       }\endxy} 
\end{array}
\;
\stackrel{ii.}{\leadsto}
\quad \; 
\begin{array}{c}
\vcenter{\xy\xygraph{[]!{0;<1.1pc,0pc>:}
 {                             \true
 }-@{=}^<>(.5){  
                \,  \{ \, \unitonedown \, , \, \unittwodown \}  
                                  } _<>(.5){ \Delta_4}
                                   [dd] {
                                   R_4
                                 }
         -@{=}^<>(.5){ 
             \,  \{ \, \ruleaidown\,  \} 
                       } _<>(.5){  \Delta_2}
                                 [dd] {
                                 R'_3
                                    }
         -@{=}^<>(.5){ 
             \,  \{ \, \rulewdown\,  \} 
                       } _<>(.5){  \Delta_3}
                                 [dd] {
                                 R_2
                                 }
         -@{=}^<>(.5){ 
             \,  \{ \, \swir \,  \} 
                       } _<>(.5){\Delta_{1,b}}
                                 [dd] {
                                 R_1
                                 }
         -@{=}^<>(.5){\, 
         \{ \, \rulecdown \,\}
             }_<>(.5){\Delta_{1,a}} [dd] {R}
       }\endxy} 
\end{array}
\;
\stackrel{iii.}{\leadsto}
\quad \; 
\begin{array}{c}
\vcenter{\xy\xygraph{[]!{0;<1.1pc,0pc>:}
 {                             \true
 }-@{=}^<>(.5){  
                \,  \{ \, \unitonedown \, , \, \unittwodown \}  
                                  } _<>(.5){ \Delta_4}
                                   [dd] {
                                   R_4
                                 }
         -@{=}^<>(.5){ 
             \,  \{ \, \ruleaidown\,  \} 
                       } _<>(.5){  \Delta_2}
                                 [dd] {
                                 R'_3
                                    }
         -@{=}^<>(.5){ 
             \,  \{ \, \swir\,  \} 
                       } _<>(.5){\Delta'_{1,b} }
                                 [dd] {
                                 R'_2
                                 }
         -@{=}^<>(.5){ 
             \,  \{ \,  \rulewdown \,  \} 
                       } _<>(.5){ \Delta'_3}
                                 [dd] {
                                 R'_1
                                 }
         -@{=}^<>(.5){\, 
         \{ \, \rulecdown \,\}
             }_<>(.5){\Delta_{1,a}} [dd] {R}
       }\endxy} 
\end{array}
\\[8pt]
$$ 
\end{thm}

\proof 
We can derive the  rule, that we call \emph{distributive}($\mathsf{d}$),
as follows:

$$
\dernote{\rulecdown}{}{S\pars{\aprs{R \land T} \lor U}}{
\rootr{\swir\;}{}{S\pars{\aprs{R \land T} \lor U \lor U}}{
\rootr{\swir\;}{}{S\pars{\aprs{\pars{R \lor U} \land T} \lor U}}{
\leaf{S\aprs{\pars{R \lor U} \land \pars{T \lor U}}}}}}
\\[8pt]
$$
By applying this rule exhaustively to the structure $R$ bottom up, 
we obtain the derivation  $\Delta_1$ with the premise $R_2$, which 
is in conjunctive normal form. Because $R_2$ is provable, each disjunction
in $R_2$ must have an atom $a$ and its dual $\neg{a}$. By applying 
the rule $\ruleaidown$ bottom  up to each one of these pairs of dual atoms, 
we obtain the derivation $\Delta_2$ with the premise $R_3$, where each 
disjunction has an instance of the unit $\true$. By applying the rule 
$\rulewdown$ exhaustively to all the remaining structures in each disjunction
which are different from the unit $\true$, we obtain the derivation $\Delta_3$. 
By applying the rules $\unitonedown$ and $\unittwodown$ exhaustively
we obtain $\Delta_4$.\\

(i) 
With structural induction on $R$, we obtain the derivations
$\Delta_{1,a}$ and $\Delta_{1,b}$ from the derivation 
$\Delta_1$.
 If $R$ is an atom or the unit 
$\true$ or $\false$, then it is already in conjunctive normal 
form.  If $R = \aprs{T \land U}$ or $R = \pars{T \lor U}$ then we have 
the derivations $(1.)$ and $(2.)$ below  by induction hypothesis
where $T_2$ and $U_2$ are in conjunctive normal form.
Let $n$ be the number of disjunctions in $U_2$. We assume 
that $n$ is greater than one. Otherwise, we can exchange 
$T_2$ with $U_2$, or if in both $T_2$ and $U_2$, there are 
less than 2 disjunctions, then they are 
already in conjunctive normal form. We construct the derivations
for $R = \aprs{T \land U}$ and $R = \pars{T \lor U}$, respectively,  
as in $(3.)$ and $(4.)$ below:

$$
\begin{array}{cccc}
(1.) \qquad  & \qquad (2.) \qquad &  \quad \quad \; (3.) \qquad &  \qquad \qquad (4.) \quad \\[6pt]
%%%%%%%%%%%%%%%%%
\;
\vcenter{\xy\xygraph{[]!{0;<1.2pc,0pc>:}
         {T_2}-@{=}^<>(.5){     
                   \{ \swir  \}  
                                  }  _<>(.5){\Delta'_{T}} [dd] {\;\,T_1}
           -@{=}^<>(.5){\{ \rulecdown \}
                                  } _<>(.5){\Delta_{T}} [dd] {T}
       }\endxy}  
\qquad 
&
\qquad
\vcenter{\xy\xygraph{[]!{0;<1.2pc,0pc>:}
         {U_2}-@{=}^<>(.5){     
                   \{ \swir  \}  
                                  }  _<>(.5){\Delta'_{U}} [dd] {\;\,U_1}
           -@{=}^<>(.5){\{ \rulecdown \}
                                  } _<>(.5){\Delta_{U}} [dd] {U}
       }\endxy}  
%%%%%%%%%%%%%%%%%%%%%%%%%%
\qquad 
&
\quad 
\vcenter{\xy\xygraph{[]!{0;<1.2pc,0pc>:}
         {\aprs{T_2 \land U_2}}-@{=}^<>(.5){     
                   \{ \, \swir \,  \}  
                                  }  _<>(.5){
                                             \pars{\Delta'_{T} ,  \Delta'_{U}}
                } [dd] {\; \aprs{T_1 \land U_1} }
           -@{=}^<>(.5){\{ \, \rulecdown \, \}
                                  } _<>(.5){\pars{\Delta_{T} ,  \Delta_{U}}
               } [dd] {\aprs{T \land U}}
       }\endxy}  
%\qquad \qquad
\qquad 
&
\qquad 
\vcenter{\xy\xygraph{[]!{0;<1.2pc,0pc>:}
 {R_2}-@{=}^<>(.5){  
                \, \{ \, \swir \, \}  
                                  } _<>(.5){}
                                   [dd] {\;\pars{T_2 \lor \ldots \lor T_2 \lor U_2}  }
         -@{=}^<>(.5){ 
             \,  \{ \, \swir \,  \} 
                       } _<>(.5){\pars{\Delta'_{T} , \ldots , \Delta'_{T} , \Delta'_{U}}}
                                 [dd] {\;\pars{T_1 \lor \ldots \lor T_1 \lor U_1} }
         -@{=}^<>(.5){ 
             \,  \{ \, \rulecdown \,  \} 
                       } _<>(.5){\pars{\Delta_{T} ,  \ldots , \Delta_{T} , \Delta_{U}}}
                                 [dd] {\;\pars{
             % \underbrace{
             T \lor \ldots \lor T
            %}_{\times \, n} 
                         ,U}}
         -@{=}^<>(.5){\, \{ \, \rulecdown \,\}} [dd] {\pars{T \lor U}}
       }\endxy} 
\end{array}
\\[8pt]
$$ 

(ii) We trivially permute each instance of  $\rulewdown$ under the instances 
of  $\ruleaidown$.\\

(iii)
We  permute the instances of the rule $\swir$ over the 
rule $\rulewdown$: Other cases being trivial, we consider the following:\\
\begin{enumerate}[label=\({\alph*}]
%%%%%%%%%%%%%%%%%
\item%[$(a.)$] 
The redex is of $\weakr$ is inside the contractum of $\swir$.
%%%%%%%%%%%%%%%%%

$$
\vcenter{
  \dernote{\swir}{}{S\rdx{\pars{\aprs{R \land T} \lor U}}}{
  \rootr{\weakr}{}{S\aprs{\rdx{\pars{R \lor U}} \land T}}{
  \leaf{S\aprs{\false \land T}}
  }}
}
\qquad
\stackrel{a.}{\leadsto}
\qquad
\vcenter{
  \dernote{\weakr}{}{S\pars{\aprs{R \land T} \lor \rdx{U}}}{
  \rootr{\weakr}{}{S\aprs{\rdx{R} \land T}}{
  \leaf{S\aprs{\false \land T}}
  }}
}
\\[8pt]
$$
%%%%%%%%%%%%%%%%%
\item%[$(b.)$] 
The contractum of $\swir$ is inside the redex of $\weakr$.
%%%%%%%%%%%%%%%%%

$$
\vcenter{
  \dernote{\swir}{}{S\aprs{\pars{\aprs{R \land T} \lor U} \land P}}{
  \rootr{\weakr}{}{S\aprs{\pars{R \lor U} \land T \land P}}{
  \leaf{S\aprs{\pars{R \lor U} \land \false}}
  }}
}
\qquad
 \stackrel{b.}{\leadsto}
\qquad
\vcenter{
  \dernote{\weakr}{}{S\aprs{\pars{\aprs{R \land T} \lor  U} \land  P}}{
  \rootr{\weakr}{}{ S\aprs{\pars{\aprs{R \land T} \lor  U} \land  \false} }{
   \rootr{\swir}{}{ S\aprs{\pars{\aprs{R \land \false} \lor  U} \land \false} }{
     \leaf{S\aprs{\pars{R \lor U} \land \false}}
  }}}
}\eqno{\qEd}
\\[8pt]
$$
%%%%%%%%%%%%%%%%%
%\qed
\end{enumerate}\newpage

The proof scheme given by the theorem above simulates the construction 
of the sequent calculus proofs, where
in the worst case, an exponentially branching proof tree is generated. 
The transformation into conjunctive normal form provides a strategy for proof 
construction, and this results in  an exponential cost for some classes of
formulae. However, besides these proofs that simulate sequent calculus, deep inference 
provides proofs that are exponentially shorter than those  that are 
available in the sequent calculus \cite{BG08}.
Below we use 
Theorem \ref{theorem:semantic:cut:elim}
as a tool for proving that  $\KSdli$ is complete
by exploiting the intermediate stages between proof phases.

\begin{thm} \label{theorem:deep:KSgdli}
The systems $\KSu$ and $\KSdli$ are equivalent.
\end{thm}

\proof
Every proof in $\KSdli$ is a proof in $\KSu$.
For the other direction, we have by Theorem \ref{theorem:semantic:cut:elim} that 
if a structure $R$ has a proof in $\KSu$, it has proof of the form on the left below.
In proof $\Pi$, we apply the map 

$$
\{ \,
\land \leftrightarrow \ltens \, , \; 
\lor \leftrightarrow \lpar \, , \; 
\true \leftrightarrow \lone \, , \;
\false \leftrightarrow \bot \, \}
\\[8pt]
$$ 
and we transform $\Pi$ into the proof $\Pi'$ on the right by applying Theorem 
\ref{MLSdli:equivalent}.

$$
\begin{array}{c}
\vcenter{\xy\xygraph{[]!{0;<1.1pc,0pc>:}
 {                             \true
 }-@{=}^<>(.5){  
                \,  \{ \, \ruleaidown \, , \,\swir \, , \,\unitonedown \, , \, \unittwodown \}  
                                  } _<>(.5){ \Pi}
                                   [dd] {
                                   R_1
                                 }
         -@{=}^<>(.5){\, 
         \{ \, \rulecdown \, , \,\rulewdown \,\}
             }_<>(.5){\Delta} [dd] {R}
       }\endxy} 
\end{array}
\qquad
\leadsto
\qquad
\begin{array}{c}
\vcenter{\xy\xygraph{[]!{0;<1.1pc,0pc>:}
 {                             \true
 }-@{=}^<>(.5){  
                \,  \{ \, \ruleaidown \, , \,\deeplazyintswir \, , \,\unitonedown \, , \, \unittwodown \}  
                                  } _<>(.5){ \Pi'}
                                   [dd] {
                                   R_1
                                 }
         -@{=}^<>(.5){\, 
         \{ \, \rulecdown \, , \,\rulewdown \,\}
             }_<>(.5){\Delta} [dd] {R}
       }\endxy} 
\end{array}\eqno{\qEd}
\\[8pt]
$$ 
%\qed

\begin{cor}
The systems $\KSu$, $\KSd$, 
$\KSl$, $\KSi$, 
$\KSdi$, $\KSli$, 
$\KSdl$ and $\KSdli$ are equivalent.
\end{cor}

%%%%%%%%%%%%%%%%%%%%%%%%%%%%
%%%%%%%%%%%%%%%%%%%%%%%%%%%%
%%%%%%%%%%%%%%%%%%%%%%%%%%%%
%%%%%%%%%%%%%%%%%%%%%%%%%%%%
%%%%%%%%%%%%%%%%%%%%%%%%%%%%
%%%%%%%%%%%%%%%%%%%%%%%%%%%%
%%%%%%%%%%%%%%%%%%%%%%%%%%%%
%%%%%%%%%%%%%%%%%%%%%%%%%%%%
%%%%%%%%%%%%%%%%%%%%%%%%%%%%
%%%%%%%%%%%%%%%%%%%%%%%%%%%%
%%%%%%%%%%%%%%%%%%%%%%%%%%%%
%%%%%%%%%%%%%%%%%%%%%%%%%%%%
%%%%%%%%%%%%%%%%%%%%%%%%%%%%
%%%%%%%%%%%%%%%%%%%%%%%%%%%%

We have shown that $\KSdli$ is complete for classical logic. 
We conclude the discussion on classical logic by showing that $\KSdli$ 
polynomially simulates the sequent calculus and also provides 
exponentially shorter proofs for certain formulae.

\begin{defi}
The sequent-calculus proof system \textsf{Analytic Gentzen}
 is defined by the inference rules in Figure \ref{figure:Gentzen}, where $\phi$ and $\psi$ 
stand for multisets of formulae and the symbol `,' represents the multiset union. 
We interpret multisets of formulae
as their disjunction. Derivations in \textsf{Analytic Gentzen}, denoted by $\Phi$, are
trees obtained by composing instances of these inference rules. 
The leaves of a derivation are
its premises and the root is its conclusion. 
A derivation with no premises is a proof. The size $|\Phi|$ of a derivation
$\Phi$ is the number of unit and atom occurrences appearing in it. 
\end{defi}

\begin{nota}
For convenience, we denote the structures and the formulae with the same notation. 
However, the congruence relation in Definition \ref{definition:KSg:structures} 
applies only to structures, and they do not hold 
within the proofs in \textsf{Analytic Gentzen} system.
\end{nota}\smallskip

\renewcommand{\rulebox}[1]{\mbox{$#1$}}
\begin{figure}[b]
%    \fbox{
%      \parbox{0.99\textwidth}{
%%%%%%%%%%%%%%%%%%%%%%%%%%%%%%%%%%%%%%%%%%%%%%
\begin{center}
{
\small
\framebox[5.4in]
{
\begin{minipage}[t]{5.4in}

\begin{center}
\vspace{2mm}
%%%%%%%%%%%%%%%%%%%%%%%%%%%%%%%%%%%%%%%%%%%%%%%
$$
\begin{array}{ll}
  \rulebox{\inf{\axiom}{\vdash A, \neg A} {}}
\quad \; \;
 \rulebox{\inf{\true}{\sqn{\true}}{} }
\quad \; \;
 \rulebox{\inf{\mathsf{w}}{\sqn{\phi, A}}{\sqn{\phi}} }
\quad \; \;
 \rulebox{\inf{\mathsf{c}}{\sqn{\phi, A}}{\sqn{\phi, A, A}} }
 \quad \; \; 
 \rulebox{\inf{\lor}{\sqn{\phi, \pars{A \lor B}}}{\sqn{\phi, A, B}} }
\quad \; \;
\rulebox{\iinf{\land} {\sqn{ \phi,  \aprs{A \land B}, \psi}}
  {\sqn{\phi, A}}{\sqn{B, \psi}}} 
\end{array}
$$
%%%%%%%%%%%%%%%%%%%%%%%%%%%%%%%%%%%%%%%%%%%%%%%%%%
\vspace{2mm}

\end{center}
\end{minipage}
}  }

\end{center}
%%%%%%%%%%%%%%%%%%%%%%%%%%%%%%%%%%%%%%%%%%%%%%%%%%%
    \caption{\textsf{Analytic Gentzen} system in the sequent calculus.}
    \label{figure:Gentzen}
\end{figure}

\begin{thm}
\label{theorem:KSu:poly}
For every \textsf{Analytic Gentzen} proof $\Phi$ with  conclusion $P$,
there is a proof $\Pi$ in $\KSu$ such that \\
\begin{enumerate}[label=(\roman*)]
\item%[i]
if $n$ is the size of $\Phi$,
the size of $\Pi$ is $\mathcal{O}(n^2)$;
\item%[ii]
and $\Pi$ is of the following form:

$$
\begin{array}{c}
\vcenter{\xy\xygraph{[]!{0;<1.1pc,0pc>:}
 {                             \true
 }-@{=}^<>(.5){  
                \,  \{ \, \ruleaidown \, , \,\swir \, , \,\unitonedown \, , \, \unittwodown \}  
                                  } _<>(.5){ \Pi}
                                   [dd] {
                                   R_1
                                 }
         -@{=}^<>(.5){\, 
         \{ \, \rulecdown \, , \,\rulewdown \,\}
             }_<>(.5){\Delta} [dd] {R}
       }\endxy} 
\end{array}
\\[8pt]
$$
\end{enumerate}
\end{thm}

\proof
(i) is as in the proof of a more a general result in \cite{BG08}, and  it proceeds by induction on the tree structure of $\Phi$. 
The base case is given with the translation of the rule $\mathsf{id}$ as in the proof of Theorem \ref{theorem:equivalent:MLS:MLSu}
and the translation of the instances of  rule $\true$ in Figure \ref{figure:Gentzen}
to the instances of the rule $\unittwodown$. The inductive cases are translated into derivations in 
$\KSu$ as in \cite{BG08}, where the rules $\unitonedown$ and $\unittwodown$ are explicitly 
applied when required, and the size of the resulting proof is given by the larger cases as $\mathcal{O}(n^2)$.  
For the proof of (ii), we permute up all the instances of the rules $\swir$, $\ruleaidown$, $\unitonedown$
and $\unittwodown$ as in Theorem \ref{theorem:semantic:cut:elim}, 
whereby the size of the proof in (i) is preserved by these permutations.
\qed

\begin{cor}
The system $\KSdli$ polynomially simulates \textsf{Analytic Gentzen} system.
\end{cor}

\proof
We first apply Theorem \ref{theorem:KSu:poly} and then Corollary \ref{corollary:p:simulates} 
as in the proof of  Theorem \ref{theorem:deep:KSgdli}.
\qed

\begin{rem}
In \cite{BG08}, Bruscoli and Guglielmi show that Statman 
tautologies have quadratic size proofs in the size of the proved 
tautologies in $\KSu$,
in contrast to their exponential size proofs in the sequent calculus. 
It is straight-forward to see that $\KSdli$ preserves these 
quadratic size proofs of Statman tautologies. 
\end{rem}

\begin{rem}
By replacing the rule $\rulecdown$ and $\deeplazyintswir$ in  
$\KSdli$ with the rule derived by combining these rules as in

$$
   \vcenter{
         \dernote{\rulecdown}{}
        {S\pars{U \lor \aprs{R \land T}   }}
        {\rootr{\deeplazyintswir \;}{}{S\pars{U \lor U \lor \aprs{R \land T}   }}{
\leaf{S\pars{\aprs{\pars{U \lor  R} \land T}  \lor U }}}}}
\qquad 
\leadsto
\qquad
        \vcenter{\dernote{\rulecdown ; \, \deeplazyintswir }{}
        {S\pars{U \lor \aprs{R \land T}   }}
        {\leaf{S\pars{\aprs{\pars{U \lor  R} \land T}  \lor U }}}}
\\[8pt]
$$
and integrating the rule $\rulewdown$ to the rule $\ruleaidown$ as in

$$
   \vcenter{
         \dernote{\ruleaidown}{}
        {S\pars{a \lor \neg a \lor R   }}
        {\rootr{\rulewdown \;}{}{S\pars{\true \lor R   }}{
         \rootr{\unitonedown \;}{}{S\pars{\true \lor \false   }}{
\leaf{
S\cons{\true}
}}}}}
\qquad 
\leadsto
\qquad
   \vcenter{
         \dernote{\ruleaidown ; \, \rulewdown ; \, \unitonedown}{\quad ,}
        {S\pars{a \lor \neg a \lor R   }}{
\leaf{
S\cons{\true}
}}}
\\[8pt]
$$
we obtain a complete deep inference system for  
classical logic, which consists of only invertible rules. 
\end{rem}

%%%%%%%%%%%%%%%
%%%%%%%%%%%%%%%
\section{Discussion}

%%%%%%%%%%%%%%%%%%%%%%%%%%%%%%%%
%%%%%%%%%%%%%%%%%%%%%%%%%%%%%%%%
%%%%%%%%%%%%%%%%%%%%%%%%%%%%%%%%
%%%%%%%%%%%%%%%%%%%%%%%%%%%%%%%%
%%%%%%%%%%%%%%%%%%%%%%%%%%%%%%%%
%%%%%%%%%%%%%%%%%%%%%%%%%%%%%%%%
%%%%%%%%%%%%%%%%%%%%%%%%%%%%%%%%
%%%%%%%%%%%%%%%%%%%%%%%%%%%%%%%%
%%%%%%%%%%%%%%%%%%%%%%%%%%%%%%%%
%%%%%%%%%%%%%%%%%%%%%%%%%%%%%%%%
%%%%%%%%%%%%%%%%%%%%%%%%%%%%%%%%
%%%%%%%%%%%%%%%%%%%%%%%%%%%%%%%%
%%%%%%%%%%%%%%%%%%%%%%%%%%%%%%%%
%%%%%%%%%%%%%%%%%%%%%%%%%%%%%%%%

We have introduced a  technique on deep inference systems
for reducing nondeterminism in proof search.
The restrictions that we impose on the switch rule provide a
reduction in non-determinism in proof search without sacrificing
proof theoretic cleanliness as these  restrictions 
do not break the cut elimination property.
Our technique permits the construction of proofs 
 with the capability of polynomially simulating shallow 
 inference of  the sequent calculus, 
while preserving shorter proofs that are available due 
to deep inference. For example,  polynomial size proofs 
of Statman tautologies \cite{BG08} can be built in $\KSdli$ 
in contrast to their exponential size sequent calculus proofs.

Our technique exploits an interaction pattern on the switch rule, which is 
the rule responsible for context management. This interaction pattern is
realized by a condition given in 
Definition \ref{definition:deep:lazy:int:swir}. 
In this respect, further refining the condition
of the interaction switch rule without sacrificing 
proof theoretic cleanliness  is a topic of 
future investigation, which is however difficult 
(see, for example,  Remark \ref{remark:counter:example}).

The switch rule is the rule responsible for 
context management  in deep inference systems 
for different fragments 
of  linear logic \cite{Str03a,Str02}, 
modal logics \cite{GT06,Sto07},
intuitionistic logic \cite{TiuINT05}, the logic $\BV$ \cite{Gug02}
and its extension with the exponentials of linear logic, that is, the logic $\NEL$ 
\cite{GugStr02,GS11,SG11}. We believe that the technique here 
generalizes to these other systems. For instance, 
in \cite{Kah05c,KahTh}, we have shown that in $\BV$ 
the  switch rule can be replaced with the rule $\lazyintswir$
without losing completeness. However, in  
$\BV$ and its extensions, the non-trivial interactions
between commutative and non-commutative contexts 
make it difficult to extend these ideas to non-commutative 
context management rule of this logic. 
Some preliminary ideas along these lines 
can be found in \cite{KahTh}.

The key to extending our technique to the deep inference 
systems for other logics is provided by the splitting theorem 
(Theorem \ref{theorem:split:MLSdi}). Given that splitting theorems 
for linear logic and  $\NEL$, given in \cite{StraTh,GS11}, 
can be modified to corporate the restrictions that we have introduced
on the switch rule, the technique introduced here can be carried over 
also to the setting of these logics. 
Another related topic that we leave for future work is extending 
the scope of the  interaction pattern to the 
other inference rules of these other logics. 
For example, for the case of linear logic \cite{Str03a,Str02} 
and modal logics \cite{GT06,Sto07}, the interaction 
pattern that we exploit in the rule $\deeplazyintswir$
can be used also for the other rules of these logics. For example, 
when we consider the promotion rule of linear logic ($\promr$) 
and the rule ${\mathsf{k}{\downarrow}}$ of modal logic $\mathsf{K}$
below, we observe that these rules are subject to restrictions 
with respect to the interaction patterns between the structures  
$R$ and $T$.

$$
\dernote{\promr}{}{S\pars{!R,?T}}{\leaf{S\cons{!\pars{R,T}}}} 
\qquad \qquad
\dernote{{\mathsf{k}{\downarrow}}}{}{S\pars{\Box R, \Diamond T}}{\leaf{S\cons{ \Box\pars{R,T}}}} 
 \\[8pt]
$$
Introducing an interaction condition for the instances of these
rules, that is,  $\at \, \neg{R} \cap \at \, T \neq \emptyset$, should result in a reduced 
nondeterminism for proof search in these systems.

Our long term goal is developing analytic deep inference  \cite{BG09} theorem 
provers for different logics.   Multiplicative linear logic provides an 
excellent theoretical framework for these investigations due to its 
complexity \cite{Ka91,LiWi94} and simplicity \cite{Gue99} under the same hood.  
In this respect,
the ideas presented in this paper in combination with the focusing technique 
should provide effective means to benefit from short proofs that deep 
inference makes available for different logics.

%%%%%%%%%%%%%%%%%%%%%%%%%%%%%%%%
%%%%%%%%%%%%%%%%%%%%%%%%%%%%%%%%
%%%%%%%%%%%%%%%%%%%%%%%%%%%%%%%%
%%%%%%%%%%%%%%%%%%%%%%%%%%%%%%%%
%%%%%%%%%%%%%%%%%%%%%%%%%%%%%%%%
%%%%%%%%%%%%%%%%%%%%%%%%%%%%%%%%
%%%%%%%%%%%%%%%%%%%%%%%%%%%%%%%%

\bibliographystyle{plain}
\bibliography{ozan}

\begin{thebibliography}{10}

\bibitem{And92}
Jean-Marc Andreoli.
\newblock Logic programming with focusing proofs in linear logic.
\newblock {\em J. Log. Comput.}, 2(3):297--347, 1992.

\bibitem{And01}
Jean-Marc Andreoli.
\newblock Focussing and proof construction.
\newblock {\em Ann. Pure Appl. Logic}, 107(1-3):131--163, 2001.

\bibitem{Bru03a}
Kai Br{\" u}nnler.
\newblock Atomic cut elimination for classical logic.
\newblock In M.~Baaz and J.~A. Makowsky, editors, {\em CSL 2003}, volume 2803
  of {\em LNCS}, pages 86--97. Springer, 2003.

\bibitem{BruTh}
Kai Br\"unnler.
\newblock {\em Deep Inference and Symmetry in Classical Proofs}.
\newblock PhD thesis, Technische Universit\"at Dresden, 2003.

\bibitem{Bru06b}
Kai Br{\"u}nnler.
\newblock Deep inference and its normal form of derivations.
\newblock In Arnold Beckmann, Ulrich Berger, Benedikt L{\"o}we, and John~V.
  Tucker, editors, {\em Computability in Europe 2006}, volume 3988 of {\em
  LNCS}, pages 65--74. Springer, July 2006.

\bibitem{Bru06}
Kai Br{\"u}nnler.
\newblock Locality for classical logic.
\newblock {\em Notre Dame Journal of Formal Logic}, 47(4):557--580, 2006.

\bibitem{PaolaBVL02}
Paola Bruscoli.
\newblock A purely logical account of sequentiality in proof search.
\newblock In Peter~J. Stuckey, editor, {\em Logic Programming, 18th
  International Conference}, volume 2401 of {\em Lecture Notes in Computer
  Science}, pages 302--316. Springer, 2002.

\bibitem{BG09}
Paola Bruscoli and Alessio Guglielmi.
\newblock On analyticity in deep inference.
\newblock Available on the web at \texttt{http://cs.bath.ac.uk/ag/p/ADI.pdf},
  2009.

\bibitem{BG08}
Paola Bruscoli and Alessio Guglielmi.
\newblock On the proof complexity of deep inference.
\newblock {\em ACM Transactions on Computational Logic}, 2(14):1--34, 2009.

\bibitem{CHP00}
Iliano Cervesato, Joshua~S. Hodas, and Frank Pfenning.
\newblock Efficient resource management for linear logic proof search.
\newblock {\em Theoretical Computer Science}, 232:133--163, 2000.

\bibitem{Cha08}
Kaustuv Chaudhuri.
\newblock Focusing strategies in the sequent calculus of synthetic connectives.
\newblock In {\em Proceedings of LPAR'08, Logic for Programming, Artificial
  Intelligence, and Reasoning, 15th International Conference}, volume 5330 of
  {\em LNCS}, pages 467--481. Springer, 2008.

\bibitem{CGS12}
Kaustuv Chaudhuri, Nicolas Guenot, and Lutz Stra{\ss}burger.
\newblock The focused calculus of structures.
\newblock In Marc Bezem, editor, {\em Computer Science Logic (CSL'11) - 25th
  International Workshop/20th Annual Conference of the EACSL}, volume~12, pages
  159--173. LIPICS, 2011.

\bibitem{Das11}
Anupam Das.
\newblock On the proof complexity of cut-free bounded deep inference.
\newblock In {\em Proceedings of Tableaux'11}, volume 6793 of {\em LNCS}, pages
  134--148. Springer, 2011.

\bibitem{Das12}
Anupam Das.
\newblock Complexity of deep inference via atomic flows.
\newblock In {\em Proceedings of CiE'12}, volume 7318 of {\em LNCS}, pages
  139--150. Springer, 2012.

\bibitem{GT06}
Rajeev Gor{\'e} and Alwen Tiu.
\newblock Classical modal display logic in the calculus of structures and
  minimal cut-free deep inference calculi for {S5}.
\newblock {\em Journal of Logic and Computation}, 17(4):767--794, 2007.

\bibitem{Gue99}
Stefano Guerrini.
\newblock Correctness of multiplicative proof nets is linear.
\newblock In {\em In Fourteenth Annual IEEE Symposium on Logic in Computer
  Science}, pages 454--463, 1999.

\bibitem{Gug03}
Alessio Guglielmi.
\newblock Mismatch.
\newblock Available on the web at \texttt{http://cs.bath.ac.uk/ag/p/AG9.pdf},
  2003.

\bibitem{Gug02}
Alessio Guglielmi.
\newblock A system of interaction and structure.
\newblock {\em ACM Transactions on Computational Logic}, 8(1):1--64, 2007.

\bibitem{GG08}
Alessio Guglielmi and Tom Gundersen.
\newblock Normalisation control in deep inference via atomic flows.
\newblock {\em Logical Methods in Computer Science}, 2008.
\newblock In press.

\bibitem{GGP10}
Alessio Guglielmi, Tom Gundersen, and Michel Parigot.
\newblock A proof calculus which reduces syntactic bureaucracy.
\newblock In {\em Proceedings of the International Conference on Rewriting
  Techniques and Applications 2010 (Edinburgh)}, pages 135--150. Schloss
  Dagstuhl - Leibniz-Zentrum fuer Informatik 2010 LIPIcs, 2010.

\bibitem{GGS10}
Alessio Guglielmi, Tom Gundersen, and Lutz Stra{\ss}burger.
\newblock Breaking paths in atomic flows for classical logic.
\newblock In {\em Proceedings of LICS'10}, pages 284--293. IEEE, 2010.

\bibitem{GugStr02}
Alessio Guglielmi and Lutz Stra{\ss}burger.
\newblock A non-commutative extension of {MELL}.
\newblock In M.~Baaz and A.~Voronkov, editors, {\em LPAR 2002}, volume 2514 of
  {\em LNAI}, pages 231--246. Springer, 2002.

\bibitem{GS11}
Alessio Guglielmi and Lutz Strassburger.
\newblock A system of interaction and structure {V}: The exponentials and
  splitting.
\newblock {\em Mathematical Structures in Computer Science}, 21(3):563--584,
  2010.

\bibitem{Kah08}
Ozan Kahramano{\u g}ullar{\i}.
\newblock System {BV} is {NP}-complete.
\newblock {\em Annals of Pure and Applied Logic}, 152(1--3):107--121, 2008.

\bibitem{Kah09}
Ozan Kahramano{\u g}ullar{\i}.
\newblock On linear logic planning and concurrency.
\newblock {\em Information and Computation}, 207(11):1229--1258, 2009.

\bibitem{Kah04}
Ozan {Kahramano\u gullar\i}.
\newblock System {BV} without the equalities for unit.
\newblock In {\em Proceedings of the 19th International Symposium on Computer
  and Information Sciences, ISCIS'04}, volume 3280 of {\em LNCS}, pages
  986--995, Antalya, Turkey, 2004. Springer.

\bibitem{KahTh}
Ozan {Kahramano\u gullar\i}.
\newblock {\em Nondeterminism and Language Design in Deep Inference}.
\newblock PhD thesis, TU Dresden, 2006.

\bibitem{Kah05c}
Ozan {Kahramano\u gullar\i}.
\newblock Reducing nondeterminism in the calculus of structures.
\newblock In Miki Hermann and Andrei Voronkov, editors, {\em Logic for
  Programming, Artificial Intelligence, and Reasoning, Proceedings of the 13th
  International Conference, LPAR 2006, Phnom Penh, Cambodia}, volume 4246 of
  {\em LNCS}, pages 272--286. Springer, 2006.

\bibitem{Kah07b}
Ozan {Kahramano{\u g}ullar\i}.
\newblock Maude as a platform for designing and implementing deep inference
  systems.
\newblock In {\em Proceedings of the Eighth International Workshop on
  Rule-Based Programming, {RULE'07}}, ENTCS. Elsevier, 2008.
\newblock In press.

\bibitem{Ka91}
Max Kanovich.
\newblock The multiplicative fragment of linear logic is {NP}-complete.
\newblock Technical Report X-91-13, Institute for Language, Logic, and
  Information, 1991.

\bibitem{LiWi94}
Patrick Lincoln and Timothy~C. Winkler.
\newblock Constant-only multiplicative linear logic is {NP}-complete.
\newblock In {\em Theoretical Computer Science}, volume 135(1), pages 155--169.
  1994.

\bibitem{MM96}
Narciso Mart{\'\i}-Oliet and Jos{\` e} Meseguer.
\newblock Rewriting logic as a logical and semantic framework.
\newblock In J.~Meseguer, editor, {\em Proc. 1st Internat Workshop on Rewriting
  Logic and its Application, WRLA' 96}, volume~4 of {\em Electronic Notes in
  Theoretical Computer Science}, pages 189--224. Elsevier, 1996.

\bibitem{Sta78}
Richard Statman.
\newblock Bounds for proof-search and speed-up in the predicate calculus.
\newblock {\em Annals of Mathematical Logic}, 15:225--287, 1978.

\bibitem{Sto07}
Phiniki Stouppa.
\newblock A deep inference system for the modal logic {S5}.
\newblock {\em Studia Logica}, 85(2):199--214, 2007.

\bibitem{Str02}
Lutz Stra{\ss}burger.
\newblock A local system for linear logic.
\newblock In M.~Baaz and A.~Voronkov, editors, {\em LPAR 2002}, volume 2514 of
  {\em LNAI}, pages 388--402. Springer, 2002.

\bibitem{StraTh}
Lutz Stra{\ss}burger.
\newblock {\em Linear Logic and Noncommutativity in the Calculus of
  Structures}.
\newblock PhD thesis, TU Dresden, 2003.

\bibitem{Str03a}
Lutz Stra{\ss}burger.
\newblock {MELL} in the calculus of structures.
\newblock {\em Theoretical Computer Science}, 309:213--285, 2003.

\bibitem{StraUnd}
Lutz Stra{\ss}burger.
\newblock System {NEL} is undecidable.
\newblock In Ruy De~Queiroz, Elaine Pimentel, and Luc{\'\i}lia Figueiredo,
  editors, {\em 10th Workshop on Logic, Language, Information and Computation
  (WoLLIC)}, volume~84 of {\em Electronic Notes in Theoretical Computer
  Science}, 2003.

\bibitem{SG11}
Lutz Strassburger and Alessio Guglielmi.
\newblock A system of interaction and structure {IV}: The exponentials and
  decomposition.
\newblock {\em ACM Transactions on Computational Logic}, 12(4), 2011.

\bibitem{TiuINT05}
Alwen~Fernanto Tiu.
\newblock A local system for intuitionistic logic.
\newblock In Miki Hermann and Andrei Voronkov, editors, {\em Logic for
  Programming, Artificial Intelligence, and Reasoning, Proceedings of the 13th
  International Conference, LPAR 2006, Phnom Penh, Cambodia}, volume 4246 of
  {\em LNCS}, pages 242--256. Springer, 2006.

\bibitem{TiuTh01}
Alwen~Fernanto Tiu.
\newblock A system of interaction and structure {II}: the need for deep
  inference.
\newblock {\em Logical Methods in Computer Science}, 2 (2:4):1--24, April 2006.

\end{thebibliography}
\end{document}